\documentclass[symmetry,article,accept,moreauthors,10pt,a4paper]{mdpi}

\usepackage{amsmath,amssymb}
\usepackage[T1]{fontenc}
\usepackage{bm}
\usepackage{microtype}

\usepackage{color}

\usepackage{chemfig}
\usepackage{rotating}
\usepackage{physics}
\usepackage[version=4]{mhchem}
\usepackage{tikz-3dplot}
\usepackage{pgfplots}
\usepackage{siunitx}

\usepackage{verbatim}

\DeclareMathAlphabet{\mathpzc}{OT1}{pzc}{m}{it}

\newcommand\mydots{\hbox to 1em{.\hss.\hss.}}

\newcommand{\ai}{\textit{ab initio}}

\newcommand{\vect}[1]{\boldsymbol{#1}}
\newcommand{\normvec}[1]{\langle #1 \rangle}

\newcommand{\fig}[1]{Fig. \ref{#1}}
\newcommand{\sect}[1]{Section \ref{#1}}
\newcommand{\tabl}[1]{Table \ref{#1}}

\newcommand{\hcch}{$^{12}$C$_2$H$_2$}
\newcommand{\ethane}{H$_3$$^{12}$C$^{12}$CH$_3$}

\newcommand{\Cv}[1]{{\itshape\bfseries C}$_{#1{\rm v}}$}
\newcommand{\Dh}[1]{{\itshape\bfseries D}$_{#1{\rm h}}$}

\newcommand{\Td}{{\itshape\bfseries T}$_{\rm d}$}
\newcommand{\GTS}{{\itshape\bfseries G}$_{36}$}

\newcommand{\Trove}{{\sc TROVE}}

\newcommand{\twobytwo}[4]{$\left(\begin{array}{cc} #1 & #2 \\ #3 & #4 \end{array} \right)$}
\newcommand{\zwobyzwo}[4]{\left(\begin{array}{cc} #1 & #2 \\ #3 & #4 \end{array} \right)}

\firstpage{1}
\articlenumber{x}
\doinum{10.3390/------}
\pubvolume{xx}
\pubyear{2018}
\copyrightyear{2018}
\externaleditor{Academic Editor: name}
\history{Received: date; Accepted: date; Published: date}

\Title{Transformation Properties under the Operations of  the Molecular Symmetry Groups {\itshape\bfseries G}$_{36}$ and {\itshape\bfseries G}$_{36}$(EM) of Ethane H$_3$CCH$_3$}

\Author{Thomas M. Mellor$^{1}$, Sergei N. Yurchenko$^{1}$\orcidB{},  Barry P. Mant$^{1,2}$, Per Jensen$^{3}$*\orcidA{}}

\AuthorNames{Thomas M. Mellor, Sergei N. Yurchenko, Barry Mant, Per Jensen}

\address{%
$^{1}$ \quad Department of Physics and Astronomy,
University College London,
London WC1E 6BT, UK \\
$^{2}$ \quad  Institut f{\"u}r Ionenphysik und Angewandte Physik, Universit{\"a}t Innsbruck, A-6020 Innsbruck, Tirol, Austria \\
$^3$ \quad Physikalische und Theoretische Chemie, Fakult{\"a}t f{\"u}r Mathematik und Naturwissenschaften, Bergische Universit{\"a}t Wuppertal, D-42097 Wuppertal, Germany
}

\corres{Correspondence: jensen@uni-wuppertal.de}

\abstract{
In the present work, we report a detailed description of the symmetry properties of the eight-atomic molecule ethane, with the aim of facilitating the variational calculations of rotation-vibration spectra of ethane and related molecules. Ethane consists of two methyl groups CH$_3$ where the internal rotation (torsion) of one CH$_3$ group relative to the other is of large amplitude and involves tunneling between multiple minima of the potential energy function. The molecular symmetry group of ethane is the 36-element group {\itshape\bfseries G}$_{36}$, but the construction of symmetrized basis functions is most conveniently done in terms of the 72-element extended molecular symmetry group {\itshape\bfseries G}$_{36}$(EM). This group can subsequently be used in the construction of block-diagonal matrix representations of the ro-vibrational Hamiltonian for ethane. The derived transformation matrices associated with {\itshape\bfseries G}$_{36}$(EM) have been implemented in the variational nuclear motion program TROVE (Theoretical ROVibrational Energies). TROVE variational calculations will be used as a practical example of a {\itshape\bfseries G}$_{36}$(EM) symmetry adaptation for large systems with a non-rigid, torsional degree of freedom.
We present the derivation of irreducible transformation matrices for all 36 (72) operations of {\itshape\bfseries G}$_{36}$(M) ({\itshape\bfseries G}$_{36}$(EM)) and also describe algorithms for a numerical construction of these matrices based on a set of  four (five) generators. The methodology presented is illustrated on the construction of the symmetry-adapted representations both of the potential energy function of ethane and of the rotation, torsion and vibration basis set functions.
}

\keyword{ro-vibrational;  point groups; molecular symmetry groups; ethane}

\begin{document}

	\section{Introduction}\label{sec:intro}
		
\begin{figure}[htbp!]
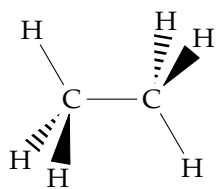

		\centering
		\chemfig{{C}([:120]-{H})([:-100]<{H})([:-130]<:{H})-C([:-60]-H)([:80]<:H)([:50]<H)}
		\caption{The structure of ethane in the staggered configuration.}
		\label{fig:ethane_equilibrium}
	\end{figure}

	In variational calculations of molecular rotation-vibration energies and wavefunctions, a matrix representation
	of the molecular rotation-vibration Hamiltonian, constructed in terms of suitable basis functions, is diagonalized
	numerically. It is well known that this type of calculation can be facilitated by the introduction of symmetrized
	basis functions, i.e., basis functions that generate irreducible representations of a symmetry group for the molecule
	in question (see, for example, Ref.~\cite{98BuJexx}). Even though the calculation of energies and wavefunctions
	can, in principle, be carried out without the use of symmetry, the subsequent simulation of molecular spectra, involving
	the computation of transition intensities, would be impossible in practice without the consideration of symmetry~\cite{18ChJeYu}.
	We implement the symmetry information in the variational calculation by defining groups of matrices that are
	isomorphic or homomorphic to the original symmetry group of the molecule and that are associated with the various
	irreducible representations (irreps) of this symmetry group~\cite{98BuJexx}. When suitable matrix group elements are known, projection
	operator techniques are employed to obtain the desired symmetrized basis functions for the variational calculation~\cite{98BuJexx}.
	In a recent paper~\cite{18ChJeYu}, we described an example of such a symmetry adaption, in that we
	presented character tables and irreducible representation transformation matrices  for \Dh{n} groups with arbitrary $n$-values.
	With these results, we could practically utilize the  linear-molecule symmetry properties described by the   \Dh{\infty} point group
	in numerical calculations with  the variational nuclear motion program {\Trove}~\cite{07YuThJe.method}, the acetylene molecule \hcch\ serving as an example.
	In the present work, we report an analogous analysis for the molecular symmetry (MS) group~\GTS\ and the extended molecular symmetry (EMS) group \GTS(EM) \cite{98BuJexx} of the ethane molecule \ethane, shown in \fig{fig:ethane_equilibrium}, with the aim of facilitating the solution of the rotation-vibration Schr\"{o}dinger equation for ethane and related molecules.

The MS group \GTS\  has been the subject of a number of studies (see, for example, \citep{81SmBexx,12CaAlSe}). The EMS group \GTS(EM)  was studied in detail  by Di Lauro and Lattanzi~\citep{10LaDixx,05LaDixx}. Examples of the application of the \GTS(EM) group include studies of the rotation-torsional spectra of various ethane-type molecules \citep{80Hougen,99LaLaxx,96LaLaVa,04LaDiHo,06LaDiHo}. \Trove\ (Theoretical ROVibrational Energies) \cite{07YuThJe.method,15YaYuxx.method} is a general variational program for computing ro-vibrational spectra and properties for small to medium-size polyatomic molecules of arbitrary structure. It has been applied to a large number of polyatomic species \cite{09YuBaYa.NH3,11YaYuJe.H2CO,14SoHeYu.PH3,15SoAlTe.PH3,14UnYuTe.SO3,15AlYuTe.H2CO,14YuTexx.CH4,15YaYuxx.method,15AlOvPo.H2O2,15OwYuYa.CH3Cl,15OwYuYa.SiH4,15AdYaYuJe.CH3,15OwYuTh.NH3,16AlPoOv.HOOH,16UnTeYu.SO2,16OwYuYa,16OwYuYa1.CH4,18OwYaTh,18MaYaTe,15OwYuYa1.SiH4,19CoYuKo,19MaChYa}, most of which have considerable symmetry [e.g., with  molecular symmetry groups~\cite{98BuJexx} such as \Cv{3}(M), \Dh{n}(M) and \Td(M)]. TROVE is an efficient  computer program for simulating hot spectra of polyatomic molecules; it is one of the main tools of the ExoMol project \cite{12TeYuxx.db,17TeYuxx}. The most recent updates of TROVE have been reported in Refs.~\citenum{17TeYuxx,GAIN}. TROVE uses an automatic approach for constructing a symmetry-adapted basis set to be used in setting up a matrix representation of the molecular rotation-vibration Hamiltonian. As mentioned above, we require towards this end, for each irrep of the symmetry group in question~\cite{98BuJexx}, a group of matrices constituting that irrep. However,  for \GTS\ or \GTS(EM), these matrices have not been available in the literature thus far. The present work aims at describing fully the groups \GTS\ and \GTS(EM) and, in particular, determining matrix groups that define the irreps. The matrix groups obtained are used for symmetrizing the potential energy surface and the rotation-vibration  basis functions for ethane.

	\section{The Structure of the {\itshape\bfseries G}$_{36}$ Group}
\begin{figure}[htbp!]
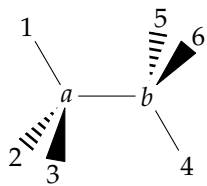

		\centering
		\chemfig{{\it a}([:120]-{1})([:-100]<{3})([:-130]<:{2})-{\it b}([:-60]-4)([:80]<:5)([:50]<6)}
		\caption{The labelling of the ethane nuclei.}
		\label{fig:ethane_labelling}
	\end{figure}

	Longuet-Higgins~\cite{63Longuet.linear} has shown that the group {\itshape\bfseries G}$_{36}$
	can be written as a direct product of two smaller groups {\itshape\bfseries C}$_{3{\rm v}}^{(-)}$ and  {\itshape\bfseries C}$_{3{\rm v}}^{(+)}$
	\begin{equation}
	 \mbox{\itshape\bfseries G}_{36} =
	\mbox{\itshape\bfseries C}_{3{\rm v}}^{(-)} \times  \mbox{\itshape\bfseries C}_{3{\rm v}}^{(+)};
	\end{equation}
	both of these groups are of order 6 and isomorphic to the
	{\itshape\bfseries C}$_{3{\rm v}}$ point group.
	The top row and leftmost column of \tabl{tab:multiplicationtable} define  the elements of these two groups. For convenience, we label
	the elements of {\itshape\bfseries C}$_{3{\rm v}}^{(\pm)}$ as
	$R_j^{(\pm)}$, $j$ $=$ 1, 2, \dots, 6; this notation is also defined in \tabl{tab:multiplicationtable}. The nuclei are labelled as in \fig{fig:ethane_labelling}.
	
	In \tabl{tab:multiplicationtable} we list the products
 $R_j^{(-)} \, R_k^{(+)}$ $=$  $R_k^{(+)} \, R_j^{(-)}$, where
 $R_j^{(-)}$ $\in$ {\itshape\bfseries C}$_{3{\rm v}}^{(-)}$ and $R_k^{(+)}$ $\in$ {\itshape\bfseries C}$_{3{\rm v}}^{(+)}$.
 Since \GTS\ $=$ {\itshape\bfseries C}$_{3{\rm v}}^{(-)}$ $\times$ {\itshape\bfseries C}$_{3{\rm v}}^{(+)}$, the 36 possible products
 $R_j^{(-)} \, R_k^{(+)}$ $=$  $R_k^{(+)} \, R_j^{(-)}$ constitute the complete group
\GTS.
	
Each of the groups
 {\itshape\bfseries C}$_{3{\rm v}}^{(-)}$ and  {\itshape\bfseries C}$_{3{\rm v}}^{(+)}$ has three
classes,
${\mathcal C}_1^{(\pm)}$ $=$ $\{ E \}$ $=$ $\{ R_1^{(\pm)} \}$,
${\mathcal C}_2^{(\pm)}$ $=$ $\{ R_2^{(\pm)}, R_3^{(\pm)} \}$,  and
${\mathcal C}_3^{(\pm)}$ $=$ $\{ R_4^{(\pm)}, R_5^{(\pm)}, R_6^{(\pm)} \}$.
Since {\itshape\bfseries G}$_{36}$ is the direct product of
{\itshape\bfseries C}$_{3{\rm v}}^{(-)}$ and  {\itshape\bfseries C}$_{3{\rm v}}^{(+)}$ (where, as discussed by
Longuet-Higgins~\cite{63Longuet.linear}, each element of
{\itshape\bfseries C}$_{3{\rm v}}^{(-)}$ commutes with
each element of  {\itshape\bfseries C}$_{3{\rm v}}^{(+)}$), its classes are obtained as
${\mathcal C}_i^{(-)} \times {\mathcal C}_j^{(+)}$, that is, a class of   {\itshape\bfseries G}$_{36}$ contains all elements
$RS$ where $R$ $\in$ ${\mathcal C}_i^{(-)}$ and $S$ $\in$ ${\mathcal C}_j^{(+)}$.
In the top row and leftmost column of \tabl{tab:multiplicationtable}, we indicate the class structures of {\itshape\bfseries C}$_{3{\rm v}}^{(+)}$ and  {\itshape\bfseries C}$_{3{\rm v}}^{(-)}$, respectively. A class
${\mathcal C}_i^{(+)}$ of {\itshape\bfseries C}$_{3{\rm v}}^{(+)}$ is simultaneously, in the form ${\mathcal C}_1^{(-)} \times {\mathcal C}_i^{(+)}$, a class of \GTS. Similarly, the classes of {\itshape\bfseries C}$_{3{\rm v}}^{(-)}$ are simultaneously classes of \GTS.
In \tabl{tab:multiplicationtable}, we have indicated the complete set of \GTS\ classes
 ${\mathcal C}_i^{(-)} \times {\mathcal C}_j^{(+)}$.

\begin{table}[hbtp!]
	\caption{\label{tab:multiplicationtable}The class structure of \GTS.$^a$ }
{\scriptsize
\begin{center}
\renewcommand{\arraystretch}{1.5}
\begin{tabular}{l|ll|lll}
\hline
\hline
 \begin{sideways}$R_1^{(-)}$ = $R_1^{(+)}$ = $E$ \end{sideways}  &
	\begin{sideways}$R_2^{(-)}$ = (132)(456) \end{sideways} &
	\begin{sideways}$R_3^{(-)}$ = (123)(465) \end{sideways} &
	\begin{sideways}$R_4^{(-)}$ = (14)(25)(36)($ab$)\phantom{X} \end{sideways} &
	\begin{sideways}$R_5^{(-)}$ = (16)(24)(35)($ab$) \end{sideways} &
	\begin{sideways}$R_6^{(-)}$ = (15)(26)(34)($ab$) \end{sideways} \\
\hline
$R_2^{(+)}$ = $(123)(456)$ & $(465)$ &$(132)$ & $(153426)(ab)$ & $(143625)(ab)$  & $(163524)(ab)$ \\
$R_3^{(+)}$ = (132)(465) & (123) & (456) & (162435)$(ab)$ &   (152634)$(ab)$ & (142536)$(ab)$   \\
\hline
$R_4^{(+)}$ = (14)(26)(35)$(ab)^*$&  (152436)$(ab)^*$ & (163425)$(ab)^*$& (23)(56)$^*$&  (12)(46)$^*$ &(13)(45)$^*$\\
$R_5^{(+)}$ = (16)(25)(34)$(ab)^*$&  $(142635)(ab)^*$ & $(153624)(ab)^*$& (13)(46)$^*$&  (23)(45)$^*$ &(12)(56)$^*$\\
$R_6^{(+)}$ = (15)(24)(36)$(ab)^*$&  (162534)$(ab)^*$ & (143526)$(ab)^*$& (12)(45)$^*$ &(13)(56)$^*$ & (23)(46)$^*$ \\
\hline
\hline
\end{tabular}
\end{center}}
{\small
\begin{flushleft}
{$^a$The top row and leftmost column contain the \GTS\ elements that also belong to the {\itshape\bfseries C}$_{3{\rm v}}^{(-)}$ or {\itshape\bfseries C}$_{3{\rm v}}^{(+)}$ group, respectively. The remaining entries are the products  $R_j^{(-)} \, R_k^{(+)}$ $=$ $R_k^{(+)} \, R_j^{(-)}$, where $R_j^{(-)}$ $\in$ {\itshape\bfseries C}$_{3{\rm v}}^{(-)}$ is at
the top of the column and
$R_k^{(+)}$ $\in$ {\itshape\bfseries C}$_{3{\rm v}}^{(+)}$ is at the left end
of the row.
The horizontal and vertical lines denote the separation of the classes.
}
\end{flushleft}}
\end{table}

\section{Irreducible Representations of {\itshape\bfseries G}$_{36}$}
\label{sect:irrep_g36}
Again, since {\itshape\bfseries G}$_{36}$ is the direct product of
{\itshape\bfseries C}$_{3{\rm v}}^{(-)}$ and  {\itshape\bfseries C}$_{3{\rm v}}^{(+)}$, we could in principle label
its irreps as $(\Gamma^{(-)}, \Gamma^{(+)})$, where
$\Gamma^{(-)}$ is an irrep of {\itshape\bfseries C}$_{3{\rm v}}^{(-)}$ and
$\Gamma^{(+)}$ is an irrep of {\itshape\bfseries C}$_{3{\rm v}}^{(+)}$.
{\itshape\bfseries C}$_{3{\rm v}}^{(-)}$ and  {\itshape\bfseries C}$_{3{\rm v}}^{(+)}$ both have the one-dimensional
irreps $A_1$ and $A_2$ together with the two-dimensional irrep $E$ (Table~\ref{tab:C3vchartab} in \sect{sec:appendixC3vgroups}). 
Customarily, the irreps of  {\itshape\bfseries G}$_{36}$
 are labelled as in Table~12 of Longuet-Higgins~\cite{63Longuet.linear}. Table~\ref{tab:chartab} is the character
table for {\itshape\bfseries G}$_{36}$, obtained from
Table~12 of Longuet-Higgins~\cite{63Longuet.linear}, with the irreducible representations labelled by
their customary labels $\Gamma_{36}$  and the combination labels
$(\Gamma^{(-)}, \Gamma^{(+)})$. In the bottom row of the table, we indicate the ${\mathcal C}_i^{(-)} \times {\mathcal C}_j^{(+)}$ label of each class. For each of the nine {\itshape\bfseries G}$_{36}$ classes,
we give in Table~\ref{tab:chartab} a representative element together with the number of group elements in the class.  The complete classes are obtained from \tabl{tab:multiplicationtable}.

In the present work, we are particularly concerned with the
doubly-degenerate irreducible representations
$E_1$ $=$  $(E,A_1)$,
$E_2$ $=$  $(E,A_2)$,
$E_3$ $=$  $(A_1,E)$, and
$E_4$ $=$  $(A_2,E)$ together with the
four-dimensional irreducible representation $G$ $=$ $(E, E)$.

\begin{table}[hbtp!]
\caption{\label{tab:chartab} Character table of {\itshape\bfseries G}$_{36}$.}
\begin{center}
\renewcommand{\arraystretch}{1.5}
\begin{tabular}{lcrrrrrrrrr}
\hline\hline
$\Gamma_{36}$    & $(\Gamma^{(-)}, \Gamma^{(+)})$
 & $E$ &
\begin{sideways}(123)(456) \end{sideways} &
\begin{sideways}(14)(26)(35)($ab$)$^*$\phantom{xxx} \end{sideways} &
\begin{sideways}(123)(465) \end{sideways} &
\begin{sideways}(123) \end{sideways} &
\begin{sideways}(142635)($ab$)$^*$  \end{sideways} &
\begin{sideways}(14)(25)(36)($ab$) \end{sideways} &
\begin{sideways}(142536)($ab$)  \end{sideways} &
\begin{sideways}(12)(45)$^*$ \end{sideways} \\
& & 1 & 2 & 3 & 2 & 4 & 6 & 3 & 6 & 9 \\
\hline
$A_1$ & $(A_1,A_1)$ & 1 &  1 & 1 & 1 & 1 & 1 & 1 & 1 & 1 \\
$A_2$ & $(A_2,A_1)$ & 1 &  1 & 1 & 1 & 1 & 1 & $-$1 & $-$1 & $-$1 \\
$A_3$ & $(A_1,A_2)$  & 1 &  1 & $-$1 & 1 & 1 & $-$1 & 1 & 1 & $-$1 \\
$A_4$ & $(A_2,A_2)$ & 1 &  1 & $-$1 & 1 & 1 & $-$1 & $-$1 & $-$1 & 1 \\
$E_1$ &  $(E,A_1)$ & 2 & 2    &    2 & $-$1 & $-$1 & $-$1 &    0 & 0 & 0 \\
$E_2$ &  $(E,A_2)$ & 2 & 2    & $-$2 & $-$1 & $-$1 &    1 &    0 & 0 & 0 \\
$E_3$ &  $(A_1,E)$ & 2 & $-$1 &    0 &    2 & $-$1 &    0 &    2 & $-$1 & 0 \\
$E_4$ &  $(A_2,E)$ & 2 & $-$1 &    0 &    2 & $-$1 &    0 & $-$2 & 1 & 0 \\
$G$   &$(E,E)$ & 4 &  $-$2 & 0 & $-$2 & 1 & 0 & 0 & 0 & 0 \\
\hline
  &  &
  \begin{sideways} ${\mathcal C}_1^{(-)} \times {\mathcal C}_1^{(+)}$\phantom{X}  \end{sideways}  &
  \begin{sideways} ${\mathcal C}_1^{(-)} \times {\mathcal C}_2^{(+)}$  \end{sideways}  &
  \begin{sideways} ${\mathcal C}_1^{(-)} \times {\mathcal C}_3^{(+)}$  \end{sideways}  &
  \begin{sideways} ${\mathcal C}_2^{(-)} \times {\mathcal C}_1^{(+)}$  \end{sideways}  &
  \begin{sideways} ${\mathcal C}_2^{(-)} \times {\mathcal C}_2^{(+)}$  \end{sideways}  &
  \begin{sideways} ${\mathcal C}_2^{(-)} \times {\mathcal C}_3^{(+)}$  \end{sideways}  &
  \begin{sideways} ${\mathcal C}_3^{(-)} \times {\mathcal C}_1^{(+)}$  \end{sideways}  &
  \begin{sideways} ${\mathcal C}_3^{(-)} \times {\mathcal C}_2^{(+)}$  \end{sideways}  &
  \begin{sideways} ${\mathcal C}_3^{(-)} \times {\mathcal C}_3^{(+)}$  \end{sideways}  \\[5pt]
\hline\hline
\end{tabular}
\end{center}
\end{table}

In Section~12.4 of Ref.~\cite{98BuJexx}, and in
Section~3.1 of Ref.~\cite{18ChJeYu}, it is discussed how the point group
{\itshape\bfseries C}$_{3{\rm v}}$ and any group isomorphic to it can be defined in terms of two
generating operations, one of which belongs to the two-member class ${\mathcal C}_2^{(\pm)}$  of {\itshape\bfseries C}$_{3{\rm v}}$ and
the other to the three-member class ${\mathcal C}_3^{(\pm)}$ . For the group
{\itshape\bfseries C}$_{3{\rm v}}^{(-)}$, we choose the generating operations as
$R_2^{(-)}$ $=$ (123)(465)  and
$R_4^{(-)}$ $=$ (14)(25)(36)($ab$), whereas for {\itshape\bfseries C}$_{3{\rm v}}^{(+)}$,
we choose
$R_2^{(+)}$ $=$  (123)(456) and
$R_4^{(+)}$ $=$  (14)(26)(35)($ab$)$^*$. We then have $R_{3}^{(\pm)} = (R_2^{(\pm)})^2$, $R_6^{(\pm)} = R_2^{(\pm)}R_4^{(\pm)}$, and $R_5^{(\pm)} = R_2^{(\pm)}R_6^{(\pm)}.$

\section{Representation Matrices for the $E_i$ Irreducible Representations of {\itshape\bfseries G}$_{36}$}
The representation matrices for the non-degenerate irreps
$A_1$, $A_2$, $A_3$, and $A_4$
 of \GTS\ are uniquely defined as equal to the representation
characters; these can be found in Table~\ref{tab:chartab}. The $2 \times 2$ representation matrices of
the irreps  $E_1$, $E_2$, $E_3$, and $E_4$ are, however, not uniquely defined. Having found one such set of matrices ${\mathbf M}_{E_i} [ O_i ]$, where $O_i$ $\in$ \GTS, we can generate infinitely many equivalent representations with representation matrices
${\mathbf M}_{E_i}^{\prime} [ O_i ]$ $=$
${\mathbf V}\,{\mathbf M}_{E_i} [ O_i ]\,{\mathbf V}^{-1}$,
where ${\mathbf V}$ is an arbitrary and invertible $2 \times 2$ matrix.
In practice, we want to keep the representation matrices real and orthogonal, and that drastically limits the possible choices. We select one particular set of representation matrices here
for $E_1$ by initially choosing the representation matrix for $R_2^{(-)}$ $=$ (123)(465), one of the generating operations
of {\itshape\bfseries C}$_{3{\rm v}}^{(-)}$. We set
\begin{equation}{\mathbf M}_{E} [ (123)(465)   ] =
\zwobyzwo{\cos\left( \frac{2\,\pi}{3} \right)}{-\sin\left( \frac{2\,\pi}{3} \right)}{\sin\left( \frac{2\,\pi}{3} \right)}{\cos\left( \frac{2\,\pi}{3} \right)} =
\zwobyzwo{-\frac{1}{2}}{-\frac{\sqrt{3}}{2} }{\frac{\sqrt{3}}{2} }{-\frac{1}{2}}.
 \label{eq:genmat1}
\end{equation}
The $2 \times 2$ orthogonal matrix
${\mathbf M}_{E} [ (123)(465)   ]$ satisfies the relation
${\mathbf M}_{E} [ (123)(465)   ]^3$ $=$ ${\mathbf E}$, the $2 \times 2$ unit matrix, imposed by the fact that
$[ (123)(465)   ]^3$ $=$ $E$. An alternative choice would be the matrix with the signs of the $\sin(2\pi/3)$ terms reversed,
The other generating operation of {\itshape\bfseries C}$_{3{\rm v}}^{(-)}$,
$R_4^{(-)}$ $=$ [(14)(25)(36)($ab$)], is self-inverse:
$(R_4^{(-)})^2$ $=$ [(14)(25)(36)($ab$)]$^2$ $=$ $E$.
The $2 \times 2$ orthogonal matrix representing $R_4^{(-)}$ is also self-inverse and we can choose it as
\begin{equation}
{\mathbf M}_{E} [ (14)(25)(36)(ab)  ] =
\zwobyzwo{\cos \theta}{\sin \theta}{\sin \theta}{-\cos \theta}
\end{equation}
where $\theta$ is arbitrary. All such matrices satisfy
${\mathbf M}_{E} [ (14)(25)(36)(ab)  ]^2$ $=$ ${\mathbf E}$.
We choose $\theta =0$, so that
\begin{equation}{\mathbf M}_{E} [ (14)(25)(36)(ab)  ] =
\zwobyzwo{1}{0}{0}{-1}.
 \label{eq:genmat2}
\end{equation}
The two matrices  ${\mathbf M}_{E} [ (123)(465)   ]$  and ${\mathbf M}_{E} [ (14)(25)(36)(ab)  ] $ have traces of
$-1$ and 0, respectively, and it is seen from Table~\ref{tab:C3vchartab}  that they generate the irrep $E$ of {\itshape\bfseries C}$_{3{\rm v}}^{(-)}$.
We can now use the relations in Section~12.4 of Ref.~\cite{98BuJexx} or, equivalently, in
Section~3.1 of Ref.~\cite{18ChJeYu} to determine, by matrix multiplication involving
${\mathbf M}_{E} [ (123)(465)   ]$  and ${\mathbf M}_{E} [ (14)(25)(36)(ab)  ] $, the representation matrices
for all operations in {\itshape\bfseries C}$_{3{\rm v}}^{(-)}$.

We have discussed above how the irrep $E_1$ of \GTS\ can be described as
$(\Gamma^{(-)}, \Gamma^{(+)})$ $=$  $(E,A_1)$, where $\Gamma^{(-)}$ and $\Gamma^{(+)}$ are irreps of
{\itshape\bfseries C}$_{3{\rm v}}^{(-)}$ and
{\itshape\bfseries C}$_{3{\rm v}}^{(+)}$, respectively.
We have already determined a group of representation matrices belonging to the $E$ irrep of
{\itshape\bfseries C}$_{3{\rm v}}^{(-)}$, and to obtain one for the $A_1$ (totally symmetric) irrep of
{\itshape\bfseries C}$_{3{\rm v}}^{(+)}$, we introduce the $1 \times 1$  representation matrices
\begin{equation}{\mathbf M}_{A_1} [ (123)(456) ] =
{\mathbf M}_{A_1} [ (14)(26)(35)(ab)^*  ] = 1
 \label{eq:genmat3}
\end{equation} for its generating operations
$R_2^{(+)}$ $=$  (123)(456) and
$R_4^{(+)}$ $=$  (14)(26)(35)($ab$)$^*$. Again, we can use the relations in Section~12.4 of Ref.~\cite{98BuJexx} or in
Section~3.1 of Ref.~\cite{18ChJeYu} to determine the representation matrices
for all operations on {\itshape\bfseries C}$_{3{\rm v}}^{(+)}$. It is rather trivial here since these representation matrices
all are the $1 \times 1$  matrix 1.

We now have $E$ representation matrices for the six operations in {\itshape\bfseries C}$_{3{\rm v}}^{(-)}$ and
$A_1$ representation matrices for the six operations in {\itshape\bfseries C}$_{3{\rm v}}^{(+)}$, and we can form $E_1$ representation
matrices for the 36 operations in \GTS\ by forming the 36 products of  ${\mathbf M}_{E} [ R   ]$,
$R$ $\in$  {\itshape\bfseries C}$_{3{\rm v}}^{(-)}$, with the constant (=1 always in this case)
${\mathbf M}_{A_1} [ S  ]$,
$S$ $\in$  {\itshape\bfseries C}$_{3{\rm v}}^{(+)}$. The products are formed with the
computer-algebra program {\tt maxima}~\cite{maxima}.
The resulting $2 \times 2$ transformation matrices   ${\mathbf M}_{E_1} [ R  ]$,
   $R$ $\in$  \GTS, are included in \ref{sec:appendixa} (Section~\ref{sec:Einalmatrices}).

For
$E_2$ $=$  $(E,A_2)$ we obtain the representation matrices as just described for $E_1$. The only difference is that we
replace the {\itshape\bfseries C}$_{3{\rm v}}^{(+)}$ representation matrices by
\begin{equation}{\mathbf M}_{A_2} [ (123)(456) ] = 1\;\; \mbox{\rm and} \;\;
{\mathbf M}_{A_2} [ (14)(26)(35)(ab)^*  ] = -1.
 \label{eq:genmat4}
\end{equation}
For
$E_3$ $=$  $(A_1,E)$ and
$E_4$ $=$  $(A_2,E)$, the representation matrices are determined by interchanging the
{\itshape\bfseries C}$_{3{\rm v}}^{(-)}$ and
{\itshape\bfseries C}$_{3{\rm v}}^{(+)}$ representation matrices in the determination made for
$E_1$ and $E_2$, respectively.
The $2 \times 2$ transformation matrices   ${\mathbf M}_{E_2} [ R  ]$,
${\mathbf M}_{E_3} [ R  ]$, and ${\mathbf M}_{E_4} [ R  ]$,
   $R$ $\in$  \GTS, are included in \ref{sec:appendixa} (Section~\ref{sec:Einalmatrices}).

\section{Representation Matrices for the $G$ Irreducible Representation of {\itshape\bfseries G}$_{36}$}

We now aim at determining a set of $4 \times 4$ matrices constituting the irrep $G$ of {\itshape\bfseries G}$_{36}$.
In principle, we could continue as outlined in the preceding section by utilizing that
$G$ $=$ $(\Gamma^{(-)}, \Gamma^{(+)})$ $=$
$(E, E)$. However, this would imply that we use the $2 \times 2$ $E$-representation matrices for
{\itshape\bfseries C}$_{3{\rm v}}^{(-)}$ and
{\itshape\bfseries C}$_{3{\rm v}}^{(+)}$  to generate
 $4 \times 4$ $G$-representation matrices of \GTS\ as the `super-product' matrices discussed
in Section~6.4 of Ref.~\cite{98BuJexx}: Each of the 16 elements of the  $G$-representation matrix
is a product of two $E$-representation matrix elements, one of the four elements in
the
{\itshape\bfseries C}$_{3{\rm v}}^{(-)}$ $E$-representation matrix times
 one of the four elements in
the
{\itshape\bfseries C}$_{3{\rm v}}^{(+)}$  $E$-representation matrix (see, in particular, Eq.~(6-97) of
Ref.~\cite{98BuJexx} and the discussion of it).
The formation of such a super-product is not a standard linear-algebra operation and not
easily programmable in {\tt maxima}~\cite{maxima}.
Consequently, we make use of a more straightforward approach: We initially note
 that the six CH-bond lengths $r_k$, $k$ $=$ 1, 2, 3, \dots, 6,
of H$_3$CCH$_3$ span the representation
$A_1$ $\oplus$ $A_4$ $\oplus$ $G$ of {\itshape\bfseries G}$_{36}$.
This is easily established as the transformation properties of the $r_k$
under the operations of  \GTS\ are simple permutations.
 We define the CH bond length $r_k$ as the one involving proton $k$, with the protons labelled  as in \fig{fig:ethane_labelling}. We now proceed to
use representation-theory methods to determine the symmetrized linear combinations
of the $r_k$ and, subsequently, the transformation properties of the resulting $G$-symmetry coordinates.

As just mentioned,
it is straightforward to determine the effect of the operations $O_i$ in {\itshape\bfseries G}$_{36}$ on the six bond lengths
$r_k$. In general we can write
\begin{equation}
O_i\, \left( \begin{array}{l}
r_{1} \\ r_{2} \\
r_{3} \\
r_{4} \\
r_{5} \\
r_{6}
\end{array} \right) =
\left( \begin{array}{l}
r_{1}^{\prime} \\ r_{2}^{\prime} \\
r_{3}^{\prime} \\
r_{4}^{\prime} \\
r_{5}^{\prime} \\
r_{6}^{\prime}
\end{array} \right) = {\mathbf M}^{\prime} [ O_i ] \,
\left( \begin{array}{l}
r_{1} \\ r_{2} \\
r_{3} \\
r_{4} \\
r_{5} \\
r_{6}
\end{array} \right),\end{equation}
where the elements of the $6 \times 6$ matrix ${\mathbf M}^{\prime} [ O_i ]$ are determined from
the general idea that after the operation $O_i$ has been carried out, the proton 1, say, occupies the position
previously occupied by the proton $k$ (and the C nucleus to which it is bound has
moved with it), and therefore $r_{1}^{\prime}$ $=$ $r_k$, with similar considerations
made for the bond lengths $r_2$, \dots, $r_6$.

Having obtained the matrices ${\mathbf M}^{\prime} [ O_i ]$, $i$ $=$ 1, 2, \dots, 36, for all operations in {\itshape\bfseries G}$_{36}$, we form the linear combination
\begin{equation}
{\mathbf O}_{G} = \sum_{i=1}^{36} \chi_G[O_i]\,  {\mathbf M}^{\prime} [ O_i ],
\end{equation}
where $\chi_G(O_i)$ is the character of the operation $O_i$ for the irrep $G$. The $6 \times 6$ matrix
 ${\mathbf O}_{G}$ projects out the $G$-symmetry part of its argument and we multiply it on to the $6 \times 1$ column vector with the $r_k$, thus obtaining six linear combinations of the $r_k$ that are all, in principle, of $G$ symmetry.
However, only four independent $G$ coordinates exist and so the linear combinations are linearly dependent. We discard two of them and the remaining four are linearly independent. We then orthogonalize
(with the Gram-Schmidt technique)
 the four 6-component vectors with the coefficients of the four linearly independent combinations, thus obtaining the rows 3, \dots, 6 of the
$6 \times 6$ matrix ${\mathbf V}$  describing the transformation to symmetrized combinations of the $r_k$:
\begin{equation}
\left( \begin{array}{l}
S_{A_1} \\ S_{A_4} \\
S_{G;1} \\
S_{G;2} \\
S_{G;3} \\
S_{G;4} \end{array} \right) =
\left( \begin{array}{c}
r_1 + r_2 + r_3 + r_4 + r_5 + r_6 \\
r_1 + r_2 + r_3 - r_4 - r_5 - r_6 \\
2\,r_1 - r_2 - r_3 + 2\, r_4 - r_5 - r_6 \\
         r_2 - r_3           + r_5 - r_6 \\
-2\,r_1 + r_2 + r_3 + 2\, r_4 - r_5 - r_6 \\
        -r_2 + r_3           + r_5 - r_6
 \end{array} \right) =
{\mathbf V} \,
\left( \begin{array}{l}
r_{1} \\ r_{2} \\
r_{3} \\
r_{4} \\
r_{5} \\
r_{6}
\end{array} \right) \label{eq:gtransform} \end{equation}
with
\begin{equation}
{\mathbf V} =
\left( \begin{array}{rrrrrr}
 1&1&1&1&1&1 \\
 1&1&1&-1&-1&-1 \\
   2 & -1& -1&  2&  -1& - 1 \\
                         0&  1& -1&  0&   1& -1 \\
                        -2&  1&  1&  2&  -1& -1 \\
                         0& -1&  1&  0&   1& -1
\end{array} \right). \end{equation}
Rows 1 and 2 of ${\mathbf V}$ describe symmetrized coordinates of $A_1$ and $A_4$ symmetry, respectively, the remaining
four rows produce $G$-symmetry coordinates.

We now transform, with the help of the computer algebra program {\tt maxima}~\cite{maxima}, the 36 matrices
${\mathbf M}^{\prime} [ O_i ]$, $i$ $=$ 1, 2, \dots, 36, to be expressed in terms of the symmetrized $S$ coordinates
\begin{equation}
 {\mathbf Q}_i =
{\mathbf V} \, {\mathbf M}^{\prime} [ O_i ] \, {\mathbf V}^{-1}
 \end{equation}
and it is checked that each of the resulting matrices ${\mathbf Q}_i $ is block diagonal with two $1 \times 1$ blocks
containing the characters of the irreps $A_1$ and $A_4$, respectively, and a $4 \times 4$ block whose trace is the $G$ character.

Finally, we subject the ${\mathbf Q}_i $ matrices to the final transformations to obtain the transformation matrices
of the $G$-symmetry coordinates:
\begin{equation}
{\mathbf M}_G\left[ O_i \right] =
{\mathbf T}_{\rm norm} \,
{\mathbf T}_{4 \times 6} \, {\mathbf Q}_i \, {\mathbf T}_{4 \times 6}^{\rm T}\, {\mathbf T}_{\rm norm}^{-1}
\end{equation}
where
\begin{equation}
{\mathbf T}_{4 \times 6}
=
\left(
\begin{array}{cccccc}
 0 & 0 & 1 & 0 & 0 & 0 \\
0 & 0 & 0 & 1 & 0 & 0 \\
0 & 0 & 0 & 0 & 1 & 0 \\
0 & 0 & 0 & 0 & 0 & 1
\end{array}
\right),
\end{equation}
and
\begin{equation}
{\mathbf T}_{\rm norm} =
\left(
\begin{array}{cccc}
\frac{1}{2\sqrt{3}} & 0 & 0 & 0\\
 0 & \frac{1}{2} & 0 & 0\\
0 & 0 & \frac{1}{2\sqrt{3}} & 0 \\
0 & 0 & 0 & \frac{1}{2} \\
 \end{array}
\right),
\end{equation}
and the superscript T denotes transposition. The transformation with ${\mathbf T}_{4 \times 6}$ cuts out the $G$-block of
each transformation matrix and that with $ {\mathbf T}_{\rm norm}$ ``normalizes'' the symmetrized coordinates so that the resulting
$G$ transformation matrices ${\mathbf M}_G\left[ O_i \right]$
 become orthogonal. The normalization could in principle just as well have been incorporated in the
matrix ${\mathbf V}$, but that made the output from {\tt maxima}~\cite{maxima} rather unreadable, and so the transformations were split into several steps as described here.

The orthogonal matrices ${\mathbf M}_G\left[ O_i \right]$ are listed in
\ref{sec:appendixa} (Section~\ref{sec:finalmatrices}).

\section{Generation of a Symmetry Adapted Basis Set for Ethane}
\label{sec:symmetrisation}

As mentioned in \sect{sec:intro}, the analysis in the present paper is analogous to previous work~\cite{18ChJeYu} on the point group \Dh{\infty} of acetylene and other centrosymmetric linear molecules. Also in the present work, we use the variational nuclear-motion program \Trove\ in the generation of a symmetry-adapted ro-vibrational basis set for ethane.

The general method used by \Trove\ is thoroughly explained  in \cite{07YuThJe.method,16YuYaOv.methods,18ChYaTe}. The basic idea is to take advantage of the useful fact that in a symmetry-adapted basis set $\{\Psi\}$,  the ro-vibrational Hamiltonian matrix is block diagonal with the form
\begin{equation}
\bra{\Psi_{\mu,n_s}^{J,\Gamma_s}} \hat{H}_{\rm rv} \ket{\Psi_{\mu',n_t}^{J,\Gamma_t}} = H_{\mu, \mu'}\, \delta_{t,s}\, \delta_{n_s,n_t}
\end{equation}
where the indices $\mu$, $\mu'$ label the basis functions,  $\Gamma_s$ and $\Gamma_t$  denote irreducible representations of the symmetry group, and $n_s$($n_t$) labels the components of the irrep $\Gamma_s$($\Gamma_t$). To  construct a symmetrized basis set, one can diagonalise an operator which  commutes with all operations of the MS group. Of course, the ro-vibrational Hamiltonian is one such operator, by definition, so \Trove\ utilises this by constructing, from the full ro-vibrational Hamiltonian, reduced Hamiltonians that also commute with the group operations, each reduced Hamiltonian depending on fewer coordinates than the full Hamiltonian. The reduced Hamiltonians are diagonalised separately and products of their -- by necessity symmetrised -- eigenfunctions are  again symmetrised. The result is a symmetry-adapted basis set appropriate for the complete ro-vibrational coordinate space. We describe this symmetrization procedure for ethane in the following sections.

\subsection{Definition of the internal coordinates used for ethane}

To construct the reduced Hamiltonians mentioned in the preceding section, one first separates the internal coordinates into subsets which transform into each other under the symmetry  group operations. We use the convention in Section~{1.2} of \cite{98BuJexx} for transforming functions of Cartesian coordinates. The basis set associated with each coordinate subset is obtained in terms of
one-dimensional primitive functions $\phi_{n_i}(q_i)$, where $n_i$ is the number of quanta for coordinate $q_i$. By averaging the Hamiltonian over the `ground state' ($n_i=0$;  $\phi_0(q_i) \equiv \ket{0}_{q_i}$) primitives associated with the coordinates in all subsets not under consideration, i.e.
\begin{equation}
\bra{0}_{q_1} \ldots \bra{0}_{q_{k-1}} \bra{0}_{q_{l+1}} \ldots \hat{H}_{\rm rv} \ldots \ket{0}_{q_{l+1}} \ket{0}_{q_{k-1}} \ldots \ket{0}_{q_1}
\end{equation}
where $\{q_{k}\ldots q_{l}\}$ are the coordinates in the subset of interest, one obtains the reduced Hamiltonian which depends only on these coordinates; it commutes with the operations in the MS group. Product functions of the form $\phi_{n_k}(q_k) \ldots \phi_{n_l}(q_l)$ are then used as a basis set for diagonalizing this reduced Hamiltonian.

\begin{figure}[htbp!]
\centering
\begin{tikzpicture}[scale=1.7]
\draw (0,0) circle(2pt)[fill] -- (1,0) circle(2pt)[fill] -- ++(300:1) circle(2pt)[fill];
\draw (1,0) -- ++(80:1) circle(2pt)[fill];
\draw (1,0) -- ++(50:1) circle(2pt)[fill];
\draw (0,0) -- ++(120:1) circle(2pt)[fill];
\draw (0,0) -- ++(260:1) circle(2pt)[fill];
\draw (0,0) -- ++(230:1) circle(2pt)[fill];
\draw[<->] (0,0.2) -- (0.5,0.2) node[above] {$R$} -- (1,0.2);
\draw[<->] (-0.2,-0.1) -- ++ (120:0.5) node[below left] {$r$} -- ++(120:0.5);
\draw [domain=180:240] plot  ({1 + 0.3*cos(\x)}, {0.3*sin(\x)}) node[below] {$\alpha$};
\draw [domain=240:300] plot  ({1 + 0.3*cos(\x)}, {0.3*sin(\x)}) ;
\end{tikzpicture}
\caption{Representative members of three of the vibrational coordinate subsets. Here $R$ is the \ce{C-C} bond length, $r$ is one of the six C--H$_k$ bond lengths $r_k$, and $\alpha$ is one of the six $\angle$({H$_k$-C-C}) bond angles $\alpha_k$.}
\label{fig:three_vibrational_classes}
\end{figure}
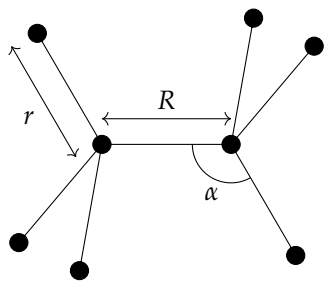

In the case of ethane, there are $3 \times 8 - 7 = 17$ small-amplitude vibrational coordinates, one torsional coordinate (describing the independent rotations of the \ce{CH3} groups), and three rotational coordinates.  \fig{fig:three_vibrational_classes} shows representative members of three of the vibrational coordinate subsets: the \ce{C-C} bond length denoted by $R$, one of six \ce{C-H$_k$} bonds denoted by $r$, and one of six bond angles $\angle$({H$_k$--C--C})  denoted by $\alpha$. The small-amplitude vibrational coordinates $R$, $r_k$, and  $\alpha_k$ ($k=1\ldots 6$) measure the displacements of the respective internal coordinates from their equilibrium values, that is the coordinates actually used are $R-R_{\rm e}$,  $r_i-r_{\rm e}$ ($i=0\ldots 6$) and $\alpha_k-\alpha_{\rm e}$ ($k=1\ldots 6$). The $r_k$($\alpha_k$) coordinates are equivalent and so they have a common equilibrium value $r_{\rm e}$($\alpha_{\rm e}$).

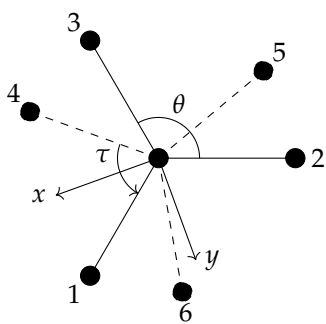
\begin{figure}[htbp!]
\centering
\begin{tikzpicture}[scale=1.8]
\draw (0,0) circle(2pt)[fill] -- ++ (0:1) circle(2pt)[fill] node[right] {\ 2};
\draw (0,0) -- ++(120:1) circle(2pt)[fill] node[above left] {3};
\draw (0,0) -- ++(240:1) circle(2pt)[fill] node[below left] {1};

\draw [dashed] (0,0)  -- ++ (40:1) circle(2pt)[fill] node[above right] {5};
\draw [dashed] (0,0) -- ++(160:1) circle(2pt)[fill] node[above left] {4};
\draw [dashed] (0,0) -- ++(280:1) circle(2pt)[fill]  ++ (280:0.15) circle(0pt) node {6};
\draw[->] (0,0) -- ++ (200:0.8) node[left] {$x$};
\draw[->] (0,0) -- ++ (200+90:0.8) node[right] {$y$};

\draw [domain=0:60] plot  ({ 0.3*cos(\x)}, {0.3*sin(\x)}) node[above] {$\theta$};
\draw [domain=60:120] plot  ({ 0.3*cos(\x)}, {0.3*sin(\x)});

\draw[-] [domain=160:200] plot  ({ 0.3*cos(\x)}, {0.3*sin(\x)}) node[above left] {$\tau$};
	\draw[->] [domain=200:240] plot  ({ 0.3*cos(\x)}, {0.3*sin(\x)});
\end{tikzpicture}
	\caption{A Newman projection of ethane, with the CH$_3$ group containing protons 1, 2, 3, indicated by solid \ce{C-H} bonds, being closest to the viewer. One of the dihedral angles used in the vibrational subsets is labelled by $\theta$ and the torsional angle is labelled by $\tau$ and is measured in the counterclockwise direction. The $x$ axis halves the dihedral angle between the \ce{H$_1$-C-C} and \ce{H$_4$-C-C} planes.
}
\label{fig:dihedrals}
\end{figure}

The last vibrational subset is obtained form
six dihedral angles
$\theta_{12}$, $\theta_{23}$, $\theta_{31}$,
$\theta_{45}$, $\theta_{56}$, $\theta_{64}$,
one of which is labelled $\theta$ in \fig{fig:dihedrals}.
$\theta_{ij}$ is the angle between
the  \ce{H$_i$-C-C} and \ce{H$_j$-C-C} planes, where protons $i$ and $j$ belong to the same \ce{CH3} group. Only four of the six angles are linearly independent due to the constraints
$\theta_{12} + \theta_{23} + \theta_{31}$ $=$
$\theta_{45} + \theta_{56} +\theta_{64}$ $=$
$2 \pi$. The positive directions of rotation for the $\theta_{ij}$ angles
are from proton 1 $\rightarrow$ 2 $\rightarrow$ 3 for
the one \ce{CH3} group, and from proton 4 $\rightarrow$ 5 $\rightarrow$ 6 for
the other.
The independent coordinates constructed from the six $\theta_{ij}$ angles are
\begin{equation}
\begin{split}
{\gamma}_1 &= \frac{1}{\sqrt{6}}(2\theta_{23}- \theta_{31} - \theta_{12}), \\
{\gamma}_2 &= \frac{1}{\sqrt{2}}(\theta_{31} - \theta_{12}), \\
{\delta}_1 &= \frac{1}{\sqrt{6}}(2\theta_{56} - \theta_{64} - \theta_{45}), \\
{\delta}_2 &=  \frac{1}{\sqrt{2}}(\theta_{64} - \theta_{45}),
\end{split}
\label{eq:gamma-delta}
\end{equation}
which transform as the $G$ representation of \GTS.

\begin{figure}[htbp!]
\centering
\begin{tikzpicture}[scale=1.8]
\draw (0,0) circle(2pt)[fill] -- ++ (0:1) circle(2pt)[fill] node[right] {\ 2};
\draw (0,0) -- ++(120:1) circle(2pt)[fill] node[above left] {3};
\draw (0,0) -- ++(240:1) circle(2pt)[fill] node[below left] {1};

	\draw [dashed] (0,0)  -- ++ (40+20:1) circle(2pt)[fill] node[above right] {5};
	\draw [dashed] (0,0) -- ++(160+20:1) circle(2pt)[fill] node[above left] {4};
\draw [dashed] (0,0) -- ++(280+20:1) circle(2pt)[fill]  ++ (280:0.15) circle(0pt) node {6};
\draw[->] (0,0) -- ++ (240:0.8) node[left] {$x_a$};
\draw[->] (0,0) -- ++ (240+90:0.8) node[right] {$y_a$};
\draw[->] (0,0) -- ++ (160+20:0.8) node[above] {$x_b$};
	\draw[->] (0,0) -- ++ (160+90+20:0.8) node[below] {$y_b$};

\end{tikzpicture}
	\caption{A Newman projection of ethane, with the CH$_3$ group containing protons 1, 2, and 3, indicated by solid \ce{C-H} bonds, being closest to the viewer. The $x$ and $y$ components of the coordinate axes attached to each \ce{CH3} group is shown, the subscript $a$ signifying that the coordinate axes are for the \ce{C_$a$H3} group. To ensure the coordinate system is right handed, the $z$ axis (the same for both groups) points from \ce{C_$b$} to \ce{C_$a$}. With this construction, the $\theta$ and $\phi$ Euler describing the direction of the $z$ axis are the same for \ce{CH3} group are the same while the $\chi$ angle describing the rotation about the $z$ axis different and are denoted by $\chi_a$ and $\chi_b$. These increase in the counterclockwise direction due to the right hand rule.}
\label{fig:coordinate_axes}
\end{figure}
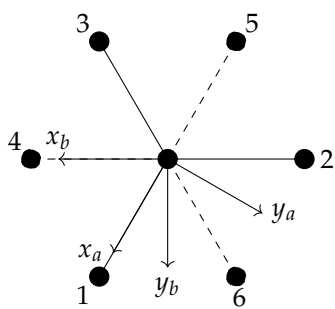

To describe the orientation of the ethane molecule with Euler angles, one approach is to attach coordinate axes on each \ce{CH3} group. Here, the $z$ axis is the same for both and points from \ce{C_$b$} to \ce{C_$a$}, while the $x_a$ and $x_b$ axes point in the direction of \ce{C_$a$}-\ce{H1} and \ce{C_$b$}-\ce{H4}, respectively, when the molecule is viewed in the Newman projection of \fig{fig:coordinate_axes}. The $y$ axes ensure that the Cartesian axes are right handed. With this construction, the $\theta$ and $\phi$ Euler angles describing the direction of the $z$ axis for \ce{CH3} group are the same while the $\chi$ angles describing the rotation about the $z$ axis are different and are denoted by $\chi_a$ and $\chi_b$. These increase in the counterclockwise direction due to the right hand rule. This essentially follows Section 15.4.4 of \cite{98BuJexx} in the definition of the Euler angles according to the convention described in Section 10.1.1 of \cite{98BuJexx}.

\begin{figure}[htbp!]
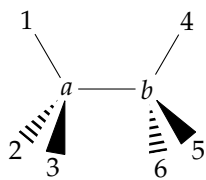

		\centering
		\chemfig{{\it a}([:120]-{1})([:-100]<{3})([:-130]<:{2})-{\it b}([:60]-4)([:-50]<5)([:-80]<:6)}
		\caption{Ethane in the eclipsed configuration.}
		\label{fig:ethane_eclipsed}
	\end{figure}

As described in \cite{98BuJexx}, to achieve maximum separation of the torsional and rotational motion, it is expedient to define two new coordinates from $\chi_a$ and $\chi_b$. We set the rotational coordinate $\chi$ to
\begin{equation}\label{eq:chichiachib}
	\chi = \frac12 ( \chi_a + \chi_b)
\end{equation}
and hence our $x$ axis shown in \fig{fig:dihedrals} halves the angle between \ce{H1-C-C} and \ce{H4-C-C} and increases in the counterclockwise direction. The torsional angle $\tau$ could be defined as
\begin{equation}\label{eq:tau=chia-chib}
\tau = \chi_a - \chi_b
\end{equation}
and hence, as indicated in \fig{fig:dihedrals}, is the angle from \ce{H4-C-C} to \ce{H1-C-C} in the counterclockwise direction.
In the TROVE calculations, we use a different choice and define $\tau$ as the average of three dihedral angles. To define these, we form three pairs of protons $(4,1)$, $(6,2)$, and $(5,3)$ with the two members belonging to different \ce{CH3} groups.  The pairs are chosen such that the protons $i$ and $j$ in each $(i,j)$ pair form a dihedral angle $\tau_{ij}$ of $\pi$ in the staggered equilibrium geometry of \fig{fig:ethane_equilibrium} and $\tau_{ij}=0$ for the eclipsed geometry (see \fig{fig:ethane_eclipsed}), using the labelling of \fig{fig:ethane_labelling}. In a general instantaneous geometry, each $(i,j)$ pair defines a dihedral angle $\tau_{ij}$ (where the positive direction of rotation for the $\tau_{ij}$ angles
is from proton 1 $\rightarrow$ 2 $\rightarrow$ 3)
and the torsional angle is then given by a symmetric combination
\begin{equation}
\tau =
\frac13 ( \tau_{41} + \tau_{62} + \tau_{53}).
\label{eq:taudef}
\end{equation}
With this definition, $\tau=0$, $2\pi/3$ and $4\pi/3$ correspond to eclipsed configurations, while at $\tau = \pi/3$, $\pi$, $5\pi/3$ the molecule is in one of its three equilibrium geometries. The torsional angle $\tau$ has definite  transformation
properties under the operations of \GTS\ (see also \ref{sect:appenix_b}). As discussed in Section~\ref{sec:extenedGTS}, the two  coordinate-pair values ($\tau$, $\chi$) and ($\tau + 2\pi$, $\chi + \pi$) describe the same physical situation. However, the coordinate which $\tau$ is based on, $\chi_a - \chi_b$, has a range of $4\pi$. Although a given geometry can be described by a value of $\tau$ in the interval $[0, 2\pi]$, we must allow $\tau$ to range over $[0, 4\pi]$  to obtain a correct correlation with $\chi_a - \chi_b$ (see Section~\ref{sec:extenedGTS} below).

In conclusion, the coordinate subsets, for which we initially diagonalize reduced Hamiltonians, are
\begin{enumerate}
\item the \ce{C-C} bond length $R$,
\item  six \ce{C-H} bond lengths $r_k$, $k$ $=$ 1, 2, \ldots, 6,
\item six bond angles $\angle$({H$_k$-C-C}) $=$ $\alpha_k$, $k$ $=$ 1, 2, \ldots, 6,
\item four dihedral-angle coordinates
${\gamma}_1$,
${\gamma}_2$,
${\delta}_1$, and
${\delta}_2$,
\item the torsional angle $\tau$, and
\item the three rotational angles $(\theta, \phi,  \chi)$.
\end{enumerate}
The primitive basis functions for subsets 1, 2, 3, and 5 are obtained using the Numerov-Cooley approach \citep{07YuThJe.method,24Numerov.method,61Coxxxx.method},  while we use harmonic-oscillator eigenfunctions~\cite{98BuJexx} for subset 4.
The primitive basis set 5 is obtained by solving the 1D torsional Schr\"{o}dinger equation
\begin{equation}
\label{eq:torsion-schr-eq}
 \left[-\frac{1}{2}\frac{\partial}{\partial \tau} g_{\tau,\tau}^{\rm 1D}(\tau) \frac{\partial}{\partial \tau} + V^{\rm 1D}(\tau) \right] | n \rangle = E_n | n \rangle
\end{equation}
using the basis set constructed from the normalised, $4\pi$-periodic Fourier series functions $\sqrt{1/2\pi} \cos(k \tau / 2 )$ and $\sqrt{1/2\pi} \sin(k \tau / 2 )$.  Here $g_{\tau,\tau}^{\rm 1D}(\tau)$ is the purely torsional element of the TROVE kinetic-energy $g$-matrix (see \cite{07YuThJe.method}) and $V^{\rm 1D}(\tau)$ is the 1-dimensional, $2\pi/3$-periodic torsional part of the 18D potential energy function of ethane with the remaining 17 vibrational coordinates set to their equilibrium values. For both the kinetic-energy function $g_{\tau,\tau}^{\rm 1D}(\tau)$ and the potential function $V^{\rm 1D}(\tau)$ we must consider $\tau$-values in the extended interval $[0, 4\pi]$ as detailed in Section~\ref{sec:extenedGTS} below.

\subsection{Transformation of the vibrational coordinates under \GTS\ }

As described in \sect{sect:irrep_g36}, one can construct every operation of \GTS\ with only four generating operations. In the \Trove\ calculations, we use the generators $R_2^{(-)} = (132)(456)$, $R_2^{(+)} = (123)(456)$, $R_3^{(+)} = (14)(26)(35)(ab)^*$ and $R_4^{(-)} = (14)(25)(36)(ab)$. Then, to describe the transformation properties of a coordinate for all elements in \GTS, we need only determine these properties for the four generating operations. We have implemented in \Trove\ the procedure to generate the irreducible representations of \GTS\ based on these four group generators   and the multiplication rules from Table~\ref{tab:multiplicationtable}. The latter can be conveniently represented as the following recursive rule:
\begin{equation}
T_i = T_j \, T_k,\
\end{equation}
where the operations $T_i$,  $T_j$ and $T_k$ ($i,j,k = 1,\ldots,36$) are as organised in Table~\ref{tab:recurrence_prod}. 

\begin{table}[hbtp!]
\caption{\label{tab:recurrence_prod} The recursive rules to generate the elements of \GTS\ using the four generators  $T_2 = R_2^{(+)} = (123)(465)$, $T_{4} = R_4^{(-)}$ = $(14)(25)(36)(ab)$, $T_7 = R_2^{(-)}$ $=$ $(132)(456)$, and $ T_{19} = R_4^{(+)}$ $=$ $(14)(26)(35)(ab)^*$. See also Table~\protect\ref{tab:multiplicationtable} for the class structure of \GTS\ and  Fig.~\protect\ref{fig:4pi} for an illustration of the effects of the generators.}
\begin{center}
\renewcommand{\arraystretch}{1.35}
\begin{tabular}{lllclll}
\hline\hline
$T_{1\phantom{0}}=$  & $E$      & \phantom{\huge l}                     &&   $T_{19}=$   &
$(14)(25)(36)(ab)$ & $=R_4^{(-)}$                  \\
$T_{2\phantom{0}}=$  &   $(123)(456)$     & $=R_2^{(+)}$     &&          $T_{20}=$   & (16)(24)(34)$(ab)$   &   $= T_{7} \, T_{21}$    \\
$T_{3\phantom{0}}=$  & $(132)(465)$  &    $={T_2}^2 $    &&     $T_{21}=$   & (15)(26)(34)$(ab)$  & $= T_{7}\, T_{19}$  \\
$T_{4\phantom{0}}=$  &  $(14)(26)(35)(ab)^*$  &                   $=R_4^{(+)}$            &&   $T_{22}=$  & (153426)$(ab)$  & $= T_{19} \, T_2$   \\
$T_{5\phantom{0}}=$  &  $(16)(25)(34)(ab)^*$   &  $=T_2   \, T_{6}$    &&      $T_{23}=$   & (162435)$(ab)$    &  $=T_{19} \, T_3 $   \\
$T_{6\phantom{0}}=$  & $(15)(24)(36)(ab)^*$ &    $=T_{2} \, T_{4}$    && $T_{24}=$   & (143624)$(ab)$   &    $= T_{20}  \, T_2  $   \\
$T_{7\phantom{0}}=$  &   $(132)(456)$  &        $=R_2^{(-)}$              &&   $T_{25}=$  & (152634)$(ab)$  &       $=   T_{20}  \, T_3$   \\
$T_{8\phantom{0}}=$  &  $(123)(465)$  &   $={T_7}^2 $    &&   $T_{26}=$  & (163524)$(ab)$  &       $=  T_{21} \,T_2  $   \\
$T_{9\phantom{0}}=$  &   $(465)$ &    $=T_7   \, T_{2}$    &&   $T_{27}=$   & (142536)$(ab)$   &    $= T_{21} \, T_3  $   \\
$T_{10}=$ & (123)  &    $=T_7   \, T_{3}$    &&   $T_{28}=$   & (23)(56)$^*$  &    $=T_{19}  \, T_4   $   \\
$T_{11}=$ &    $(132)$ &   $=T_8   \, T_{2}$    &&   $T_{29}=$   &  (13)(46)$^*$  &   $=  T_{19}  \, T_{5}  $   \\
$T_{12}=$ &  $(456)$ &     $=T_8   \, T_{3}$    &&   $T_{30}=$   & (12)(45)$^*$   &     $=  T_{19}  \, T_{6}  $   \\
$T_{13}=$ & (152436)$(ab)^*$   &    $=T_7   \, T_{4}$    &&   $T_{31}=$   & (12)(46)$^*$   &   $= T_{20} \, T_4   $   \\
$T_{14}=$ & (142635)$(ab)^*$   &   $=T_{7}   \, T_{5}$    &&   $T_{32}=$   & (23)(45)$^*$  &    $=T_{20} \, T_{5} $      \\
$T_{15}=$ & (162534)$(ab)^*$    &  $=T_{7}   \, T_{6}$    &&   $T_{33}=$   & (13)(56)$^*$ &   $= T_{20}  \, T_{6}  $   \\
$T_{16}=$ & (163425)$(ab)^*$    &  $=T_8   \, T_{4}$    &&   $T_{34}=$   & (13)(45)$^*$   &   $= T_{21} \, T_4    $   \\
$T_{17}=$ & (153624)$(ab)^*$    &  $=T_{8}   \, T_{5}$    &&   $T_{35}=$   & (12)(56)$^*$ &   $= T_{21} \, T_{5}   $   \\
$T_{18}=$ &  (143526)$(ab)^*$  &   $=T_{8}   \, T_{6}$    &&   $T_{36}=$   & (23)(46)$^*$ &     $=T_{21} \, T_{6}    $   \\
\hline
\hline
\end{tabular}
\end{center}
\end{table}

Each of the vibrational coordinates and the torsional coordinate can be expressed as a function of the nuclear Cartesian coordinates. In the following, and in \ref{sec:appendixB}, we describe the procedure in determining the transformation of these coordinates under the \GTS\ operations in a systematic way, although in the majority of cases the result is intuitive.

The \ce{C-C} bond length $R$ is invariant
under all \GTS\ operations. We label the six \ce{C-H_k} bond lengths $r_k$,
and the six bond angles $\angle${H$_k$-C-C} $=$ $\alpha_k$, by the generic labels $\beta_k$,  as the two subsets transform identically. The transformation properties are determined by recognizing
 that after the operation $O_i$ $\in$ \GTS\ has been carried out, the proton 1, say, occupies the position
previously occupied by the proton $k$ (and the C nucleus to which it is bound has
moved with it), and therefore the transformed value of the \ce{C-H$_1$} bond length is $r_{1}^{\prime}$ $=$ $r_k$, the original value of the \ce{C-H$_k$} bond length,  with analogous considerations
 for the bond lengths $r_2$, \dots, $r_6$, and the $\alpha_i$ angles.

For ($\beta_1$, $\beta_2$, $\beta_3$), we obtain the following transformation properties under the generating operations:
\begin{center}
\begin{tabbing}
\parbox[l]{7.4truecm}{
$
\begin{pmatrix}
\beta_1^\prime \\
\beta_2^\prime \\
\beta_3^\prime
\end{pmatrix} =
\begin{pmatrix}
0 & 0 & 1  \\
1 & 0 & 0 \\
0 & 1 & 0
\end{pmatrix}
\begin{pmatrix}
\beta_1 \\
\beta_2 \\
\beta_3
\end{pmatrix}
	$
for (123)(456),}
\parbox[l]{8.1truecm}{
$
\begin{pmatrix}
\beta_1^\prime \\
\beta_2^\prime \\
\beta_3^\prime
\end{pmatrix} =
\begin{pmatrix}
1 & 0 & 0  \\
0 & 0 & 1 \\
0 & 1 & 0
\end{pmatrix}
\begin{pmatrix}
\beta_4 \\
\beta_5 \\
\beta_6
\end{pmatrix}
$
for $(14)(26)(35)(ab)^*,$}
\end{tabbing}
\begin{tabbing}
\parbox[l]{7.4truecm}{
$
\begin{pmatrix}
\beta_1^\prime \\
\beta_2^\prime \\
\beta_3^\prime
\end{pmatrix} =
\begin{pmatrix}
0 & 1 & 0  \\
0 & 0 & 1 \\
1 & 0 & 0
\end{pmatrix}
\begin{pmatrix}
\beta_1\\
\beta_2 \\
\beta_3
\end{pmatrix}
$
for $(132)(456)$,}
\parbox[l]{8.0truecm}{
$
\begin{pmatrix}
\beta_1^\prime \\
\beta_2^\prime \\
\beta_3^\prime
\end{pmatrix} =
\begin{pmatrix}
1 & 0 & 0  \\
0 & 1 & 0 \\
0 & 0 & 1
\end{pmatrix}
\begin{pmatrix}
\beta_4 \\
\beta_5 \\
\beta_6
\end{pmatrix}
$
for $(14)(25)(36)(ab)$}
\end{tabbing}
\end{center}
while for $(\beta_4, \beta_5, \beta_6)$ they are given by
\begin{center}
\begin{tabbing}
\parbox[l]{7.4truecm}{
	$
\begin{pmatrix}
\beta_4^\prime \\
\beta_5^\prime \\
\beta_6^\prime
\end{pmatrix} =
\begin{pmatrix}
0 & 0 & 1  \\
1 & 0 & 0 \\
0 & 1 & 0
\end{pmatrix}
\begin{pmatrix}
\beta_4 \\
\beta_5 \\
\beta_6
\end{pmatrix}
$
	for $(123)(456)$,}
\parbox[l]{8.1truecm}{
	$
\begin{pmatrix}
\beta_4^\prime \\
\beta_5^\prime \\
\beta_6^\prime
\end{pmatrix} =
\begin{pmatrix}
1 & 0 & 0  \\
0 & 0 & 1 \\
0 & 1 & 0
\end{pmatrix}
\begin{pmatrix}
\beta_1 \\
\beta_2 \\
\beta_3
\end{pmatrix}
$
for $(14)(26)(35)(ab)^*,$}
\end{tabbing}

\begin{tabbing}
\parbox[l]{7.4truecm}{
$
\begin{pmatrix}
\beta_4^\prime \\
\beta_5^\prime \\
\beta_6^\prime
\end{pmatrix} =
\begin{pmatrix}
0 & 0 & 1  \\
1 & 0 & 0 \\
0 & 1 & 0
\end{pmatrix}
\begin{pmatrix}
\beta_4 \\
\beta_5 \\
\beta_6
\end{pmatrix}
	$
for $(132)(456)$,}
\parbox[l]{8.0truecm}{
$
\begin{pmatrix}
\beta_4^\prime \\
\beta_5^\prime \\
\beta_6^\prime
\end{pmatrix} =
\begin{pmatrix}
1 & 0 & 0  \\
0 & 1 & 0 \\
0 & 0 & 1
\end{pmatrix}
\begin{pmatrix}
\beta_1 \\
\beta_2 \\
\beta_3
\end{pmatrix}
	$
	for $(14)(25)(36)(ab)$.}
\end{tabbing}
\end{center}

The dihedral-angle coordinates
$({\gamma}_1, {\gamma}_2, {\delta}_1,
{\delta}_2)$ are mixed by the \GTS\ operations. The transformation properties are most easily determined by using the correspondence with the transformation of the
$r_k$ and $\alpha_k$ coordinates.
For $({\gamma}_1, {\gamma}_2)$ we write
\begin{equation}
\begin{pmatrix}
{\gamma}_1 \\
{\gamma}_2 \\
2\pi \\
\end{pmatrix} =
{\mathbf Z} \,
\begin{pmatrix}
\theta_{12} \\ \theta_{23} \\ \theta_{31} \\
\end{pmatrix}
\end{equation}
with
\begin{equation}
{\mathbf Z} =
\begin{pmatrix}
-\frac{1}{\sqrt{6}}& \frac{2}{\sqrt{6}} &-\frac{1}{
 \sqrt{6}}\\ -{\frac{1}{\sqrt{2}}}&0&{\frac{1}{\sqrt{2}}}
 \\
1 & 1 & 1 \\
\end{pmatrix}
\end{equation}
where we have taken into account the constraint $\theta_{12} + \theta_{23} + \theta_{31}$ $=$
$2 \pi$ in the third row of ${\mathbf Z}$.

After we have carried out the  operation $(123)(456)$, the protons 1, 2, and 3 are found at the positions initially occupied by the protons 3, 1, and 2, respectively, and so
the angles
 $\theta_{12}$,
$\theta_{23}$, $\theta_{31}$
are permuted as follows
\begin{equation}
\begin{pmatrix}
\theta_{12}' \\ \theta_{23}' \\ \theta_{31}' \\
\end{pmatrix} =
\begin{pmatrix}
\theta_{31} \\ \theta_{12} \\ \theta_{23} \\
\end{pmatrix} =
{\mathbf S} \,
\begin{pmatrix}
\theta_{12} \\ \theta_{23} \\ \theta_{31} \\
\end{pmatrix}
= {\mathbf S}\, {\mathbf Z}^{-1} \,
\begin{pmatrix}
{\gamma}_1 \\
{\gamma}_2 \\
2\pi \\
\end{pmatrix}
\end{equation}
with
\begin{equation}
{\mathbf S} =
\begin{pmatrix}
0 & 0 & 1 \\
1 & 0  & 0 \\
0 &  1  & 0 \\
\end{pmatrix}
\end{equation}
and we can finally calculate the transformed values of
$({\gamma}_1, {\gamma}_2)$ as
\begin{equation}
\begin{pmatrix}
{\gamma}_1' \\
{\gamma}_2' \\
2\pi \\
\end{pmatrix} =
{\mathbf Z}\,
\begin{pmatrix}
\theta_{12}' \\ \theta_{23}' \\ \theta_{31}' \\
\end{pmatrix} =
{\mathbf Z}\, {\mathbf S}\, {\mathbf Z}^{-1} \,
\begin{pmatrix}
{\gamma}_1 \\
{\gamma}_2 \\
2\pi \\
\end{pmatrix}
\end{equation}
where
\begin{equation}
{\mathbf Z}\, {\mathbf S}\, {\mathbf Z}^{-1} =
\begin{pmatrix}
-\frac12 & -\frac{\sqrt{3}}{2} & 0  \\
\frac{\sqrt{3}}{2} & -\frac12 & 0 \\
0 & 0 & 1 \\
\end{pmatrix}.
\end{equation}
With the upper 2~$\times$~2 corner of this matrix we can express the transformed values $({\gamma}_1', {\gamma}_2')$ in terms of $({\gamma}_1, {\gamma}_2)$.
 The transformation matrix for $({\gamma}_1, {\gamma}_2)$ for $(123)(456)$, and the ones for the other generating operations, obtained in a similar manner, are
\begin{center}
\begin{tabbing}
\parbox[l]{7.3truecm}{
$
\begin{pmatrix}
{\gamma}_1^\prime \\
{\gamma}_2^\prime \\
\end{pmatrix} =
\begin{pmatrix}
-\frac12 & -\frac{\sqrt{3}}{2}   \\
\frac{\sqrt{3}}{2} & -\frac12 \\
\end{pmatrix}
\begin{pmatrix}
{\gamma}_1 \\
{\gamma}_2\\
\end{pmatrix}
$ for $(123)(456)$,}
\parbox[l]{8.2truecm}{
$
\begin{pmatrix}
{\gamma}_1^\prime \\
{\gamma}_2^\prime \\
\end{pmatrix} =
\begin{pmatrix}
1 & 0   \\
0 & -1 \\
\end{pmatrix}
\begin{pmatrix}
{\delta}_1 \\
{\delta}_2\\
\end{pmatrix}
$ for $(14)(26)(35)(ab)^*,$}
\end{tabbing}
\begin{tabbing}
\parbox[l]{7.3truecm}{
$
\begin{pmatrix}
{\gamma}_1^\prime \\
{\gamma}_2^\prime \\
\end{pmatrix} =
\begin{pmatrix}
-\frac12 & \frac{\sqrt{3}}{2}   \\
-\frac{\sqrt{3}}{2} & -\frac12 \\
\end{pmatrix}
\begin{pmatrix}
{\gamma}_1 \\
{\gamma}_2\\
\end{pmatrix}
$ for $(132)(456)$,}
\parbox[l]{8.2truecm}{
$
\begin{pmatrix}
{\gamma}_1^\prime \\
{\gamma}_2^\prime \\
\end{pmatrix} =
\begin{pmatrix}
1 & 0   \\
0 & 1 \\
\end{pmatrix}
\begin{pmatrix}
{\delta}_1 \\
{\delta}_2\\
\end{pmatrix}
$ for $(14)(25)(36)(ab),$}
\end{tabbing}
\end{center}
and those for ($\delta_1$, $\delta_2$) are given by
\begin{center}
\begin{tabbing}
\parbox[l]{7.3truecm}{
$
\begin{pmatrix}
{\delta}_1^\prime \\
{\delta}_2^\prime \\
\end{pmatrix} =
\begin{pmatrix}
-{\frac{1}{2}}&{-\frac{\sqrt{3}}{2}}\\ {\frac{
 \sqrt{3}}{2}}&-{\frac{1}{2}}\\
\end{pmatrix}
\begin{pmatrix}
{\delta}_1 \\
{\delta}_2\\
\end{pmatrix}
$ for $(123)(456)$,}
\parbox[l]{8.2truecm}{
$
\begin{pmatrix}
{\delta}_1^\prime \\
{\delta}_2^\prime \\
\end{pmatrix} =
\begin{pmatrix}
1 & 0   \\
0 & -1 \\
\end{pmatrix}
\begin{pmatrix}
{\gamma}_1 \\
{\gamma}_2\\
\end{pmatrix}
$ for $(14)(26)(35)(ab)^*,$}
\end{tabbing}

\begin{tabbing}
\parbox[l]{7.3truecm}{
$
\begin{pmatrix}
{\delta}_1^\prime \\
{\delta}_2^\prime \\
\end{pmatrix} =
\begin{pmatrix}
-{\frac{1}{2}}&{-\frac{\sqrt{3}}{2}}\\ {\frac{
 \sqrt{3}}{2}}&-{\frac{1}{2}}\\
\end{pmatrix}
\begin{pmatrix}
{\delta}_1 \\
{\delta}_2\\
\end{pmatrix}
$ for $(132)(456)$,}
\parbox[l]{8.2truecm}{
$
\begin{pmatrix}
{\delta}_1^\prime \\
{\delta}_2^\prime \\
\end{pmatrix} =
\begin{pmatrix}
1 & 0   \\
0 & 1 \\
\end{pmatrix}
\begin{pmatrix}
{\gamma}_1 \\
{\gamma}_2\\
\end{pmatrix}
$ for $(14)(25)(36)(ab).$}
\end{tabbing}
\end{center}

\subsection{The extended molecular symmetry group \GTS(EM) and the transformation of the torsional coordinate}
\label{sec:extenedGTS}
As explained in Section~{15.4.4} of \cite{98BuJexx}, our separation of the rotational and torsional degrees of freedom has led $\chi$ and $\tau$ being double-valued. That is, there are two sets of
$(\chi,\tau)$ values associated with the same physical situation.
This is most straightforwardly seen
by considering Eq.~(\ref{eq:chichiachib}) and the three angles
$\chi_a$, $\chi_b$, and $\chi$ appearing in it. The angle $\chi_a$ is determined entirely by the positions in space of the protons 1, 2, 3 and their carbon nucleus C$_a$; $\chi_b$ is determined by the positions of the protons 4, 5, 6 and their carbon nucleus C$_b$; and $\chi$ is the average of the two. Due to the $2\pi$ periodicity of $\chi_a$, increasing it by $2\pi$ does not change the positions in space of the nuclei in \ce{C_$a$H3}, however in this case $\chi$ $\rightarrow$ $\chi+\pi$ and $\tau \rightarrow \tau + 2\pi$.
That is, the two coordinate pairs ($\chi$,$\tau$) and ($\chi+\pi$,$\tau+2\pi$) describe identical physical situations.

One way of avoiding the ambiguity described above would be to use a molecule-fixed axis system with, for example, $\chi$ $=$ $\chi_a$. This molecule-axis system is attached to the \ce{CH3} group with the protons 1, 2, 3, and not influenced by the other \ce{CH3} group. A $\chi$ coordinate chosen in this manner has no ambiguity. However, as already mentioned it is advantageous to choose $\chi$ according to Eq.~(\ref{eq:chichiachib}) since this choice
(called the Internal Axis Method (IAM), see Section~{15.2.2} of \cite{98BuJexx}) yields a particularly useful form of the rotation-torsion-vibration Hamiltonian with minimized coupling between rotation and torsion. It is desirable to keep the definition of $\chi$ from Eq.~(\ref{eq:chichiachib}) and find a way of treating the ambiguity. We use here the procedure first proposed by Hougen~\cite{64Hougen} in 1964, primarily for dimethylacetylene H$_3$CCCCH$_3$, whose internal rotation is essentially free, but generally applicable to other molecules such as ethane H$_3$CCH$_3$, hydrogen peroxide HOOH, and disulfane HSSH with two identical moieties carrying out internal rotation. Recently, these ideas have been extended to molecules with two different moieties and applied to oxadisulfane (a.k.a. hydrogen thio-peroxide) HSOH~\cite{04YAMADA,09YAMADA}, and more recently they have been applied to computation of the rotation-torsion-vibrational spectra of hydrogen peroxide HOOH in Ref.~\cite{15AlOvPo.H2O2,16AlPoOv.HOOH}, where the HOOH kinetic energy operator was built using the $x$-axis chosen to halve the dihedral angle between the \ce{H$_1$-O-O} and \ce{H$_4$-O-O} planes, as well as in Ref.~\cite{18SzViRe}.

\begin{figure}[htbp!]
\centering

\parbox{3.5cm}{(A)
\begin{tikzpicture}[scale=1.1]
\draw (0,0) circle(2pt)[fill] -- ++ (0:1) circle(2pt)[fill] node[right] {\ 2};
\draw (0,0) -- ++(120:1) circle(2pt)[fill] node[above left] {3};
\draw (0,0) -- ++(240:1) circle(2pt)[fill] node[below left] {1};

\draw [dashed] (0,0)  -- ++ (40:1) circle(2pt)[fill] node[above right] {5};
\draw [dashed] (0,0) -- ++(160:1) circle(2pt)[fill] node[above left] {4};
\draw [dashed] (0,0) -- ++(280:1) circle(2pt)[fill] node[below] {6};
\draw[->] (0,0) -- ++ (200:1) node[left] {$x$};

\draw [domain=0:60] plot  ({ 0.3*cos(\x)}, {0.3*sin(\x)}) node[above] {$\theta$};
\draw [domain=60:120] plot  ({ 0.3*cos(\x)}, {0.3*sin(\x)});

\draw [domain=160:180] plot  ({ 0.3*cos(\x)}, {0.3*sin(\x)}) node[left] {$\tau$};
\draw[->] [domain=180:240] plot  ({ 0.3*cos(\x)}, {0.3*sin(\x)});
\end{tikzpicture}
}
 \parbox{3.5cm}{(B)
 \begin{tikzpicture}[scale=1.1]
\draw (0,0) circle(2pt)[fill] -- ++ (0:1) circle(2pt)[fill] node[right] {\ 3};
\draw (0,0) -- ++(120:1) circle(2pt)[fill] node[above left] {1};
\draw (0,0) -- ++(240:1) circle(2pt)[fill] node[below left] {2};

\draw [dashed] (0,0)  -- ++ (40:1) circle(2pt)[fill] node[above right] {6};
\draw [dashed] (0,0) -- ++(160:1) circle(2pt)[fill] node[above left] {5};
\draw [dashed] (0,0) -- ++(280:1) circle(2pt)[fill] node[below] {4};
\draw[->] (0,0) -- ++ (200:1) node[left] {$x$};

\draw [domain=280:400] plot  ({ e^((\x+200)/(3000))*0.3*cos(\x)}, {e^((\x+200)/(3000))*0.3*sin(\x)}) node[above] {$\tau$};
\draw[->] [domain=400:840] plot  ({ e^((\x+200)/(3000))*0.3*cos(\x)}, {e^((\x+200)/(3000))*0.3*sin(\x)});
\end{tikzpicture}
}
\parbox{3.5cm}{(C)
\begin{tikzpicture}[scale=1.1]
\draw (0,0) circle(2pt)[fill] -- ++ (0:1) circle(2pt)[fill] node[right] {\ 1};
\draw (0,0) -- ++(120:1) circle(2pt)[fill] node[above left] {2};
\draw (0,0) -- ++(240:1) circle(2pt)[fill] node[below left] {3};

\draw [dashed] (0,0)  -- ++ (40:1) circle(2pt)[fill] node[above right] {6};
\draw [dashed] (0,0) -- ++(160:1) circle(2pt)[fill] node[above left] {5};
\draw [dashed] (0,0) -- ++(280:1) circle(2pt)[fill]  node[below] {4};
\draw[->] (0,0) -- ++ (320:1) node[right] {$x$};

\draw [domain=280:320] plot  ({ e^((\x+200)/(3000))*0.3*cos(\x)}, {e^((\x+200)/(3000))*0.3*sin(\x)}) node[right] {$\tau$};
\draw[->] [domain=320:360] plot  ({ e^((\x+200)/(3000))*0.3*cos(\x)}, {e^((\x+200)/(3000))*0.3*sin(\x)});
\end{tikzpicture}
}
\parbox{3.5cm}{(D)
\begin{tikzpicture}[scale=1.1]
\draw[dashed] (0,0) circle(2pt)[fill] -- ++ (0+180:1) circle(2pt)[fill] node[left] {\ 6};
\draw[dashed]  (0,0) -- ++(120+180:1) circle(2pt)[fill] node[below right] {5};
\draw[dashed]  (0,0) -- ++(240+180:1) circle(2pt)[fill] node[above right] {4};

\draw [] (0,0)  -- ++ (40+180:1) circle(2pt)[fill] node[below left] {3};
\draw [] (0,0) -- ++(160+180:1) circle(2pt)[fill] node[below right] {1};
\draw [] (0,0) -- ++(280+180:1) circle(2pt)[fill] node[above] {2};
\draw[->] (0,0) -- ++ (200+180:1) node[right] {$x$};

\draw [domain=60:200] plot  ({ 0.3*cos(\x)}, {0.3*sin(\x)}) node[left] {$\tau$};
\draw[->] [domain=200:340] plot  ({ 0.3*cos(\x)}, {0.3*sin(\x)}) ;
\end{tikzpicture}
}

\parbox{3.5cm}{(E)
\begin{tikzpicture}[scale=1.1]
\draw (0,0) circle(2pt)[fill] -- ++ (0:1) circle(2pt)[fill] node[right] {\ 5};
\draw (0,0) -- ++(120:1) circle(2pt)[fill] node[above left] {6};
\draw (0,0) -- ++(240:1) circle(2pt)[fill] node[below left] {4};

\draw [dashed] (0,0)  -- ++ (40:1) circle(2pt)[fill] node[above right] {2};
\draw [dashed] (0,0) -- ++(160:1) circle(2pt)[fill] node[above left] {1};
\draw [dashed] (0,0) -- ++(280:1) circle(2pt)[fill] node[below] {3};
\draw[->] (0,0) -- ++ (200:1)  node[left] {$x$};

\draw[<-] [domain=160:180] plot  ({ 0.3*cos(\x)}, {0.3*sin(\x)}) node[left] {$\tau$};
\draw [domain=180:240] plot  ({ 0.3*cos(\x)}, {0.3*sin(\x)}) ;
\end{tikzpicture}
}
\parbox{3.5cm}{(F)
\begin{tikzpicture}[scale=1.1]
\draw (0,0) circle(2pt)[fill] -- ++ (0:1) circle(2pt)[fill] node[right] {\ 2};
\draw (0,0) -- ++(120:1) circle(2pt)[fill] node[above left] {3};
\draw (0,0) -- ++(240:1) circle(2pt)[fill] node[below left] {1};

\draw [dashed] (0,0)  -- ++ (40:1) circle(2pt)[fill] node[above right] {5};
\draw [dashed] (0,0) -- ++(160:1) circle(2pt)[fill] node[above left] {4};
\draw [dashed] (0,0) -- ++(280:1) circle(2pt)[fill] node[below] {6};
\draw[->] (0,0) -- ++ (200+180:1) node[right] {$x$};

\draw [domain=160:330] plot  ({ e^((\x+200)/(3000))*0.3*cos(\x)}, {e^((\x+200)/(3000))*0.3*sin(\x)}) node[right] {$\tau$};Fa
\draw[->][domain=330:240+360] plot  ({ e^((\x+200)/(3000))*0.3*cos(\x)}, {e^((\x+200)/(3000))*0.3*sin(\x)});
\end{tikzpicture}
}

\caption{Newman projections of ethane showing the effects of the \GTS(EM) generators (and their \GTS\ partners; see Table~\protect\ref{tab:torsion_transformations})
on $\tau$ and $\chi$; the change of $\tau$ is represented by the curved arrow
encircling the C--C axis  and the change of $\chi$ is illustrated by the change in $x$-axis orientation. (A): Starting configuration with $\tau = 4\pi/9$ and the $x$-axis forming the clockwise angle $8\pi/9$ with the horizontal.  (B): The effect of $ab$ $\sim$ (123)(465) which causes  $\tau$ to decrease by $4\pi/3$ (equivalent to an increase of $8\pi/3$ ) and $\chi$ to remain constant. (C): The effect of $a^5 \, b$ $\sim$ (132)(456) with $\tau$ remaining constant and $\chi$ changing by $+2\pi/3$.  (D): The effect of $c$ $\sim$ (14)(26)(35)$(ab)^*$ with $\tau$ $\rightarrow$ $2\pi - \tau$ and $\chi$ $\rightarrow$ $\chi +  \pi$.  (E): The effect of $d\, c$ $\sim$ (14)(25)(36)$(ab)$ under which $\tau$ and $\chi$ are both invariant. (F): The effect of $E'$ which has no \GTS\ partner,  $E'\, \tau$ $=$  $\tau+2\pi$ and $E'\, \chi$ $=$ $\chi+\pi$. }
\label{fig:4pi}

\end{figure}
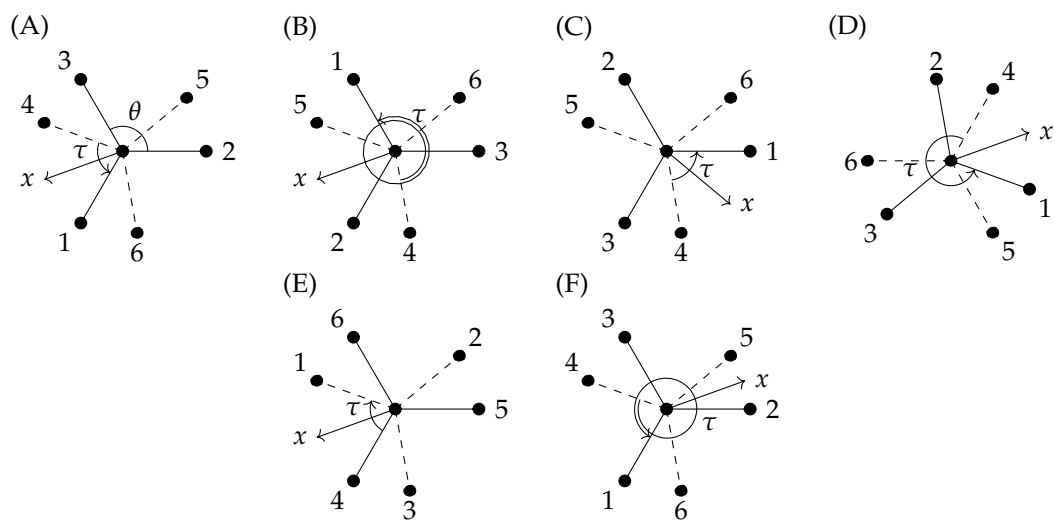

We consider the effect of \GTS\ operations on $\chi_a$, described thoroughly in Section 12.1.1 of \cite{98BuJexx}, where the axes are rotated in such a way that the coordinates of the nuclei -- measured in a space-fixed axis system -- remain the same after the operation, assuming the \ce{C_$a$H3} group is in the equilibrium configuration. Considering the operation (123), we see that, using our convention, $\chi_a$ becomes $\chi_a + 4\pi/3$ or, equivalently, $\chi_a - 2\pi/3$. This would correspond to the pair $(\tau, \chi)$ becoming $(\tau + 4\pi/3, \chi + 2\pi/3)$ or $(\tau - 2\pi/3, \chi - \pi/3)$, which are no longer equivalent but correspond to the same physical situation as noted before. For $(123)^3 = E$, the changes become $(\tau + 4\pi, \chi + 2\pi) = (\tau, \chi)$ or $(\tau -2\pi, \chi - \pi) = (\tau + 2\pi, \chi + \pi)$. These are illustrated in \fig{fig:4pi}(A) and \fig{fig:4pi}(F), respectively.  

In order to deal with the double-valuedness of ($\tau$, $\chi$), we extend the symmetry description in the manner first introduced by Hougen~\cite{64Hougen}, by extending \GTS\ to the extended molecular symmetry group \GTS(EM) as explained in Section~15.4.4  of \cite{98BuJexx}.
To appreciate the definition of \GTS(EM), we note that the
internal coordinates $R$, $r_k$ and $\alpha_k$ ($k$ $=$ 1, 2, \ldots, 6),
${\gamma}_1$,
${\gamma}_2$,
${\delta}_1$,
${\delta}_2$,
  $\theta$, and $ \phi$ used for ethane in the present work are all ``space-fixed'' in the sense of Section~15.4.4 in~\cite{98BuJexx}; the instantaneous values of these coordinates can be unambiguously obtained from the instantaneous coordinate values of the nuclei in a Cartesian, space-fixed axis system~\cite{98BuJexx}, $XYZ$ say. \ref{sec:appendixB} explains in detail how the internal coordinates values are determined from the Cartesian coordinate values.

 We present here the ideas of
Hougen~\cite{64Hougen}, using the more modern notation of
Section~15.4.4 in~\cite{98BuJexx}. The extension of
\GTS\ to  \GTS(EM) involves the introduction of
 a fictitious operation $E'$ (taken to be different from the identity $E$) which, for ethane, we can think of as letting the \ce{C_$a$H3} do a full torsional revolution relative to the \ce{C_$b$H3}. That is, $E'$ has the effect of transforming $\chi$ $\rightarrow$ $\chi+\pi$ and $\tau$ $\rightarrow$ $\tau + 2\pi$. After the application of $(E')^2$  the molecule-fixed axis system is back where it started, and so we take $(E')^2$ $=$ $E$. $E'$ does not affect the space-fixed nuclear coordinates; it has the same effect as the identity operation
on the complete  rotation-torsion-vibration wavefunction of ethane.

We now define four generating operations $a$, $b$, $c$ and $d$
that transform $\chi$ and $\tau$ unambiguously and
 have the same effect on the space-fixed
coordinates [$R$, $r_k$ and $\alpha_k$ ($k$ $=$ 1, 2, \ldots, 6),
${\gamma}_1$,
${\gamma}_2$,
${\delta}_1$,
${\delta}_2$,
  $\theta$, and $ \phi$]
as the \GTS\ operations (123), (456), (14)(26)(35)($ab$)$^*$ and (23)(56)$^*$, respectively.
 Tables~A-28 and~A-33 of~\cite{98BuJexx} are the character tables of \GTS\ and \GTS(EM), respectively.
 Comparison of the two tables shows that
the four generating operations
$R_2^{(-)} = (132)(456)$, $R_2^{(+)} = (123)(456)$, $R_4^{(+)} = (14)(26)(35)(ab)^*$ and $R_4^{(-)} = (14)(25)(36)(ab)$
chosen for \GTS\ in the present work correspond to $ab$, $a^5\, b$, $c$, and $d\, c$, respectively.
Following Section~15.4.4 of~\cite{98BuJexx},
we \textit{define} in Table~\ref{tab:torsion_transformations}
the effect on $\chi$ and $\tau$ of the \GTS(EM)
generators. The approach   of~\cite{98BuJexx} to simply postulate, by definition,  the
effect  of the \GTS(EM)
generators on $\chi$ and $\tau$ may appear arbitrary. It is not, however.
We show in
Fig.~\ref{fig:4pi} how, for each of the four \GTS(EM) generators
$ab$, $a^5\, b$, $c$, and $d\, c$, the effect on
$\chi$ and $\tau$ can be  explained as the effect of the \GTS\ partner
of the \GTS(EM) generator in question (Table~\ref{tab:torsion_transformations}).

\begin{table}[hbtp!]
\caption{\label{tab:torsion_transformations} Transformation of the torsion angle $\tau$ and the rotation angle $\chi$ under the generators of \GTS(EM).}
\begin{center}
\begin{tabular}{llll}
\hline\hline
Transformed $\tau$ &Transformed $\chi$ &
\GTS(EM) generator & \GTS\ generator \\
\hline
$\tau - 4\pi/3$ & $\chi$ & $ab$ & (123)(456)  \\
$\tau$  & $\chi + 2\pi/3$ & $a^5\, b$ &(132)(456) \\
$2\pi- \tau$ &$\chi + \pi$ & $c$ & (14)(26)(35)($ab$)$^*$ \\
$\tau$ & $\chi$ & $d\, c$ &(14)(25)(36)($ab$) \\
$\tau+2\pi$ &$\chi + \pi$& $E'$ & \\
\hline\hline
\end{tabular}
\end{center}
\end{table}

The group generated by the five operations
$a$, $b$, $c$, $d$, and $E'$ has 72 elements and, as mentioned above, it is denoted
\GTS(EM), called an extended molecular symmetry (EMS)
group, and its
character table  is given as Table~A-33
of~\cite{98BuJexx}. This character table shows that we can think of \GTS(EM) as a direct product
\begin{equation}
 \text{\GTS(EM)} = \text{\GTS} \times \{ E,E' \},
\end{equation}
where the group $\mathcal{G}_{36}$ is constructed from
the generating operations $a$, $b$, $c$ and $d$; it is isomorphic to \GTS, so that it has the irreducible representations given in Table~\ref{tab:chartab}.
The two-element group $\{ E,E' \}$ is cyclic of order 2 and  has two irreps $A'$ and $A''$, both one-dimensional, with the representation matrices $1$ or $-1$, respectively, under $E'$. The irreps of \GTS(EM) are straightforwardly constructed from those of \GTS.
 Each irrep $\Gamma$ of \GTS\ in Table~\ref{tab:chartab} gives rise to two irreps of \GTS(EM),
 $\Gamma_s$ $=$ $(\Gamma,A')$ and
 $\Gamma_d$ $=$ $(\Gamma,A'')$ as given in Table~{A-33} of \cite{98BuJexx}.
 An irrep $\Gamma_s$ has identical characters for the operations $E$ and $E'$, $\chi_s [E']$ $=$ $\chi_s [E]$, whereas for the irrep
 $\Gamma_d$, $\chi_d [E']$ $=$ $-\chi_d [E]$.
As long as we pretend that $E'$ $\ne$ $E$, we must also pretend that the coordinate values $(\tau,\chi)$ and $E'\, (\tau,\chi)$ $=$ $(\tau+2\pi,\chi+\pi)$ describe different physical situations. As a consequence, we must allow $\tau$ to be periodic with a period of $4\pi$ as already mentioned in connection with Eq.~(\ref{eq:torsion-schr-eq}). The torsional potential energy function $V^{\rm 1D}(\tau)$ is periodic with period $2\pi$, $V^{\rm 1D}(\tau)$ $=$ $V^{\rm 1D}(\tau+2\pi)$ for $\tau$ $\in$ $[0, 2\pi]$, and this symmetry causes the torsional wavefunctions $| n \rangle$
from Eq.~(\ref{eq:torsion-schr-eq}) to be either symmetric (of $\Gamma_s$ symmetry) or antisymmetric (of $\Gamma_d$ symmetry) under $E'$.

We know that in reality, $E'$ $=$ $E$, and so only functions and coordinates with  $\Gamma_s$ symmetries occur in nature. Since we can form, for example, basis functions of an allowed $\Gamma_s$ symmetry as products of an even number of factors, each with a forbidden symmetry $\Gamma_d'$, say, we need to consider also the $\Gamma_d$ symmetries initially for the torsional and rotational basis functions. The final wavefunctions resulting from our theoretical calculations should be subjected to a `reality check': They must necessarily have a $\Gamma_s$ symmetry in \GTS(EM). In particular, torsional basis functions of $d$ symmetry must be combined with rotational basis functions of $d$ symmetry to produce a torsion-rotation basis function of an allowed $s$ symmetry.

\subsection{Rigid-symmetric-rotor function representations}
As mentioned above, we use the rigid-symmetric-rotor eigenfunctions $\vert J, k, m \rangle$~\cite{98BuJexx} as primitive rotational basis functions. For $K\neq 0$ (where $K = \abs{k}$), the corresponding symmetrized functions
 are defined as
\begin{equation}
\ket{JKm\eta} = \frac{i^{\eta}(-1)^\sigma}{\sqrt{2}}(\ket{J, k, m } + (-1)^{J+K}\ket{J, -k, m})
\end{equation}
where
\begin{equation}
\begin{cases}
\sigma = K \ \text{mod} \ 3 & \eta = 1 \\
\sigma = 0 & \eta = 0, \\
\end{cases}
\end{equation}
while for $K=0$ it is $\ket{J 0 0 \eta} = i^{\eta}\ket{J,0,0}$ where $\eta= J \ \text{mod} \ 2 $. Using the rigid rotor transformation properties given in Eqs. (12-46) and (12-47) of \cite{98BuJexx}, along with Tables~{12-1} and~{15-6} of \cite{98BuJexx}, we can determine the irreps of the rigid rotor basis functions;  the results are summarised in \tabl{tab:rotation_irreps}.

\begin{table}[hbtp!]
\caption{\label{tab:rotation_irreps} The irreps of the rigid rotor wavefunctions. For $K > 0$, the two functions with  $\eta$ $=$ 0, 1 generate a two-dimensional irrep or the direct sum of two one-dimensional irreps. For a given $K$ value, we list first the $\eta=0$ function  and then the $\eta=1$ one. $n$ is a positive integer.}
\begin{center}
\begin{tabular}{ll}
\hline\hline
$K$ & $\Gamma$ \\
\hline
0 ($J$ even) & $A_{1s}$ \\
0 ($J$ odd) & $A_{2s}$ \\
$K=3n$  ($K$ even) & $A_{1s} \oplus A_{2s}$ \\
$K=3n$ ($K$ odd) & $A_{4d} \oplus A_{3d}$ \\
$K=3n+1$ ($K$ even) & $E_{1s}$ \\
$K=3n+1$ ($K$ odd) & $E_{2d}$ \\
$K=3n+2$ ($K$ even) & $E_{1s}$ \\
$K=3n+2$ ($K$ odd) & $E_{2d}$ \\
\hline\hline
\end{tabular}
\end{center}
\end{table}

\subsection{Symmetrisation of the basis set}

Once a set of basis functions has been constructed for one of the subsets mentioned above as eigenfunctions of a 1D Schr\"odinger equation, the degenerate eigenfunctions (which in general transform reducibly) are symmetrised by a sampling and projection procedure described in sections 4 and 5 of \cite{17YuYaOv.methods}. Briefly, a set of geometries is sampled (typically around 50) and each MS operation is applied to the eigenfunctions to obtain an overdetermined set of linear equations which are solved for the transformation matrix effecting the operation on the degenerate eigenfunctions. Using the traces of these matrices and the irrep matrices, one can obtain the irreducible coefficients of the reducible matrix as well as projection operators onto the irreps. These operators are then applied to the eigenfunctions to generate the symmetrized basis functions.

One then produces the basis functions for the complete TROVE calculation as products of the symmetrised 1D basis functions,  one  1D basis function  from each of the sets defined above. In general, these total basis functions do not transform irreducibly, and so a further application of the symmetrisation procedure is required, although the sampling step can be omitted as the transformation properties of the `factor' functions are already known.

\section{Potential Energy Function of Ethane in a Symmetry Adapted Representation}

To represent the PES analytically, an on-the-fly symmetrization procedure has been implemented (see, e.g., \citep{15OwYuYa.CH3Cl}). We first introduce the stretching coordinates
\begin{equation}\label{eq:stretch1}
\xi_1=1-\exp\left(-a(R - R_{\mathrm{e}})\right)
\end{equation}
\begin{equation}\label{eq:stretch2}
\xi_j=1-\exp\left(-b(r_i - r_{\mathrm{e}})\right){\,};\hspace{2mm}j=2,3,4,5,6,7{\,}, \hspace{2mm} i=j-1,
\end{equation}
($a$ is used for the C--C internal coordinate $R$, and $b$ is used for the six C--H internal coordinates $r_1,r_2$, $r_3$, $r_4$, $r_5$ and $r_6$) the bending angular coordinates
\begin{equation}\label{eq:angular1}
\xi_k = (\alpha_i - \alpha_{\mathrm{e}}){\,};\hspace{2mm}k=8,9,10,11,12,13{\,}, \hspace{2mm} i=k-7,
\end{equation}
the dihedral coordinates from Eq.~(\ref{eq:gamma-delta})
\begin{equation}
\left( \xi_{14}, \xi_{15}, \xi_{16}, \xi_{17} \right) =
\left( {\gamma}_1, {\gamma}_2, {\delta}_1, {\delta}_2 \right)
\end{equation}
and, finally, the torsional term
\begin{equation}\label{eq:angular3}
\xi_{18} = 1 + \cos 3\tau
\end{equation}
where $\tau$ is defined in Eq.~(\ref{eq:taudef}).
The quantities
 $R_{\mathrm{e}}$, $r_{\mathrm{e}}$ and $\alpha_{\mathrm{e}}$ are the  equilibrium structural parameter values of C$_2$H$_6$.

Taking an initial potential term of the form
\begin{equation}\label{eq:V_i}
V_{\bm k}^{\mathrm{initial}}=\prod_{i=1}^{18} \xi_{i}^{\,k_i}
\end{equation}
with maximum expansion order $\sum_{i} k_i =6$, each symmetry operation $T_{X}$ of \GTS\ (see Table~\ref{tab:recurrence_prod}) is independently applied to $V_{\bm k}^{\mathrm{initial}}$, i.e.
\begin{equation}\label{eq:V_op}
V_{\bm k}^{T_{X}} = T_{X}{\,} V_{\bm k}^{\mathrm{initial}}(\bm \xi) = T_{X}\left(\prod_{i=1}^{18} \xi_{i}^{\,k_i}\right)
\end{equation}
where $T_{X}$ is one of the 36 operators of \GTS, to create 36 new terms. Here ${\bm k}$ denotes the 18-dimensional hyper-index ${k_1, k_2, \ldots , k_{18}}$ and $(\bm \xi)$ denotes $\{\xi_1,\xi_2,\ldots,\xi_{18}\}$. The results are summed up to produce a final term
\begin{equation}\label{eq:V_f}
V_{\bm k}^{\mathrm{final}}= \sum_{T_{X}} V_{\bm k}^{T_{X}},
\end{equation}
which is itself subjected to the 36  \GTS\ symmetry operations $T_{X}$ to check its invariance. The total potential function is then given by the expression
\begin{equation}
V_{\mathrm{total}}(\bm \xi)= \sum_{\bm k} {\,} f_{\bm k} V_{\bm k}^{\mathrm{final}}(\bm \xi)
\end{equation}
where $f_{\bm k}$ are the corresponding expansion coefficients determined through a least-squares fitting to the \ai\ data generated at the CCSD(T)-F12b/aug-cc-pVTZ level of theory. The details of the PES will be given elsewhere. The symmetry operations are represented as matrices and are generated from the four chosen generators as described above (see also~\ref{sec:appendixa}).


\section{Numerical Example}

In this section we will apply the symmetrisation procedure on a small test basis set for the ethane molecule. We will demonstrate how the method works for subsets 2 (six C$-$H$_i$ stretching  modes), 4 (four dihedral-angle modes  ${\gamma}_1$, ${\gamma}_2$, ${\delta}_1$, ${\delta}_2$), and 5 (the torsional mode $\tau$). In subset 1 (C$-$C stretch), the symmetrised wavefunctions are of $A_{1s}$ symmetry only as each \GTS\ operation leaves the bond length $R$ invariant; in subset 3 (six C$-$C$-$H$_i$ bending  modes), each \GTS\ element has the same  effect on the coordinates as in subset 2, and so the irreps obtained for subset 3 are identical to those obtained for subset 2.

In TROVE calculations, the size of the primitive basis set is determined by the maximum polyad number $P_{\text{max}}$, which occurs in the inequality
\begin{equation}
P = a_1 n_1 + a_2 n_2 + a_3 n_3 +  \ldots  \leq P_{\text{max}}
\end{equation}
where $n_i$ is the excitation number for the primitive basis function of coordinate $i$ and $a_i$ is its polyad coefficient. In our example all $a_i$ $=$ 1.

\subsection{Subset 2 symmetrisation}

For the six-member subset 2, we set the maximum polyad number to be 1, so the space is seven-dimensional. We consider here the 7 unsymmetrised eigenfunctions at lowest energy, the latter 4 being degenerate. As the first three of these eigenfunctions eigenfunctions are non-degenerate, they are already symmetrised and no further work is required. The first one-dimensional energy solution (at $\SI{0.0}{\per\centi\meter}$) is essentially the primitive ground-state wavefunction with small contributions from the other primitives:
\begin{equation}
\label{eq:subset_1:no_1}
\Psi_1 = 0.9999374\ket{000000}  + \ldots
\end{equation}
where $\ldots$ signifies small (of the order $10^{-14}$) contributions from other primitive functions (omitted in the following). For  simplicity we give only 5--7 significant digits here, while the actual calculations  are done in quadruple precision. The eigenfunction in Eq.~(\ref{eq:subset_1:no_1}) has $A_{1s}$ symmetry.

The second one-dimensional solution (at $\SI{2929.16}{\per\centi\meter}$) is
\begin{equation}
    \Psi_2 = \frac{1}{\sqrt{6}}(-\ket{000001}  -\ket{000010}  -\ket{000100} +\ket{001000} +\ket{010000} +\ket{100000})
\end{equation}
and has $A_{4s}$ symmetry.

The final one-dimensional solution (at $\SI{2940.44}{\per\centi\meter}$) is
\begin{equation}
    \Psi_3 = \frac{1}{\sqrt{6}}(\ket{000001} +\ket{000010}  +\ket{000100} +\ket{001000}  +\ket{010000} + \ket{100000})
\end{equation}
which is totally symmetric and so of $A_{1s}$ symmetry.

The four unsymmetrised, degenerate solutions (at $\SI{3007.21}{\per\centi\meter}$) are
\begin{equation}
\begin{split}
    \Psi_4 =& +0.565540\ket{000001} -0.209498\ket{000010}   -0.356041\ket{000100} -0.482230\ket{001000}  \\
    &  -4.233691\ket{010000} +0.524567\ket{100000} \\
    \Psi_5 =& -0.536788\ket{000010} +0.446199\ket{000100} +0.330258\ket{001000}  -0.578016\ket{010000} \\
    &+0.247758\ket{100000} \\
    \Psi_6 =& +0.581276\ket{000001} -0.308888\ket{000010}-0.272388\ket{000100} +0.398602\ket{001000} \\
    &+0.157199\ket{010000} -0.555801\ket{100000} \\
    \Psi_7 =&  -0.489104\ket{000010}  +0.516345\ket{000100} -0.407635\ket{001000}+ 0.553226\ket{010000}  \\
    &-0.145592\ket{100000}
\end{split}
\end{equation}
which turns out to be of $G_s$ symmetry, and thus also already symmetrised. However,
 in this case there is ambiguity in the group matrices describing the transformation properties of the degenerate wavefunctions. We therefore find and apply a unitary transformation on the wavefunctions to produce functions with transformation properties described by our `standard' $G$ representation matrices  collected in Section~\ref{sec:finalmatrices}.
 To this end, we follow the sampling procedure from Ref.~\cite{17YuYaOv.methods}, derive the transformation properties of the above wavefunctions and convert them to have our `standard' transformation properties. For this work, we modified the sampling procedure in TROVE by applying it to the five group generators only (out of 72). This simple modification led to a significant speedup of the sampling part of the code and is now the standard part of TROVE.
 The reduced symmetrised wavefunctions become
\begin{equation}
\begin{split}
    \Psi_4 &= \frac{1}{2\sqrt{3}}( -\ket{000001} -\ket{000010}  + 2\ket{000100} -\ket{001000} -\ket{010000} +  2\ket{100000}) \\
    \Psi_5 &= \frac{1}{2}(-\ket{000001} + \ket{000010} - \ket{001000} + \ket{010000})\\
    \Psi_6 &= \frac{1}{2\sqrt{3}}(-\ket{000001}-\ket{000010} +2\ket{000100} + \ket{001000} +\ket{010000} - 2\ket{100000}) \\
    \Psi_7 &= \frac12 ( -\ket{000001} +\ket{000010} +\ket{001000} - \ket{010000})
\end{split}
\end{equation}
which are recognised as the linear combinations (defined by the matrix ${\mathbf V}$) of bond lengths $r_k$ that transform as the $G_s$ irrep [Eq.~\eqref{eq:gtransform}].

\subsection{Subset 4 symmetrisation}

For subset 4, we use a maximum polyad number of 1 and thus we have 5 eigenfunctions, one non-degenerate and four degenerate ones.
$\ket{0000}$ is the first eigenfunction (at $\SI{0.0}{\per\centi\meter})$; it has $A_{1s}$ symmetry. The other four unsymmetrised eigenfunctions (at $\SI{1469.30}{\per\centi\meter})$ are
\begin{equation}
\begin{split}
\Psi_2 &= \frac{1}{\sqrt{2}}(\ket{0001}  -\ket{0100}) \\
\Psi_3 &= -\frac{1}{\sqrt{2}}(\ket{0001} + \ket{0100}) \\
\Psi_4 &= -\frac{1}{\sqrt{2}}(\ket{0010} + \ket{1000}) \\
\Psi_5 &= \frac{1}{\sqrt{2}}(-\ket{0010} + \ket{1000})
\end{split}
\end{equation}
which again turns out to be of $G_s$ symmetry. Here and in the following, $\frac{1}{\sqrt{2}}$ stands for the numerical value $0.707106781186545$ ($\pm 10^{-15})$. The desired symmetrised functions are
\begin{equation}
\begin{split}
\Psi_2 &= \frac{1}{\sqrt{2}}(
\ket{0010} + \ket{1000}) \\
\Psi_3 &=
\frac{1}{\sqrt{2}}(
\ket{0001} + \ket{0100}) \\
\Psi_4 &=
\frac{1}{\sqrt{2}}(
\ket{0010} -\ket{1000}
) \\
\Psi_5 &= \frac{1}{\sqrt{2}}(
	\ket{0001} -\ket{0100}).
\end{split}
\end{equation}
\subsection{Subset 5 symmetrisation}

For the one dimensional subset 5 we use a maximum polyad number of 29 and show, for illustrative purposes, the symmetrisation of the first three eigenfunctions. The lowest energy solution (at $\SI{0.0}{\per\centi\meter}$ with $A_{1s}$ symmetry) is:
\begin{equation}
\Psi_1 = 0.999958\ket{0}  + \ldots.
\end{equation}
  The second and third eigenfunctions (at $\SI{0.0029}{\per\centi\meter})$ are degenerate, and the unsymmetrised wavefunctions are
\begin{equation}
\begin{split}
\Psi_2 =  & +0.0272058\ket{1} + 0.999588\ket{2}  + \ldots \\
\Psi_3 = &-0.999588\ket{1} + 0.0272058\ket{2} + \ldots.
\end{split}
\end{equation}
These turn out to be of $E_{3d}$ symmetry, which we then convert to
\begin{equation}
\begin{split}
\Psi_2 =  & -0.75926\ket{1} - 0.65072\ket{2}  + \ldots \\
\Psi_3 = &-0.65072\ket{1} + 0.75926\ket{2} + \ldots
\end{split}
\end{equation}
to obey our `standard' transformation properties imposed by the matrices in~\ref{sec:appendixa}.

\subsection{Torsional basis function symmetries}

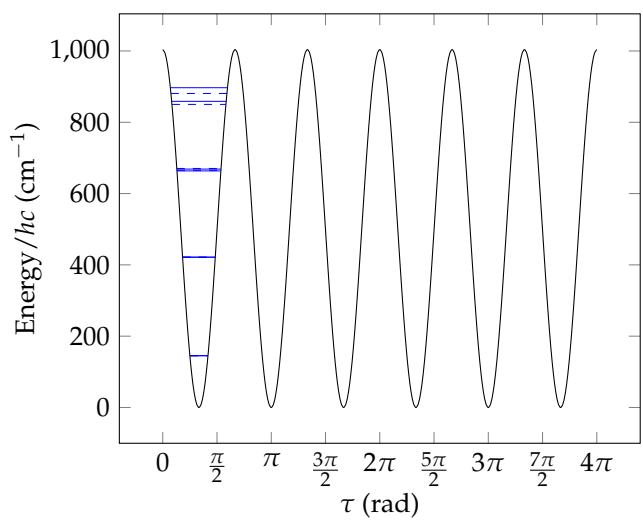
\begin{figure}[htbp!]
    \centering
    \begin{tikzpicture}
    \begin{axis}[
 xtick={
0, 1.5708, 3.14159, 4.7123889, 6.28318,
  7.85398, 9.42477, 10.9955689, 12.56636
    },
    xticklabels={
    $0$, $\frac{\pi}{2}$, $\pi$, $\frac{3\pi}{2}$, $2\pi$,
    $\frac{5\pi}{2}$, $3\pi$, $\frac{7\pi}{2}$, $4\pi$
   },
    xlabel={$\tau$ (\SI{}{\radian})},
    ylabel={Energy/$hc$ (\SI{}{\per\centi\meter})}]
    \addplot[smooth]  table[x=angle,y=energy] {
    angle energy
    0	1003.77951556
0.069813170079778	992.60301469816
0.139626340159556	959.590813837953
0.209439510239333	906.266645428049
0.279252680319111	835.078568736274
0.349065850398889	749.26978315875
0.418879020478667	652.712557344184
0.488692190558445	549.717316998068
0.558505360638222	444.8292039328
0.628318530718	342.623379896951
0.698131700797778	247.50848945125
0.767944870877556	163.545565412708
0.837758040957333	94.287710200816
0.907571211037111	42.6443703823134
0.977384381116889	10.7729746917228
1.04719755119667	4.10521118250522E-28
1.11701072127644	10.7729746917228
1.18682389135622	42.6443703823134
1.256637061436	94.287710200816
1.32645023151578	163.545565412708
1.39626340159556	247.50848945125
1.46607657167533	342.623379896951
1.53588974175511	444.8292039328
1.60570291183489	549.717316998068
1.67551608191467	652.712557344184
1.74532925199444	749.26978315875
1.81514242207422	835.078568736274
1.884955592154	906.266645428049
1.95476876223378	959.590813837953
2.02458193231356	992.60301469816
2.09439510239333	1003.77951556
2.16420827247311	992.60301469816
2.23402144255289	959.590813837953
2.30383461263267	906.266645428049
2.37364778271244	835.078568736274
2.44346095279222	749.26978315875
2.513274122872	652.712557344184
2.58308729295178	549.717316998069
2.65290046303156	444.829203932801
2.72271363311133	342.623379896951
2.79252680319111	247.50848945125
2.86233997327089	163.545565412708
2.93215314335067	94.2877102008161
3.00196631343044	42.6443703823134
3.07177948351022	10.7729746917228
3.14159265359	4.16074439846136E-28
3.21140582366978	10.7729746917228
3.28121899374956	42.6443703823133
3.35103216382933	94.287710200816
3.42084533390911	163.545565412708
3.49065850398889	247.50848945125
3.56047167406867	342.623379896951
3.63028484414844	444.8292039328
3.70009801422822	549.717316998068
3.769911184308	652.712557344184
3.83972435438778	749.26978315875
3.90953752446756	835.078568736274
3.97935069454733	906.266645428049
4.04916386462711	959.590813837953
4.11897703470689	992.60301469816
4.18879020478667	1003.77951556
4.25860337486644	992.60301469816
4.32841654494622	959.590813837953
4.398229715026	906.266645428049
4.46804288510578	835.078568736274
4.53785605518556	749.26978315875
4.60766922526533	652.712557344184
4.67748239534511	549.717316998069
4.74729556542489	444.829203932801
4.81710873550467	342.623379896951
4.88692190558445	247.50848945125
4.95673507566422	163.545565412708
5.026548245744	94.2877102008161
5.09636141582378	42.6443703823134
5.16617458590356	10.7729746917229
5.23598775598333	4.27181083037364E-28
5.30580092606311	10.7729746917228
5.37561409614289	42.6443703823133
5.44542726622267	94.2877102008159
5.51524043630244	163.545565412708
5.58505360638222	247.50848945125
5.654866776462	342.623379896951
5.72467994654178	444.8292039328
5.79449311662156	549.717316998068
5.86430628670133	652.712557344184
5.93411945678111	749.26978315875
6.00393262686089	835.078568736274
6.07374579694067	906.266645428049
6.14355896702045	959.590813837953
6.21337213710022	992.60301469816
6.28318530718	1003.77951556
6.35299847725978	992.60301469816
6.42281164733956	959.590813837953
6.49262481741933	906.266645428049
6.56243798749911	835.078568736274
6.63225115757889	749.26978315875
6.70206432765867	652.712557344184
6.77187749773844	549.717316998069
6.84169066781822	444.829203932801
6.911503837898	342.623379896951
6.98131700797778	247.50848945125
7.05113017805756	163.545565412708
7.12094334813733	94.2877102008161
7.19075651821711	42.6443703823134
7.26056968829689	10.7729746917229
7.33038285837667	4.43841047824206E-28
7.40019602845644	10.7729746917228
7.47000919853622	42.6443703823133
7.539822368616	94.2877102008159
7.60963553869578	163.545565412708
7.67944870877556	247.50848945125
7.74926187885533	342.623379896951
7.81907504893511	444.8292039328
7.88888821901489	549.717316998068
7.95870138909467	652.712557344184
8.02851455917445	749.26978315875
8.09832772925422	835.078568736274
8.168140899334	906.266645428049
8.23795406941378	959.590813837953
8.30776723949356	992.60301469816
8.37758040957333	1003.77951556
8.44739357965311	992.603014698161
8.51720674973289	959.590813837953
8.58701991981267	906.266645428049
8.65683308989244	835.078568736274
8.72664625997222	749.26978315875
8.796459430052	652.712557344184
8.86627260013178	549.717316998069
8.93608577021156	444.829203932801
9.00589894029133	342.623379896951
9.07571211037111	247.50848945125
9.14552528045089	163.545565412708
9.21533845053067	94.2877102008162
9.28515162061044	42.6443703823135
9.35496479069022	10.7729746917229
9.42477796077	4.66054334206662E-28
9.49459113084978	10.7729746917228
9.56440430092956	42.6443703823133
9.63421747100933	94.2877102008159
9.70403064108911	163.545565412708
9.77384381116889	247.50848945125
9.84365698124867	342.623379896951
9.91347015132845	444.8292039328
9.98328332140822	549.717316998068
10.053096491488	652.712557344184
10.1229096615678	749.26978315875
10.1927228316476	835.078568736274
10.2625360017273	906.266645428049
10.3323491718071	959.590813837953
10.4021623418869	992.60301469816
10.4719755119667	1003.77951556
10.5417886820464	992.603014698161
10.6116018521262	959.590813837953
10.681415022206	906.26664542805
10.7512281922858	835.078568736274
10.8210413623656	749.26978315875
10.8908545324453	652.712557344184
10.9606677025251	549.717316998069
11.0304808726049	444.829203932801
11.1002940426847	342.623379896951
11.1701072127644	247.50848945125
11.2399203828442	163.545565412708
11.309733552924	94.2877102008162
11.3795467230038	42.6443703823135
11.4493598930836	10.7729746917229
11.5191730631633	4.93820942184732E-28
11.5889862332431	10.7729746917228
11.6587994033229	42.6443703823132
11.7286125734027	94.2877102008159
11.7984257434824	163.545565412708
11.8682389135622	247.50848945125
11.938052083642	342.623379896951
12.0078652537218	444.8292039328
12.0776784238016	549.717316998068
12.1474915938813	652.712557344184
12.2173047639611	749.26978315875
12.2871179340409	835.078568736274
12.3569311041207	906.266645428049
12.4267442742004	959.590813837953
12.4965574442802	992.60301469816
12.56637061436	1003.77951556
};

 \addplot[mark=none,,blue,style=ultra thin] coordinates {(0.786555, 145.221) (1.30784, 145.221)};
 \addplot[mark=none,blue,dashed, style=ultra thin] coordinates {(0.786553, 145.224) (1.30784, 145.224)};
  \addplot[mark=none,blue,style=ultra thin] coordinates {(0.786547, 145.23) (1.30785, 145.23)};
  \addplot[mark=none,blue, dashed, style=ultra thin] coordinates {(0.786544, 145.233) (1.30785, 145.233)};

  \addplot[mark=none,blue,dashed, style=ultra thin] coordinates {(0.576294, 421.286) (1.5181, 421.286)};
    \addplot[mark=none,blue,style=ultra thin] coordinates {(0.576223, 421.391) (1.51817, 421.391)};
  \addplot[mark=none,blue, dashed, style=ultra thin] coordinates {(0.576073, 421.613) (1.51832, 421.613)};
  \addplot[mark=none,blue,style=ultra thin] coordinates {(0.575998, 421.725) (1.5184, 421.725)};

    \addplot[mark=none,blue,style=ultra thin] coordinates {(0.41252, 663.551) (1.68187, 663.551)};
    \addplot[mark=none,blue, dashed, style=ultra thin] coordinates {(0.411362, 665.193) (1.68303, 665.193)};
    \addplot[mark=none,blue,style=ultra thin] coordinates {(0.408908, 668.661) (1.68549, 668.661)};
    \addplot[mark=none,blue, dashed, style=ultra thin] coordinates {(0.407645, 670.445) (1.68675, 670.445)};

    \addplot[mark=none,blue, dashed, style=ultra thin] coordinates {(0.265027, 850.114) (1.82937, 850.114)};
    \addplot[mark=none,blue,style=ultra thin] coordinates {(0.256877, 858.736) (1.83752, 858.736)};
    \addplot[mark=none,blue,dashed,style=ultra thin] coordinates {(0.235004, 880.801) (1.85939, 880.801)};
    \addplot[mark=none,blue,style=ultra thin] coordinates {(0.217873, 896.944) (1.87652, 896.944)};

\end{axis}
    \end{tikzpicture}
     \caption{The torsional potential energy  as a function of the torsion angle $\tau$. The allowed energy values  are marked by blue horizontal lines. Each energy may correspond to more than one eigenfunction of a given irreducible representation. Dashed lines indicate states of $d$-type symmetry.}
    \label{fig:irrep_energies}
\end{figure}

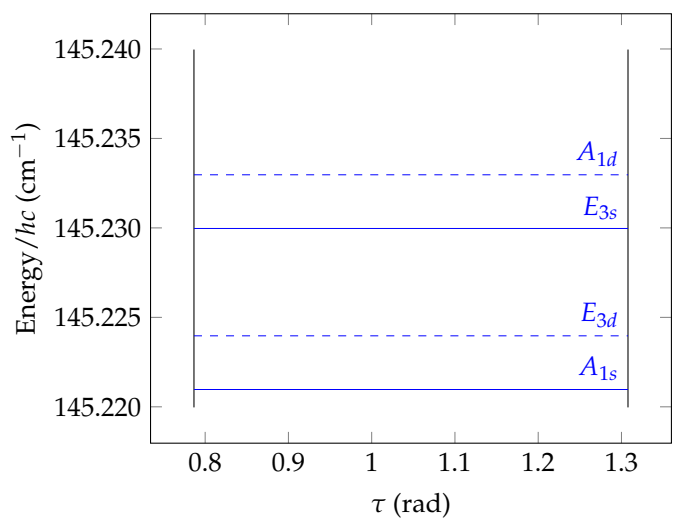
\begin{figure}[htbp!]
    \centering
    \begin{tikzpicture}
    \begin{axis}[
    ylabel style={yshift=0.4cm},
    xlabel={$\tau$ (\SI{}{\radian})},
    ylabel={Energy/$hc$ (\SI{}{\per\centi\meter})},
      yticklabel style={/pgf/number format/.cd,fixed zerofill,precision=3}]
    \addplot[smooth]  table[x=angle,y=energy] {
    angle energy
0.786557 145.22
0.786538 145.24
  };
  \addplot[smooth]  table[x=angle,y=energy] {
    angle energy
1.30784 145.22
1.30786 145.24
  };
     \addplot[mark=none,blue,style=ultra thin] coordinates {(0.786555, 145.221) (1.30784, 145.221)} node[above left]{$A_{1s}$};
 \addplot[mark=none,blue,dashed, style=ultra thin] coordinates {(0.786553, 145.224) (1.30784, 145.224)}  node[above left]{$E_{3d}$};
  \addplot[mark=none,blue,style=ultra thin] coordinates {(0.786547, 145.23) (1.30785, 145.23)}  node[above left]{$E_{3s}$};
  \addplot[mark=none,blue, dashed, style=ultra thin] coordinates {(0.786544, 145.233) (1.30785, 145.233)}  node[above left]{$A_{1d}$};
    \end{axis}
    \end{tikzpicture}
 \caption{A enlarged detail  of \protect\fig{fig:irrep_energies}, showing the lowest energy cluster with the \GTS(EM) symmetry labels indicated.}
    \label{fig:zoomed_plot}
\end{figure}

\fig{fig:irrep_energies} shows the torsional potential energy  as a function of $\tau$. In order for the torsional basis functions to transform according to the irreps of the extended MS group \GTS(EM), we obtain them as solutions of the 1D torsional Schr\"odinger equation in Eq.~\eqref{eq:torsion-schr-eq}. In solving this Schr\"odinger equation in a Fourier-series basis [see the discussion of Eq.~\eqref{eq:torsion-schr-eq}], we let $\tau$ vary from 0 to $4\pi$ so that the potential energy curve in \fig{fig:irrep_energies} has six minima.  In the figure, we indicate the lowest allowed torsional energies. Some of these energies are degenerate, i.e., associated with more than one eigenfunction of a given irreducible representation.  In the limit of an infinite barrier height, we expect the energies to be six-fold degenerate. With the actual, finite height of the barrier, the energies form near-degenerate clusters with a total multiplicity of  6 as shown in \fig{fig:zoomed_plot}. However, since our description of the torsional angle $\tau$ $\in$ $[0,  4 \pi]$  is unphysical, only three of the states in the cluster (of $s$ symmetry) exist in nature if they are combined with rigid-rotor basis functions of $s$ symmetry. The other three states (of $d$ symmetry) must be combined with
rigid-rotor basis functions of $d$ symmetry in order that the total rotation-torsion state can exist in nature.
For $J$ $=$ 0, only the $s$-type torsional states will exist since only $s$-type rigid-rotor basis functions are available.

\begin{figure}[htbp!]
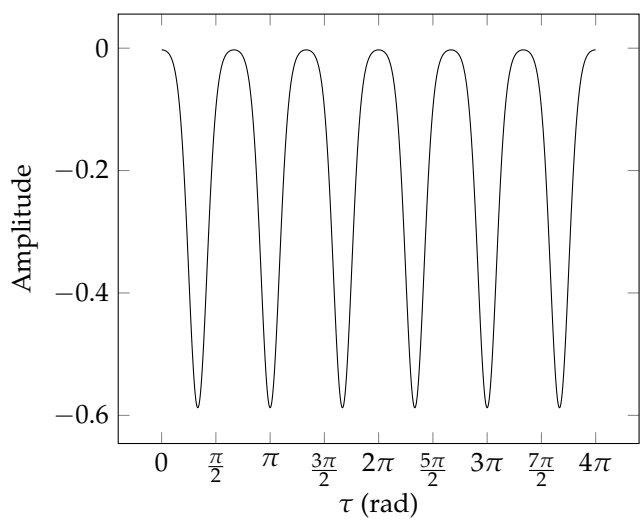

	    \centering

																							  \caption{The ground state torsional wavefunction.}
																								  \label{fig:torsion_ground}
																		  \end{figure}

\begin{figure}[htbp!]
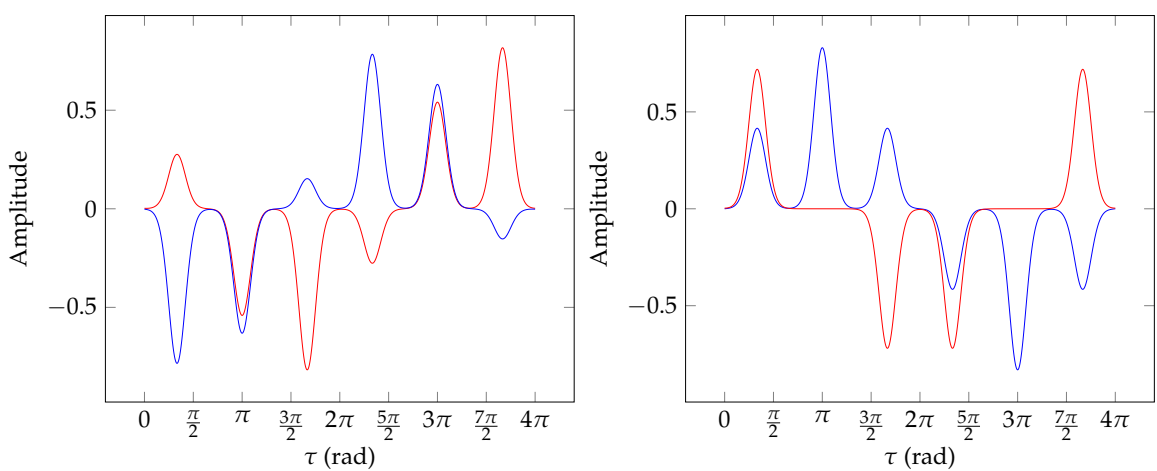

																			      \centering
																				      %
																																														\caption{The degenerate first excited state torsional wavefunctions. Left display:  The two wavefunctions which generate the $E_{3d}$ irrep, but which do not transform as given by the transformation matrices of \protect\sect{sec:Einalmatrices}, displayed as a red and a blue curve, respectively. Right display: The corresponding two $E_{3d}$ wavefunctions obtained as transforming according to the  transformation matrices of \protect\sect{sec:Einalmatrices} and displayed as a red and a blue curve, respectively.}
																																															\label{fig:torsion_first}
																								
\end{figure}

\fig{fig:torsion_ground} shows the ground state torsional wavefunction which, as always, is totally symmetric (of $A_{1s}$ symmetry) in \GTS(EM). The effect of  the \GTS(EM) generators on $\tau$ is shown in \tabl{tab:torsion_transformations}. In \fig{fig:torsion_first} (left display) we show the doubly-degenerate wavefunctions of the first excited state. They span the $E_{3d}$ irrep but must be transformed to obtain the transformation properties imposed by the $E_{3d}$ matrices described in \sect{sec:Einalmatrices}. The transformed functions, with the desired transformation properties are shown in \fig{fig:torsion_first} (right display).

The symmetry adapted ro-vibrational basis set functions are constructed as direct products of the symmetrized component
functions from the different subspaces as  $ \Psi_{\lambda_0}^{(0),\Gamma_0} \otimes
\Psi_{\lambda_1}^{(1),\Gamma_1} \otimes \Psi_{\lambda_2}^{(2),\Gamma_2} \cdots \otimes
\Psi_{\lambda_L}^{(L),\Gamma_L}$, where $L$ is the number of vibrational subspaces. The symmetry of a product basis function is the direct  product of the irreducible representations of the component functions. This symmetry must be further reduced, but this is rather straightforward  when each component function generates one of the irreps of the group and has defined, `standard' transformation properties. In this case we use the same projection operator symmetrization technique described above without further sampling of the symmetry properties of the corresponding components. The $d$-symmetry components are unphysical and must be disregarded. Some $s$ symmetry ro-vibrational basis functions will necessarily contain products of the $d$ components, underpinning the importance of the \GTS(EM) extended symmetry group. The nuclear spin statistical weights of the ro-vibrational states  can be deduced using the procedure from \cite{98BuJexx}. We obtain nuclear spin statistical weights of  6, 10, 6, 10, 4, 4, 2, 6, 12  for the \GTS(EM) symmetries $A_{1s},  A_{2s}, A_{3s}, A_{4s}, E_{1s}, E_{2s}, E_{3s}, E_{4s}$, and $G_s$, respectively.


\section{Conclusion}

We have presented a detailed description of the \GTS\ and \GTS(EM) molecular symmetry groups. A full set of irreducible representation matrices have been derived and tested for constructing symmetry-adapted potential energy functions and basis functions of ethane C$_2$H$_6$. Both the construction of the transformation matrices and the symmetry adaption can be implemented as numerical procedures as part of  computational approaches to the solution of the ro-vibrational Schr\"odinger equation. A self-consistent choice of the vibrational, torsional and rotational coordinates for ethane, satisfying the \GTS(EM) symmetry requirements have been introduced and analyzed in full detail. The irreducible representation matrices as well as the coordinate choice made for ethane in the present work have been implemented as part of
the program system TROVE, and we have discussed a few examples of symmetrized wavefunctions.   These results of the present work will be important in our planned line-list calculations for ethane.

The results of the present work are, of course, not only applicable to
ethane H$_3$CCH$_3$, but also to other molecules with MS groups \GTS\ and \GTS(EM).
A prominent molecule of this kind is dimethylacetylene
H$_3$CCCCH$_3$. Also, subgroups of the 
 matrix groups constructed can be applied to molecules whose MS groups are subgroups of \GTS\ and \GTS(EM), such as D$_3$CCH$_3$. The general ideas used for generating the matrix groups can be applied to other MS groups. In particular, generation of matrix groups in a manner similar to that of the present work will be required for MS groups with irreps of dimension 4 and higher.

\section{Acknowledgements}

This work was supported by the STFC Projects No. ST/M001334/1 and ST/R000476/1, and by the COST action MOLIM (CM1405). The authors acknowledge the use of the UCL Legion High Performance Computing Facility (Legion@UCL) and associated support services in the completion of this work, along with the Cambridge Service for Data Driven Discovery (CSD3), part of which is operated by the University of Cambridge Research Computing on behalf of the STFC DiRAC HPC Facility (www.dirac.ac.uk). The DiRAC component of CSD3 was funded by BIS capital funding via STFC capital grants ST/P002307/1 and ST/R002452/1 and STFC operations grant ST/R00689X/1. DiRAC is part of the National e-Infrastructure. The present work was supported in part by the Deutsche Forschungsgemeinschaft.

\setcounter{section}{0}
\setcounter{equation}{0}
\setcounter{table}{0}
\renewcommand\thesection{Appendix \Alph{section}}
\renewcommand\thesubsection{\Alph{section}.\arabic{subsection}}
\renewcommand\theequation{\Alph{section}.\arabic{equation}}
\renewcommand\thetable{\Alph{section}.\arabic{table}}
\renewcommand\thesubsubsection{\Alph{section}.\arabic{subsection}.\arabic{subsubsection}}

\section{Character table of the isomorphic groups
{\itshape\bfseries C}$_{3{\rm v}}^{(-)}$ and  {\itshape\bfseries C}$_{3{\rm v}}^{(+)}$.}\label{sec:appendixC3vgroups}

{\itshape\bfseries G}$_{36}$ is the direct product of
{\itshape\bfseries C}$_{3{\rm v}}^{(-)}$ and  {\itshape\bfseries C}$_{3{\rm v}}^{(+)}$, 
{\itshape\bfseries G}$_{36}$ $=$
{\itshape\bfseries C}$_{3{\rm v}}^{(-)}$ $\times$  {\itshape\bfseries C}$_{3{\rm v}}^{(+)}$,
and so the irreps of {\itshape\bfseries G}$_{36}$ are obtained from those of
{\itshape\bfseries C}$_{3{\rm v}}^{(-)}$ and  {\itshape\bfseries C}$_{3{\rm v}}^{(+)}$.
These latter irreps are given in Table~\ref{tab:C3vchartab}. 

\begin{table}[h]
\caption{\label{tab:C3vchartab}Common character tables of 
{\itshape\bfseries C}$_{3{\rm v}}^{(-)}$ and  {\itshape\bfseries C}$_{3{\rm v}}^{(+)}$. }
\begin{center}
\renewcommand{\arraystretch}{1.5}
\begin{tabular}{lrrr}
\hline\hline
$\Gamma$    &
${\mathcal C}_1^{(\pm)}$ & ${\mathcal C}_2^{(\pm)}$ & ${\mathcal C}_3^{(\pm)}$  \\
&  1 & 2 & 3  \\
\hline
$A_1$ & 1 &  1 & 1 \\
$A_2$ & 1 &  1 & $-$1 \\
$E$ & 2 & $-$1 & 0 \\
\hline\hline
\end{tabular}
\end{center}
\end{table}
We label
	the elements of {\itshape\bfseries C}$_{3{\rm v}}^{(\pm)}$ as
	$R_j^{(\pm)}$, $j$ $=$ 1, 2, \dots, 6 (see  \tabl{tab:multiplicationtable}). The nuclei are labelled as in \fig{fig:ethane_labelling}.
	 {\itshape\bfseries C}$_{3{\rm v}}^{(\pm)}$ has three
classes,
${\mathcal C}_1^{(\pm)}$ $=$ $\{ E \}$ $=$ $\{ R_1^{(\pm)} \}$,
${\mathcal C}_2^{(\pm)}$ $=$ $\{ R_2^{(\pm)}, R_3^{(\pm)} \}$,  and
${\mathcal C}_3^{(\pm)}$ $=$ $\{ R_4^{(\pm)}, R_5^{(\pm)}, R_6^{(\pm)} \}$.

\section{The \GTS\ Transformation Matrices}\label{sec:appendixa}
\setcounter{equation}{0}

In Sections~\ref{sec:Einalmatrices}
and~\ref{sec:finalmatrices}, we have  abbreviated some of the operations
$R$ $\in$  \GTS\ as $R$ $=$ $R_j^{(-)} \, R_k^{(+)}$ $=$ $R_k^{(+)} \, R_j^{(-)}$, where $R_j^{(-)}$ $\in$ {\itshape\bfseries C}$_{3{\rm v}}^{(-)}$ and
$R_k^{(+)}$ $\in$ {\itshape\bfseries C}$_{3{\rm v}}^{(+)}$ (since, if we used the complete labels such as (15)(24)(36)$(ab)^*$ and (143526)$(ab)^*$, the text would not fit on the page). All such products, and the complete classes of \GTS, are given in
 \tabl{tab:multiplicationtable}.

\subsection{$E_i$ transformation matrices in {\itshape\bfseries G}$_{36}$}\label{sec:Einalmatrices}
For each operation $R$ $\in$ \GTS, we list here the four matrices
$ {\mathbf M}_{E_1}\left[ R \right]$,
$ {\mathbf M}_{E_2}\left[ R \right]$,
$ {\mathbf M}_{E_3}\left[ R \right]$, and
$ {\mathbf M}_{E_4}\left[ R \right]$ in that order.
\subsubsection{One-member class ${\mathcal C}_1^{(-)} \times  {\mathcal C}_1^{(+)}$  containing $E$}
\begin{center}
{\small\baselineskip=32pt
\begin{tabbing}
\parbox[b]{3.5truecm}{$ {\mathbf M}_{E_i}\left[ E \right]$}
\= \parbox[b]{3.0truecm}{\twobytwo{1}{0}{0}{1}} \=
\parbox[b]{3.0truecm}{\twobytwo{1}{0}{0}{1}}  \=
\parbox[b]{3.0truecm}{\twobytwo{1}{0}{0}{1}} \=
\parbox[b]{3.0truecm}{\twobytwo{1}{0}{0}{1}} \\[5pt]
\end{tabbing}}
\end{center}
\subsubsection{Two-member class ${\mathcal C}_1^{(-)} \times  {\mathcal C}_2^{(+)}$  containing (123)(456) and (132)(465)}
\begin{center}
{\small\baselineskip=32pt
\begin{tabbing}
\parbox[b]{3.5truecm}{$ {\mathbf M}_{E_i}\left[ R_2^{(+)} \right]$}
\= \parbox[b]{3.0truecm}{\twobytwo{1}{0}{0}{1}} \=
\parbox[b]{3.0truecm}{\twobytwo{1}{0}{0}{1}}  \=
\parbox[b]{3.0truecm}{\twobytwo{-\frac{1}{2}}{-\frac{\sqrt{3}}{2}}{\frac{\sqrt{3}}{2}}{-\frac{1}{2}}}  \=
\parbox[b]{3.0truecm}{\twobytwo{-\frac{1}{2}}{-\frac{\sqrt{3}}{2}}{\frac{\sqrt{3}}{2}}{-\frac{1}{2}}}  \\[5pt]
$  {\mathbf M}_{E_i}\left[  R_3^{(+)} \right] $ \> \twobytwo{1}{0}{0}{1} \> \twobytwo{1}{0}{0}{1}  \> \twobytwo{-\frac{1}{2}}{\frac{\sqrt{3}}{2}}{-\frac{\sqrt{3}}{2}}{-\frac{1}{2}} \> \twobytwo{-\frac{1}{2}}{\frac{\sqrt{3}}{2}}{-\frac{\sqrt{3}}{2}}{-\frac{1}{2}} \\[5pt]
\end{tabbing}}
\end{center}
\subsubsection{Three-member class ${\mathcal C}_1^{(-)} \times  {\mathcal C}_3^{(+)}$  containing (14)(26)(35)($ab$)$^*$}
\begin{center}
{\small\baselineskip=32pt
\begin{tabbing}
\parbox[b]{3.5truecm}{$ {\mathbf M}_{E_i}\left[ R_4^{(+)} \right]$}
\= \parbox[b]{3.0truecm}{\twobytwo{1}{0}{0}{1}} \=
\parbox[b]{3.0truecm}{\twobytwo{-1}{0}{0}{-1}}  \=
\parbox[b]{3.0truecm}{\twobytwo{1}{0}{0}{-1}}  \=
\parbox[b]{3.0truecm}{\twobytwo{1}{0}{0}{-1}}  \\
$ {\mathbf M}_{E_i}\left[ R_5^{(+)} \right] $ \> \twobytwo{1}{0}{0}{1} \> \twobytwo{-1}{0}{0}{-1}  \> \twobytwo{-\frac{1}{2}}{-\frac{\sqrt{3}}{2}}{-\frac{\sqrt{3}}{2}}{\frac{1}{2}} \> \twobytwo{-\frac{1}{2}}{-\frac{\sqrt{3}}{2}}{-\frac{\sqrt{3}}{2}}{\frac{1}{2}} \\
$ {\mathbf M}_{E_i}\left[ R_6^{(+)} \right] $ \> \twobytwo{1}{0}{0}{1} \> \twobytwo{-1}{0}{0}{-1}   \> \twobytwo{-\frac{1}{2}}{\frac{\sqrt{3}}{2}}{\frac{\sqrt{3}}{2}}{\frac{1}{2}} \> \twobytwo{-\frac{1}{2}}{\frac{\sqrt{3}}{2}}{\frac{\sqrt{3}}{2}}{\frac{1}{2}} \\
\end{tabbing}}
\end{center}
\subsubsection{Two-member class ${\mathcal C}_2^{(-)} \times  {\mathcal C}_1^{(+)}$  containing (123)(456) and (132)(456)}
\begin{center}
{\small\baselineskip=32pt
\begin{tabbing}
\parbox[b]{3.5truecm}{$ {\mathbf M}_{E_i}\left[ R_2^{(-)} \right]$}
\= \parbox[b]{3.0truecm}{\twobytwo{-\frac{1}{2}}{-\frac{\sqrt{3}}{2}}{\frac{\sqrt{3}}{2}}{-\frac{1}{2}} } \=
\parbox[b]{3.0truecm}{\twobytwo{-\frac{1}{2}}{-\frac{\sqrt{3}}{2}}{\frac{\sqrt{3}}{2}}{-\frac{1}{2}}}  \=
\parbox[b]{3.0truecm}{\twobytwo{1}{0}{0}{1}}  \=
\parbox[b]{3.0truecm}{\twobytwo{1}{0}{0}{1}}  \\
$ {\mathbf M}_{E_i}\left[ R_3^{(-)} \right] $ \> \twobytwo{-\frac{1}{2}}{\frac{\sqrt{3}}{2}}{-\frac{\sqrt{3}}{2}}{-\frac{1}{2}} \> \twobytwo{-\frac{1}{2}}{\frac{\sqrt{3}}{2}}{-\frac{\sqrt{3}}{2}}{-\frac{1}{2}}  \> \twobytwo{1}{0}{0}{1} \> \twobytwo{1}{0}{0}{1} \\
\end{tabbing}}
\end{center}

\subsubsection{Four-member class ${\mathcal C}_2^{(-)} \times  {\mathcal C}_2^{(+)}$  containing (123) and (456)}

\begin{center}
{\small\baselineskip=32pt
\begin{tabbing}
\parbox[b]{3.5truecm}{$ {\mathbf M}_{E_i}\left[ R_2^{(-)}R_2^{(+)} \right]$}
\= \parbox[b]{3.0truecm}{\twobytwo{-\frac{1}{2}}{-\frac{\sqrt{3}}{2}}{\frac{\sqrt{3}}{2}}{-\frac{1}{2}}} \=
\parbox[b]{3.0truecm}{\twobytwo{-\frac{1}{2}}{-\frac{\sqrt{3}}{2}}{\frac{\sqrt{3}}{2}}{-\frac{1}{2}}  }  \=
\parbox[b]{3.0truecm}{\twobytwo{-\frac{1}{2}}{-\frac{\sqrt{3}}{2}}{\frac{\sqrt{3}}{2}}{-\frac{1}{2}}}  \=
\parbox[b]{3.0truecm}{\twobytwo{-\frac{1}{2}}{-\frac{\sqrt{3}}{2}}{\frac{\sqrt{3}}{2}}{-\frac{1}{2}} }  \\
$ {\mathbf M}_{E_i}\left[  R_2^{(-)}R_3^{(+)} \right] $ \> \twobytwo{-\frac{1}{2}}{-\frac{\sqrt{3}}{2}}{\frac{\sqrt{3}}{2}}{-\frac{1}{2}} \> \twobytwo{-\frac{1}{2}}{-\frac{\sqrt{3}}{2}}{\frac{\sqrt{3}}{2}}{-\frac{1}{2}}  \> \twobytwo{-\frac{1}{2}}{\frac{\sqrt{3}}{2}}{-\frac{\sqrt{3}}{2}}{-\frac{1}{2}} \> \twobytwo{-\frac{1}{2}}{\frac{\sqrt{3}}{2}}{-\frac{\sqrt{3}}{2}}{-\frac{1}{2}} \\
$ {\mathbf M}_{E_i}\left[  R_3^{(-)}R_2^{(+)}  \right] $  \> \twobytwo{-\frac{1}{2}}{\frac{\sqrt{3}}{2}}{-\frac{\sqrt{3}}{2}}{-\frac{1}{2}} \> \twobytwo{-\frac{1}{2}}{\frac{\sqrt{3}}{2}}{-\frac{\sqrt{3}}{2}}{-\frac{1}{2}}  \> \twobytwo{-\frac{1}{2}}{-\frac{\sqrt{3}}{2}}{\frac{\sqrt{3}}{2}}{-\frac{1}{2}} \> \twobytwo{-\frac{1}{2}}{-\frac{\sqrt{3}}{2}}{\frac{\sqrt{3}}{2}}{-\frac{1}{2}} \\
$ {\mathbf M}_{E_i}\left[  R_3^{(-)}R_3^{(+)}  \right] $ \> \twobytwo{-\frac{1}{2}}{\frac{\sqrt{3}}{2}}{-\frac{\sqrt{3}}{2}}{-\frac{1}{2}} \> \twobytwo{-\frac{1}{2}}{\frac{\sqrt{3}}{2}}{-\frac{\sqrt{3}}{2}}{-\frac{1}{2}}  \> \twobytwo{-\frac{1}{2}}{\frac{\sqrt{3}}{2}}{-\frac{\sqrt{3}}{2}}{-\frac{1}{2}} \> \twobytwo{-\frac{1}{2}}{\frac{\sqrt{3}}{2}}{-\frac{\sqrt{3}}{2}}{-\frac{1}{2}} \\
\end{tabbing}}
\end{center}

\subsubsection{Six-member class ${\mathcal C}_2^{(-)} \times  {\mathcal C}_3^{(+)}$  containing (142635)($ab$)$^*$ }

\begin{center}
{\small\baselineskip=32pt
\begin{tabbing}
\parbox[b]{3.5truecm}{$ {\mathbf M}_{E_i}\left[ R_2^{(-)} R_4^{(+)} \right]$}
\= \parbox[b]{3.0truecm}{ \twobytwo{-\frac{1}{2}}{-\frac{\sqrt{3}}{2}}{\frac{\sqrt{3}}{2}}{-\frac{1}{2}} } \=
\parbox[b]{3.0truecm}{ \twobytwo{\frac{1}{2}}{\frac{\sqrt{3}}{2}}{-\frac{\sqrt{3}}{2}}{\frac{1}{2}}}  \=
\parbox[b]{3.0truecm}{\twobytwo{1}{0}{0}{-1} }  \=
\parbox[b]{3.0truecm}{\twobytwo{1}{0}{0}{-1}  }  \\
$ {\mathbf M}_{E_i}\left[ R_2^{(-)} R_5^{(+)} \right] $ \>  \twobytwo{-\frac{1}{2}}{-\frac{\sqrt{3}}{2}}{\frac{\sqrt{3}}{2}}{-\frac{1}{2}} \> \twobytwo{\frac{1}{2}}{\frac{\sqrt{3}}{2}}{-\frac{\sqrt{3}}{2}}{\frac{1}{2}}  \> \twobytwo{-\frac{1}{2}}{-\frac{\sqrt{3}}{2}}{-\frac{\sqrt{3}}{2}}{\frac{1}{2}} \> \twobytwo{-\frac{1}{2}}{-\frac{\sqrt{3}}{2}}{-\frac{\sqrt{3}}{2}}{\frac{1}{2}} \\
$ {\mathbf M}_{E_i}\left[ R_2^{(-)} R_6^{(+)} \right] $ \>  \twobytwo{-\frac{1}{2}}{-\frac{\sqrt{3}}{2}}{\frac{\sqrt{3}}{2}}{-\frac{1}{2}} \> \twobytwo{\frac{1}{2}}{\frac{\sqrt{3}}{2}}{-\frac{\sqrt{3}}{2}}{\frac{1}{2}}   \> \twobytwo{-\frac{1}{2}}{\frac{\sqrt{3}}{2}}{\frac{\sqrt{3}}{2}}{\frac{1}{2}} \> \twobytwo{-\frac{1}{2}}{\frac{\sqrt{3}}{2}}{\frac{\sqrt{3}}{2}}{\frac{1}{2}} \\
$ {\mathbf M}_{E_i}\left[ R_3^{(-)} R_4^{(+)} \right] $ \>  \twobytwo{-\frac{1}{2}}{\frac{\sqrt{3}}{2}}{-\frac{\sqrt{3}}{2}}{-\frac{1}{2}} \> \twobytwo{\frac{1}{2}}{-\frac{\sqrt{3}}{2}}{\frac{\sqrt{3}}{2}}{\frac{1}{2}}  \> \twobytwo{1}{0}{0}{-1} \> \twobytwo{1}{0}{0}{-1} \\
$ {\mathbf M}_{E_i}\left[ R_3^{(-)} R_5^{(+)} \right] $ \>  \twobytwo{-\frac{1}{2}}{\frac{\sqrt{3}}{2}}{-\frac{\sqrt{3}}{2}}{-\frac{1}{2}} \> \twobytwo{\frac{1}{2}}{-\frac{\sqrt{3}}{2}}{\frac{\sqrt{3}}{2}}{\frac{1}{2}}  \> \twobytwo{-\frac{1}{2}}{-\frac{\sqrt{3}}{2}}{-\frac{\sqrt{3}}{2}}{\frac{1}{2}} \> \twobytwo{-\frac{1}{2}}{-\frac{\sqrt{3}}{2}}{-\frac{\sqrt{3}}{2}}{\frac{1}{2}} \\
$ {\mathbf M}_{E_i}\left[ R_3^{(-)} R_6^{(+)} \right] $ \>  \twobytwo{-\frac{1}{2}}{\frac{\sqrt{3}}{2}}{-\frac{\sqrt{3}}{2}}{-\frac{1}{2}} \> \twobytwo{\frac{1}{2}}{-\frac{\sqrt{3}}{2}}{\frac{\sqrt{3}}{2}}{\frac{1}{2}}   \> \twobytwo{-\frac{1}{2}}{\frac{\sqrt{3}}{2}}{\frac{\sqrt{3}}{2}}{\frac{1}{2}} \> \twobytwo{-\frac{1}{2}}{\frac{\sqrt{3}}{2}}{\frac{\sqrt{3}}{2}}{\frac{1}{2}} \\
\end{tabbing}}
\end{center}

\subsubsection{Three-member class ${\mathcal C}_3^{(-)} \times  {\mathcal C}_1^{(+)}$  containing (14)(25)(36)($ab$) }

\begin{center}
{\small\baselineskip=32pt
\begin{tabbing}
\parbox[b]{3.5truecm}{$ {\mathbf M}_{E_i}\left[ R_4^{(-)}  \right]$}
\= \parbox[b]{3.0truecm}{\twobytwo{1}{0}{0}{-1}} \=
\parbox[b]{3.0truecm}{\twobytwo{1}{0}{0}{-1}}  \=
\parbox[b]{3.0truecm}{\twobytwo{1}{0}{0}{1}} \=
\parbox[b]{3.0truecm}{\twobytwo{-1}{0}{0}{-1}}  \\
$ {\mathbf M}_{E_i}\left[ R_5^{(-)} \right] $  \> \twobytwo{-\frac{1}{2}}{-\frac{\sqrt{3}}{2}}{-\frac{\sqrt{3}}{2}}{\frac{1}{2}} \> \twobytwo{-\frac{1}{2}}{-\frac{\sqrt{3}}{2}}{-\frac{\sqrt{3}}{2}}{\frac{1}{2}}  \> \twobytwo{1}{0}{0}{1} \> \twobytwo{-1}{0}{0}{-1} \\
$ {\mathbf M}_{E_i}\left[ R_6^{(-)} \right] $ \> \twobytwo{-\frac{1}{2}}{\frac{\sqrt{3}}{2}}{\frac{\sqrt{3}}{2}}{\frac{1}{2}} \> \twobytwo{-\frac{1}{2}}{\frac{\sqrt{3}}{2}}{\frac{\sqrt{3}}{2}}{\frac{1}{2}}  \> \twobytwo{1}{0}{0}{1} \> \twobytwo{-1}{0}{0}{-1} \\
\end{tabbing}}
\end{center}

\subsubsection{Six-member class ${\mathcal C}_3^{(-)} \times  {\mathcal C}_2^{(+)}$  containing (142536)($ab$) }

\begin{center}
{\small\baselineskip=32pt
\begin{tabbing}
\parbox[b]{3.5truecm}{$ {\mathbf M}_{E_i}\left[ R_4^{(-)} R_2^{(+)} \right]$}
\= \parbox[b]{3.0truecm}{\twobytwo{1}{0}{0}{-1}} \=
\parbox[b]{3.0truecm}{\twobytwo{1}{0}{0}{-1}}  \=
\parbox[b]{3.0truecm}{\twobytwo{-\frac{1}{2}}{-\frac{\sqrt{3}}{2}}{\frac{\sqrt{3}}{2}}{-\frac{1}{2}} } \=
\parbox[b]{3.0truecm}{\twobytwo{\frac{1}{2}}{\frac{\sqrt{3}}{2}}{-\frac{\sqrt{3}}{2}}{\frac{1}{2}} } \\[5pt]
$  {\mathbf M}_{E_i}\left[ R_5^{(-)} R_2^{(+)} \right] $  \>  \twobytwo{-\frac{1}{2}}{-\frac{\sqrt{3}}{2}}{-\frac{\sqrt{3}}{2}}{\frac{1}{2}} \> \twobytwo{-\frac{1}{2}}{-\frac{\sqrt{3}}{2}}{-\frac{\sqrt{3}}{2}}{\frac{1}{2}}  \> \twobytwo{-\frac{1}{2}}{-\frac{\sqrt{3}}{2}}{\frac{\sqrt{3}}{2}}{-\frac{1}{2}} \> \twobytwo{\frac{1}{2}}{\frac{\sqrt{3}}{2}}{-\frac{\sqrt{3}}{2}}{\frac{1}{2}} \\
$  {\mathbf M}_{E_i}\left[ R_6^{(-)} R_2^{(+)} \right] $  \>  \twobytwo{-\frac{1}{2}}{\frac{\sqrt{3}}{2}}{\frac{\sqrt{3}}{2}}{\frac{1}{2}} \> \twobytwo{-\frac{1}{2}}{\frac{\sqrt{3}}{2}}{\frac{\sqrt{3}}{2}}{\frac{1}{2}}  \> \twobytwo{-\frac{1}{2}}{-\frac{\sqrt{3}}{2}}{\frac{\sqrt{3}}{2}}{-\frac{1}{2}} \> \twobytwo{\frac{1}{2}}{\frac{\sqrt{3}}{2}}{-\frac{\sqrt{3}}{2}}{\frac{1}{2}} \\
$  {\mathbf M}_{E_i}\left[ R_4^{(-)} R_3^{(+)} \right] $ \>  \twobytwo{1}{0}{0}{-1} \> \twobytwo{1}{0}{0}{-1}  \> \twobytwo{-\frac{1}{2}}{\frac{\sqrt{3}}{2}}{-\frac{\sqrt{3}}{2}}{-\frac{1}{2}} \> \twobytwo{\frac{1}{2}}{-\frac{\sqrt{3}}{2}}{\frac{\sqrt{3}}{2}}{\frac{1}{2}} \\
$ {\mathbf M}_{E_i}\left[ R_5^{(-)} R_3^{(+)} \right] $ \>  \twobytwo{-\frac{1}{2}}{-\frac{\sqrt{3}}{2}}{-\frac{\sqrt{3}}{2}}{\frac{1}{2}} \> \twobytwo{-\frac{1}{2}}{-\frac{\sqrt{3}}{2}}{-\frac{\sqrt{3}}{2}}{\frac{1}{2}}  \> \twobytwo{-\frac{1}{2}}{\frac{\sqrt{3}}{2}}{-\frac{\sqrt{3}}{2}}{-\frac{1}{2}} \> \twobytwo{\frac{1}{2}}{-\frac{\sqrt{3}}{2}}{\frac{\sqrt{3}}{2}}{\frac{1}{2}} \\
$ {\mathbf M}_{E_i}\left[ R_6^{(-)} R_3^{(+)} \right] $ \>  \twobytwo{-\frac{1}{2}}{\frac{\sqrt{3}}{2}}{\frac{\sqrt{3}}{2}}{\frac{1}{2}} \> \twobytwo{-\frac{1}{2}}{\frac{\sqrt{3}}{2}}{\frac{\sqrt{3}}{2}}{\frac{1}{2}}  \> \twobytwo{-\frac{1}{2}}{\frac{\sqrt{3}}{2}}{-\frac{\sqrt{3}}{2}}{-\frac{1}{2}} \> \twobytwo{\frac{1}{2}}{-\frac{\sqrt{3}}{2}}{\frac{\sqrt{3}}{2}}{\frac{1}{2}} \\
\end{tabbing}}
\end{center}

\subsubsection{Nine-member class ${\mathcal C}_3^{(-)} \times  {\mathcal C}_3^{(+)}$  containing (12)(45)$^*$}

\begin{center}
{\small\baselineskip=32pt
\begin{tabbing}
\parbox[b]{3.5truecm}{$ {\mathbf M}_{E_i}\left[ R_4^{(-)} R_4^{(+)} \right]$}
\= \parbox[b]{3.0truecm}{\twobytwo{1}{0}{0}{-1}} \=
\parbox[b]{3.0truecm}{\twobytwo{-1}{0}{0}{1}}  \=
\parbox[b]{3.0truecm}{\twobytwo{1}{0}{0}{-1} } \=
\parbox[b]{3.0truecm}{\twobytwo{-1}{0}{0}{1} } \\[5pt]
$ {\mathbf M}_{E_i}\left[ R_4^{(-)} R_5^{(+)} \right] $ \> \twobytwo{1}{0}{0}{-1} \> \twobytwo{-1}{0}{0}{1}  \> \twobytwo{-\frac{1}{2}}{-\frac{\sqrt{3}}{2}}{-\frac{\sqrt{3}}{2}}{\frac{1}{2}} \> \twobytwo{\frac{1}{2}}{\frac{\sqrt{3}}{2}}{\frac{\sqrt{3}}{2}}{-\frac{1}{2}} \\
$ {\mathbf M}_{E_i}\left[ R_4^{(-)} R_6^{(+)} \right] $ \>  \twobytwo{1}{0}{0}{-1} \> \twobytwo{-1}{0}{0}{1}    \> \twobytwo{-\frac{1}{2}}{\frac{\sqrt{3}}{2}}{\frac{\sqrt{3}}{2}}{\frac{1}{2}} \> \twobytwo{\frac{1}{2}}{-\frac{\sqrt{3}}{2}}{-\frac{\sqrt{3}}{2}}{-\frac{1}{2}} \\
$ {\mathbf M}_{E_i}\left[ R_5^{(-)} R_4^{(+)} \right] $ \> \twobytwo{-\frac{1}{2}}{-\frac{\sqrt{3}}{2}}{-\frac{\sqrt{3}}{2}}{\frac{1}{2}} \> \twobytwo{\frac{1}{2}}{\frac{\sqrt{3}}{2}}{\frac{\sqrt{3}}{2}}{-\frac{1}{2}}  \> \twobytwo{1}{0}{0}{-1} \> \twobytwo{-1}{0}{0}{1} \\
$ {\mathbf M}_{E_i}\left[ R_5^{(-)} R_5^{(+)} \right] $ \> \twobytwo{-\frac{1}{2}}{-\frac{\sqrt{3}}{2}}{-\frac{\sqrt{3}}{2}}{\frac{1}{2}} \> \twobytwo{\frac{1}{2}}{\frac{\sqrt{3}}{2}}{\frac{\sqrt{3}}{2}}{-\frac{1}{2}}   \> \twobytwo{-\frac{1}{2}}{-\frac{\sqrt{3}}{2}}{-\frac{\sqrt{3}}{2}}{\frac{1}{2}} \> \twobytwo{\frac{1}{2}}{\frac{\sqrt{3}}{2}}{\frac{\sqrt{3}}{2}}{-\frac{1}{2}} \\
$ {\mathbf M}_{E_i}\left[ R_5^{(-)} R_6^{(+)} \right] $ \>  \twobytwo{-\frac{1}{2}}{-\frac{\sqrt{3}}{2}}{-\frac{\sqrt{3}}{2}}{\frac{1}{2}} \> \twobytwo{\frac{1}{2}}{\frac{\sqrt{3}}{2}}{\frac{\sqrt{3}}{2}}{-\frac{1}{2}}   \> \twobytwo{-\frac{1}{2}}{\frac{\sqrt{3}}{2}}{\frac{\sqrt{3}}{2}}{\frac{1}{2}} \> \twobytwo{\frac{1}{2}}{-\frac{\sqrt{3}}{2}}{-\frac{\sqrt{3}}{2}}{-\frac{1}{2}} \\
$ {\mathbf M}_{E_i}\left[ R_6^{(-)} R_4^{(+)} \right] $ \> \twobytwo{-\frac{1}{2}}{\frac{\sqrt{3}}{2}}{\frac{\sqrt{3}}{2}}{\frac{1}{2}} \> \twobytwo{\frac{1}{2}}{-\frac{\sqrt{3}}{2}}{-\frac{\sqrt{3}}{2}}{-\frac{1}{2}}  \> \twobytwo{1}{0}{0}{-1} \> \twobytwo{-1}{0}{0}{1} \\
$ {\mathbf M}_{E_i}\left[ R_6^{(-)} R_5^{(+)} \right] $ \>  \twobytwo{-\frac{1}{2}}{\frac{\sqrt{3}}{2}}{\frac{\sqrt{3}}{2}}{\frac{1}{2}} \> \twobytwo{\frac{1}{2}}{-\frac{\sqrt{3}}{2}}{-\frac{\sqrt{3}}{2}}{-\frac{1}{2}} \> \twobytwo{-\frac{1}{2}}{-\frac{\sqrt{3}}{2}}{-\frac{\sqrt{3}}{2}}{\frac{1}{2}} \> \twobytwo{\frac{1}{2}}{\frac{\sqrt{3}}{2}}{\frac{\sqrt{3}}{2}}{-\frac{1}{2}} \\
$ {\mathbf M}_{E_i}\left[ R_6^{(-)} R_6^{(+)} \right] $ \> \twobytwo{-\frac{1}{2}}{\frac{\sqrt{3}}{2}}{\frac{\sqrt{3}}{2}}{\frac{1}{2}} \> \twobytwo{\frac{1}{2}}{-\frac{\sqrt{3}}{2}}{-\frac{\sqrt{3}}{2}}{-\frac{1}{2}}   \> \twobytwo{-\frac{1}{2}}{\frac{\sqrt{3}}{2}}{\frac{\sqrt{3}}{2}}{\frac{1}{2}} \> \twobytwo{\frac{1}{2}}{-\frac{\sqrt{3}}{2}}{-\frac{\sqrt{3}}{2}}{-\frac{1}{2}} \\
\end{tabbing}}
\end{center}

\subsection{$G$ transformation matrices in {\itshape\bfseries G}$_{36}$}
\label{sec:finalmatrices}

\subsubsection{One-member class ${\mathcal C}_1^{(-)} \times  {\mathcal C}_1^{(+)}$  containing $E$}
\begin{displaymath}
{\mathbf M}_G\left[ R_1^{(-)} R_1^{(+)} \right] =
\left( \begin{array}{rrrr} 1&0&0&0\\ 0&1&0&0\\ 0&0&1&0\\ 0&0&0&1\end{array} \right) \end{displaymath}

\subsubsection{Two-member class ${\mathcal C}_1^{(-)} \times  {\mathcal C}_2^{(+)}$  containing (123)(456) and (132)(465)}
\begin{displaymath} {\mathbf M}_G\left[ R_1^{(-)} R_2^{(+)} \right] =
 \left( \begin{array}{rrrr} -{\frac{1}{2}}&-{\frac{\sqrt{3}}{2}}&0&0\\ {\frac{\sqrt{3}
 }{2}}&-{\frac{1}{2}}&0&0\\ 0&0&-{\frac{1}{2}}&-{\frac{\sqrt{3}
 }{2}}\\ 0&0&{\frac{\sqrt{3}}{2}}&-{\frac{1}{2}}\end{array} \right) \end{displaymath}
\begin{displaymath}  {\mathbf M}_G\left[ R_1^{(-)} R_3^{(+)} \right] =
 \left( \begin{array}{rrrr}-{\frac{1}{2}}&{\frac{\sqrt{3}}{2}}&0&0\\ -{\frac{\sqrt{3}
 }{2}}&-{\frac{1}{2}}&0&0\\ 0&0&-{\frac{1}{2}}&{\frac{\sqrt{3}
 }{2}}\\ 0&0&-{\frac{\sqrt{3}}{2}}&-{\frac{1}{2}}\end{array} \right) \end{displaymath}

\subsubsection{Three-member class ${\mathcal C}_1^{(-)} \times  {\mathcal C}_3^{(+)}$  containing (14)(26)(35)($ab$)$^*$}
\begin{displaymath} {\mathbf M}_G\left[ R_1^{(-)} R_4^{(+)} \right] =
 \left( \begin{array}{cccc}1&0&0&0\\ 0&-1&0&0\\ 0&0&-1&0\\ 0&0&0&1\end{array} \right) \end{displaymath}
\begin{displaymath} {\mathbf M}_G\left[ R_1^{(-)} R_5^{(+)} \right] =
  \left( \begin{array}{cccc}-{\frac{1}{2}}&-{\frac{\sqrt{3}}{2}}&0&0\\ -{\frac{\sqrt{3}
 }{2}}&{\frac{1}{2}}&0&0\\ 0&0&{\frac{1}{2}}&{\frac{\sqrt{3}}{2
 }}\\ 0&0&{\frac{\sqrt{3}}{2}}&-{\frac{1}{2}}\end{array} \right) \end{displaymath}
\begin{displaymath} {\mathbf M}_G\left[ R_1^{(-)} R_6^{(+)} \right] =
 \left( \begin{array}{cccc}-{\frac{1}{2}}&{\frac{\sqrt{3}}{2}}&0&0\\ {\frac{\sqrt{3}
 }{2}}&{\frac{1}{2}}&0&0\\ 0&0&{\frac{1}{2}}&-{\frac{\sqrt{3}}{
 2}}\\ 0&0&-{\frac{\sqrt{3}}{2}}&-{\frac{1}{2}}\end{array} \right) \end{displaymath}

\subsubsection{Two-member class ${\mathcal C}_2^{(-)} \times  {\mathcal C}_1^{(+)}$  containing (123)(456) and (132)(456)}
\begin{displaymath} {\mathbf M}_G\left[ R_2^{(-)} R_1^{(+)} \right] =
 \left( \begin{array}{cccc}-{\frac{1}{2}}&0&0&-{\frac{\sqrt{3}}{2}}\\ 0&-{\frac{1}{2
 }}&{\frac{\sqrt{3}}{2}}&0\\ 0&-{\frac{\sqrt{3}}{2}}&-{\frac{1}{2}}&
 0\\ {\frac{\sqrt{3}}{2}}&0&0&-{\frac{1}{2}}\end{array} \right) \end{displaymath}
\begin{displaymath} {\mathbf M}_G\left[ R_3^{(-)} R_1^{(+)} \right] =
 \left( \begin{array}{cccc}-{\frac{1}{2}}&0&0&{\frac{\sqrt{3}}{2}}\\ 0&-{\frac{1}{2
 }}&-{\frac{\sqrt{3}}{2}}&0\\ 0&{\frac{\sqrt{3}}{2}}&-{\frac{1}{2}}&
 0\\ -{\frac{\sqrt{3}}{2}}&0&0&-{\frac{1}{2}}\end{array} \right) \end{displaymath}

\subsubsection{Four-member class ${\mathcal C}_2^{(-)} \times  {\mathcal C}_2^{(+)}$  containing (123) and (456)}
\begin{displaymath} {\mathbf M}_G\left[ R_2^{(-)}R_2^{(+)} \right] =
\left( \begin{array}{cccc} {\frac{1}{4}}&{\frac{\sqrt{3}}{4}}&-{\frac{3}{4}}&{\frac{\sqrt{
 3}}{4}}\\ -{\frac{\sqrt{3}}{4}}&{\frac{1}{4}}&-{\frac{\sqrt{3}
 }{4}}&-{\frac{3}{4}}\\ -{\frac{3}{4}}&{\frac{\sqrt{3}}{4}}&{\frac{1
 }{4}}&{\frac{\sqrt{3}}{4}}\\ -{\frac{\sqrt{3}}{4}}&-{\frac{3}{
 4}}&-{\frac{\sqrt{3}}{4}}&{\frac{1}{4}}\end{array} \right) \end{displaymath}
\begin{displaymath} {\mathbf M}_G\left[ R_2^{(-)}R_3^{(+)} \right] =
\left( \begin{array}{cccc} {\frac{1}{4}}&-{\frac{\sqrt{3}}{4}}&{\frac{3}{4}}&{\frac{\sqrt{
 3}}{4}}\\ {\frac{\sqrt{3}}{4}}&{\frac{1}{4}}&-{\frac{\sqrt{3}
 }{4}}&{\frac{3}{4}}\\ {\frac{3}{4}}&{\frac{\sqrt{3}}{4}}&{\frac{1
 }{4}}&-{\frac{\sqrt{3}}{4}}\\ -{\frac{\sqrt{3}}{4}}&{\frac{3}{
 4}}&{\frac{\sqrt{3}}{4}}&{\frac{1}{4}}\end{array} \right) \end{displaymath}
\begin{displaymath} {\mathbf M}_G\left[ R_3^{(-)}R_2^{(+)} \right] =
\left( \begin{array}{cccc} {\frac{1}{4}}&{\frac{\sqrt{3}}{4}}&{\frac{3}{4}}&-{\frac{\sqrt{
 3}}{4}}\\ -{\frac{\sqrt{3}}{4}}&{\frac{1}{4}}&{\frac{\sqrt{3}
 }{4}}&{\frac{3}{4}}\\ {\frac{3}{4}}&-{\frac{\sqrt{3}}{4}}&{\frac{1
 }{4}}&{\frac{\sqrt{3}}{4}}\\ {\frac{\sqrt{3}}{4}}&{\frac{3}{4
 }}&-{\frac{\sqrt{3}}{4}}&{\frac{1}{4}}\end{array} \right) \end{displaymath}
\begin{displaymath} {\mathbf M}_G\left[ R_3^{(-)}R_3^{(+)} \right] =
\left( \begin{array}{cccc} {\frac{1}{4}}&-{\frac{\sqrt{3}}{4}}&-{\frac{3}{4}}&-{\frac{
 \sqrt{3}}{4}}\\ {\frac{\sqrt{3}}{4}}&{\frac{1}{4}}&{\frac{\sqrt{3}
 }{4}}&-{\frac{3}{4}}\\ -{\frac{3}{4}}&-{\frac{\sqrt{3}}{4}}&{\frac{
 1}{4}}&-{\frac{\sqrt{3}}{4}}\\ {\frac{\sqrt{3}}{4}}&-{\frac{3
 }{4}}&{\frac{\sqrt{3}}{4}}&{\frac{1}{4}}\end{array} \right) \end{displaymath}

\subsubsection{Six-member class ${\mathcal C}_2^{(-)} \times  {\mathcal C}_3^{(+)}$  containing (142635)($ab$)$^*$ }
\begin{displaymath} {\mathbf M}_G\left[ R_2^{(-)} R_4^{(+)} \right] =
\left( \begin{array}{cccc} -{\frac{1 }{2}}&0&0&-{\frac{\sqrt{3} }{2}}\\ 0&{\frac{1 }{2
 }}&-{\frac{\sqrt{3} }{2}}&0\\ 0&{\frac{\sqrt{3} }{2}}&{\frac{1 }{2}}&0
 \\ {\frac{\sqrt{3} }{2}}&0&0&-{\frac{1 }{2}}\end{array} \right) \end{displaymath}
\begin{displaymath} {\mathbf M}_G\left[ R_3^{(-)} R_4^{(+)} \right] =
\left( \begin{array}{cccc} -{\frac{1 }{2}}&0&0&{\frac{\sqrt{3} }{2}}\\ 0&{\frac{1 }{2}}
 &{\frac{\sqrt{3} }{2}}&0\\ 0&-{\frac{\sqrt{3} }{2}}&{\frac{1 }{2}}&0
 \\ -{\frac{\sqrt{3} }{2}}&0&0&-{\frac{1 }{2}}\end{array} \right) \end{displaymath}
\begin{displaymath} {\mathbf M}_G\left[ R_2^{(-)} R_5^{(+)} \right] =
\left( \begin{array}{cccc }{\frac{1 }{4}}&{\frac{\sqrt{3} }{4}}&-{\frac{3 }{4}}&{\frac{\sqrt{
 3} }{4}}\\ {\frac{\sqrt{3} }{4}}&-{\frac{1 }{4}}&{\frac{\sqrt{3}
  }{4}}&{\frac{3 }{4}}\\ {\frac{3 }{4}}&-{\frac{\sqrt{3} }{4}}&-{\frac{1
  }{4}}&-{\frac{\sqrt{3} }{4}}\\ -{\frac{\sqrt{3} }{4}}&-{\frac{3
  }{4}}&-{\frac{\sqrt{3} }{4}}&{\frac{1 }{4}}\end{array} \right) \end{displaymath}
\begin{displaymath} {\mathbf M}_G\left[ R_3^{(-)} R_5^{(+)} \right] =
\left( \begin{array}{cccc }{\frac{1 }{4}}&{\frac{\sqrt{3} }{4}}&{\frac{3 }{4}}&-{\frac{\sqrt{
 3} }{4}}\\ {\frac{\sqrt{3} }{4}}&-{\frac{1 }{4}}&-{\frac{\sqrt{3}
  }{4}}&-{\frac{3 }{4}}\\ -{\frac{3 }{4}}&{\frac{\sqrt{3} }{4}}&-{\frac{
 1 }{4}}&-{\frac{\sqrt{3} }{4}}\\ {\frac{\sqrt{3} }{4}}&{\frac{3 }{
 4}}&-{\frac{\sqrt{3} }{4}}&{\frac{1 }{4}}\end{array} \right) \end{displaymath}
\begin{displaymath} {\mathbf M}_G\left[ R_2^{(-)} R_6^{(+)} \right] =
\left( \begin{array}{cccc }{\frac{1 }{4}}&-{\frac{\sqrt{3} }{4}}&{\frac{3 }{4}}&{\frac{\sqrt{
 3} }{4}}\\ -{\frac{\sqrt{3} }{4}}&-{\frac{1 }{4}}&{\frac{\sqrt{3}
  }{4}}&-{\frac{3 }{4}}\\ -{\frac{3 }{4}}&-{\frac{\sqrt{3} }{4}}&-
 {\frac{1 }{4}}&{\frac{\sqrt{3} }{4}}\\ -{\frac{\sqrt{3} }{4}}&{\frac{3
  }{4}}&{\frac{\sqrt{3} }{4}}&{\frac{1 }{4}}\end{array} \right) \end{displaymath}
\begin{displaymath} {\mathbf M}_G\left[ R_3^{(-)} R_6^{(+)} \right] =
\left( \begin{array}{cccc }{\frac{1 }{4}}&-{\frac{\sqrt{3} }{4}}&-{\frac{3 }{4}}&-{\frac{
 \sqrt{3} }{4}}\\ -{\frac{\sqrt{3} }{4}}&-{\frac{1 }{4}}&-{\frac{\sqrt{
 3} }{4}}&{\frac{3 }{4}}\\ {\frac{3 }{4}}&{\frac{\sqrt{3} }{4}}&-{\frac{
 1 }{4}}&{\frac{\sqrt{3} }{4}}\\ {\frac{\sqrt{3} }{4}}&-{\frac{3 }{
 4}}&{\frac{\sqrt{3} }{4}}&{\frac{1 }{4}}\end{array} \right) \end{displaymath}

\subsubsection{Three-member class ${\mathcal C}_3^{(-)} \times  {\mathcal C}_1^{(+)}$  containing (14)(25)(36)($ab$) }

\begin{displaymath} {\mathbf M}_G\left[ R_4^{(-)} \right] =
\left( \begin{array}{cccc}1&0&0&0\\ 0&1&0&0\\ 0&0&-1&0\\ 0&0&0&-1\end{array} \right) \end{displaymath}

\begin{displaymath} {\mathbf M}_G\left[  R_5^{(-)} \right] =
\left( \begin{array}{cccc}-{\frac{1}{2}}&0&0&-{\frac{\sqrt{3}}{2}}\\ 0&-{\frac{1}{2
 }}&{\frac{\sqrt{3}}{2}}&0\\ 0&{\frac{\sqrt{3}}{2}}&{\frac{1}{2}}&0
 \\ -{\frac{\sqrt{3}}{2}}&0&0&{\frac{1}{2}}\end{array} \right) \end{displaymath}
\begin{displaymath} {\mathbf M}_G\left[ R_6^{(-)} \right] =
\left( \begin{array}{cccc}-{\frac{1}{2}}&0&0&{\frac{\sqrt{3}}{2}}\\ 0&-{\frac{1}{2
 }}&-{\frac{\sqrt{3}}{2}}&0\\ 0&-{\frac{\sqrt{3}}{2}}&{\frac{1}{2}}&
 0\\ {\frac{\sqrt{3}}{2}}&0&0&{\frac{1}{2}}\end{array} \right) \end{displaymath}

\subsubsection{Six-member class ${\mathcal C}_3^{(-)} \times  {\mathcal C}_2^{(+)}$  containing (142536)($ab$) }

\begin{displaymath} {\mathbf M}_G\left[ R_4^{(-)} R_2^{(+)} \right] =
\left( \begin{array}{cccc}-{\frac{1}{2}}&-{\frac{\sqrt{3}}{2}}&0&0\\ {\frac{\sqrt{3}
 }{2}}&-{\frac{1}{2}}&0&0\\ 0&0&{\frac{1}{2}}&{\frac{\sqrt{3}}{
 2}}\\ 0&0&-{\frac{\sqrt{3}}{2}}&{\frac{1}{2}}\end{array} \right) \end{displaymath}
\begin{displaymath}  {\mathbf M}_G\left[ R_5^{(-)} R_2^{(+)} \right] =
\left( \begin{array}{cccc}{\frac{1}{4}}&{\frac{\sqrt{3}}{4}}&-{\frac{3}{4}}&{\frac{\sqrt{
 3}}{4}}\\ -{\frac{\sqrt{3}}{4}}&{\frac{1}{4}}&-{\frac{\sqrt{3}
 }{4}}&-{\frac{3}{4}}\\ {\frac{3}{4}}&-{\frac{\sqrt{3}}{4}}&-{\frac{
 1}{4}}&-{\frac{\sqrt{3}}{4}}\\ {\frac{\sqrt{3}}{4}}&{\frac{3}{
 4}}&{\frac{\sqrt{3}}{4}}&-{\frac{1}{4}}\end{array} \right) \end{displaymath}
\begin{displaymath}  {\mathbf M}_G\left[ R_6^{(-)} R_2^{(+)} \right] =
\left( \begin{array}{cccc}{\frac{1}{4}}&{\frac{\sqrt{3}}{4}}&{\frac{3}{4}}&-{\frac{\sqrt{
 3}}{4}}\\ -{\frac{\sqrt{3}}{4}}&{\frac{1}{4}}&{\frac{\sqrt{3}
 }{4}}&{\frac{3}{4}}\\ -{\frac{3}{4}}&{\frac{\sqrt{3}}{4}}&-{\frac{1
 }{4}}&-{\frac{\sqrt{3}}{4}}\\ -{\frac{\sqrt{3}}{4}}&-{\frac{3
 }{4}}&{\frac{\sqrt{3}}{4}}&-{\frac{1}{4}}\end{array} \right) \end{displaymath}
\begin{displaymath}  {\mathbf M}_G\left[ R_4^{(-)} R_3^{(+)} \right] =
\left( \begin{array}{cccc}-{\frac{1}{2}}&{\frac{\sqrt{3}}{2}}&0&0\\ -{\frac{\sqrt{3}
 }{2}}&-{\frac{1}{2}}&0&0\\ 0&0&{\frac{1}{2}}&-{\frac{\sqrt{3}
 }{2}}\\ 0&0&{\frac{\sqrt{3}}{2}}&{\frac{1}{2}}\end{array} \right) \end{displaymath}
\begin{displaymath} {\mathbf M}_G\left[ R_5^{(-)} R_3^{(+)} \right] =
\left( \begin{array}{cccc}{\frac{1}{4}}&-{\frac{\sqrt{3}}{4}}&{\frac{3}{4}}&{\frac{\sqrt{
 3}}{4}}\\ {\frac{\sqrt{3}}{4}}&{\frac{1}{4}}&-{\frac{\sqrt{3}
 }{4}}&{\frac{3}{4}}\\ -{\frac{3}{4}}&-{\frac{\sqrt{3}}{4}}&-{\frac{
 1}{4}}&{\frac{\sqrt{3}}{4}}\\ {\frac{\sqrt{3}}{4}}&-{\frac{3}{
 4}}&-{\frac{\sqrt{3}}{4}}&-{\frac{1}{4}}\end{array} \right) \end{displaymath}
\begin{displaymath} {\mathbf M}_G\left[ R_6^{(-)} R_3^{(+)} \right] =
\left( \begin{array}{cccc}{\frac{1}{4}}&-{\frac{\sqrt{3}}{4}}&-{\frac{3}{4}}&-{\frac{
 \sqrt{3}}{4}}\\ {\frac{\sqrt{3}}{4}}&{\frac{1}{4}}&{\frac{\sqrt{3}
 }{4}}&-{\frac{3}{4}}\\ {\frac{3}{4}}&{\frac{\sqrt{3}}{4}}&-{\frac{1
 }{4}}&{\frac{\sqrt{3}}{4}}\\ -{\frac{\sqrt{3}}{4}}&{\frac{3}{4
 }}&-{\frac{\sqrt{3}}{4}}&-{\frac{1}{4}}\end{array} \right) \end{displaymath}

\subsubsection{Nine-member class ${\mathcal C}_3^{(-)} \times  {\mathcal C}_3^{(+)}$  containing (12)(45)$^*$}

\begin{displaymath} {\mathbf M}_G\left[ R_4^{(-)} R_4^{(+)} \right] =
\left( \begin{array}{cccc}1&0&0&0\\ 0&-1&0&0\\ 0&0&1&0\\ 0&0&0&-1\end{array} \right) \end{displaymath}
\begin{displaymath} {\mathbf M}_G\left[ R_4^{(-)} R_5^{(+)} \right] =
 \left( \begin{array}{cccc}-{\frac{1}{2}}&-{\frac{\sqrt{3}}{2}}&0&0\\ -{\frac{\sqrt{3}
 }{2}}&{\frac{1}{2}}&0&0\\ 0&0&-{\frac{1}{2}}&-{\frac{\sqrt{3}
 }{2}}\\ 0&0&-{\frac{\sqrt{3}}{2}}&{\frac{1}{2}}\end{array} \right) \end{displaymath}
\begin{displaymath} {\mathbf M}_G\left[ R_4^{(-)} R_6^{(+)} \right] =
 \left( \begin{array}{cccc}-{\frac{1}{2}}&{\frac{\sqrt{3}}{2}}&0&0\\ {\frac{\sqrt{3}
 }{2}}&{\frac{1}{2}}&0&0\\ 0&0&-{\frac{1}{2}}&{\frac{\sqrt{3}}{
 2}}\\ 0&0&{\frac{\sqrt{3}}{2}}&{\frac{1}{2}}\end{array} \right) \end{displaymath}
\begin{displaymath} {\mathbf M}_G\left[ R_5^{(-)} R_4^{(+)} \right] =
 \left( \begin{array}{cccc}-{\frac{1}{2}}&0&0&-{\frac{\sqrt{3}}{2}}\\ 0&{\frac{1}{2
 }}&-{\frac{\sqrt{3}}{2}}&0\\ 0&-{\frac{\sqrt{3}}{2}}&-{\frac{1}{2}}
 &0\\ -{\frac{\sqrt{3}}{2}}&0&0&{\frac{1}{2}}\end{array} \right) \end{displaymath}
\begin{displaymath} {\mathbf M}_G\left[ R_5^{(-)} R_5^{(+)} \right] =
\left( \begin{array}{cccc}{\frac{1}{4}}&{\frac{\sqrt{3}}{4}}&-{\frac{3}{4}}&{\frac{\sqrt{
 3}}{4}}\\ {\frac{\sqrt{3}}{4}}&-{\frac{1}{4}}&{\frac{\sqrt{3}
 }{4}}&{\frac{3}{4}}\\ -{\frac{3}{4}}&{\frac{\sqrt{3}}{4}}&{\frac{1
 }{4}}&{\frac{\sqrt{3}}{4}}\\ {\frac{\sqrt{3}}{4}}&{\frac{3}{4
 }}&{\frac{\sqrt{3}}{4}}&-{\frac{1}{4}}\end{array} \right) \end{displaymath}
\begin{displaymath} {\mathbf M}_G\left[ R_5^{(-)} R_6^{(+)} \right] =
\left( \begin{array}{cccc}{\frac{1}{4}}&-{\frac{\sqrt{3}}{4}}&{\frac{3}{4}}&{\frac{\sqrt{
 3}}{4}}\\ -{\frac{\sqrt{3}}{4}}&-{\frac{1}{4}}&{\frac{\sqrt{3}
 }{4}}&-{\frac{3}{4}}\\ {\frac{3}{4}}&{\frac{\sqrt{3}}{4}}&{\frac{1
 }{4}}&-{\frac{\sqrt{3}}{4}}\\ {\frac{\sqrt{3}}{4}}&-{\frac{3}{
 4}}&-{\frac{\sqrt{3}}{4}}&-{\frac{1}{4}}\end{array} \right) \end{displaymath}
\begin{displaymath} {\mathbf M}_G\left[ R_6^{(-)} R_4^{(+)} \right] =
 \left( \begin{array}{cccc}-{\frac{1}{2}}&0&0&{\frac{\sqrt{3}}{2}}\\ 0&{\frac{1}{2}}
 &{\frac{\sqrt{3}}{2}}&0\\ 0&{\frac{\sqrt{3}}{2}}&-{\frac{1}{2}}&0
 \\ {\frac{\sqrt{3}}{2}}&0&0&{\frac{1}{2}}\end{array} \right) \end{displaymath}
\begin{displaymath} {\mathbf M}_G\left[ R_6^{(-)} R_5^{(+)} \right] =
 \left( \begin{array}{cccc}{\frac{1}{4}}&{\frac{\sqrt{3}}{4}}&{\frac{3}{4}}&-{\frac{\sqrt{
 3}}{4}}\\ {\frac{\sqrt{3}}{4}}&-{\frac{1}{4}}&-{\frac{\sqrt{3}
 }{4}}&-{\frac{3}{4}}\\ {\frac{3}{4}}&-{\frac{\sqrt{3}}{4}}&{\frac{1
 }{4}}&{\frac{\sqrt{3}}{4}}\\ -{\frac{\sqrt{3}}{4}}&-{\frac{3}{
 4}}&{\frac{\sqrt{3}}{4}}&-{\frac{1}{4}}\end{array} \right) \end{displaymath}
\begin{displaymath} {\mathbf M}_G\left[ R_6^{(-)} R_6^{(+)} \right] =
\left( \begin{array}{cccc}{\frac{1}{4}}&-{\frac{\sqrt{3}}{4}}&-{\frac{3}{4}}&-{\frac{
 \sqrt{3}}{4}}\\ -{\frac{\sqrt{3}}{4}}&-{\frac{1}{4}}&-{\frac{\sqrt{
 3}}{4}}&{\frac{3}{4}}\\ -{\frac{3}{4}}&-{\frac{\sqrt{3}}{4}}&
 {\frac{1}{4}}&-{\frac{\sqrt{3}}{4}}\\ -{\frac{\sqrt{3}}{4}}&{\frac{3
 }{4}}&-{\frac{\sqrt{3}}{4}}&-{\frac{1}{4}}\end{array} \right) \end{displaymath}

	\section{Derivation of the transformation of internal coordinates}\label{sec:appendixB}
\setcounter{equation}{0}

	In the following we will describe the procedure with which to calculate the transformation of the internal coordinates due to the molecular symmetry group operations. If $f$ is a function of the Cartesian coordinates $\vect{X}_1$, $\vect{X}_2,$ and $\vect{X}_3$ of three nuclei 1, 2, and 3, respectively, then, after the operation (123), nucleus 1 is at the position nucleus 3 was previously, 2 is at 1, and 3 is at 2. We thus have $(123)f(\vect{X}_1,\vect{X}_2,\vect{X}_3) = f(\vect{X}_3,\vect{X}_1,\vect{X}_2)$. Also, $E^*f(\vect{X}_1,\vect{X}_2,\vect{X}_3) = f(-\vect{X}_1,-\vect{X}_2,-\vect{X}_3)$. Since any curvilinear coordinate can be written as a function of the Cartesian coordinates, this defines the coordinate change due to the operation. We will now go through the three types of internal coordinates that TROVE uses.

	\subsection{Bond Lengths}
	\renewcommand{\vect}[1]{\mathbf{#1}}

	The bond length between nuclei 1 and 2 (with position vectors $\vect{r}_1$ and $\vect{r}_2$, respectively in the space-fixed axis system)  is defined by $|\vect{r}_1 - \vect{r}_2|.$ We see that this value is unaffected by $E^*$ and the other operations simply result in the relabelling of the bond lengths. For example, the \ce{H1-C_$a$} bond length is given by $r_1$ $=$ $|\vect{r}_1 - \vect{r}_a|$ and the operation (123) transforms $r_1$ to $r_1'$ $=$ $|\vect{r}_3 - \vect{r}_a|$ $=$ $r_3$,  and we obtain, once again, an obvious result. The operation $(14)(26)(35)(ab)^*$ yields $r_1'$ $=$ $|-\vect{r}_4 + \vect{r}_b| = |\vect{r}_4 - \vect{r}_b| = r_4$

	\subsection{Bond Angles}

	To obtain the angle between nuclei 1 and $b$ via nucleus $a$ we define $\vect{r}_{1a} = \vect{r}_1 - \vect{r}_a$ and likewise for $\vect{r}_{ba}$. Then the sought angle is
	\begin{equation}
	\alpha_1 =
		\arccos \left(\frac{\vect{r}_{1a}\cdot \vect{r}_{ba}}{|\vect{r}_{1a}|| \vect{r}_{ba}|} \right).
		\end{equation}
		Despite the expression for $\alpha_1$ being more complex than the expression for $r_1$,
		the effect of MS group operations can be determined completely analogously.
		 $E^*$ produces no effect and the other operations relabel. For $(123)$, we have
		\begin{equation}
		\alpha_1' =
			\arccos \left(\frac{\vect{r}_{3a}\cdot \vect{r}_{ba}}{|\vect{r}_{3a}|| \vect{r}_{ba}|} \right) = \alpha_3,
			\end{equation}
			while for $(14)(26)(35)(ab)^*$
			\begin{equation}
			\alpha_1' =
				\arccos \left(\frac{(-\vect{r}_{4b})\cdot (-\vect{r}_{ab})}{|-\vect{r}_{4b}|| -\vect{r}_{ab}|} \right) =\arccos \left(\frac{\vect{r}_{4b}\cdot \vect{r}_{ab}}{|\vect{r}_{4b}|| \vect{r}_{ab}|} \right) = \alpha_4.
				\end{equation}
							
				\subsection{Dihedral Angles}
\def\ez{\vect{e}_{z}}
\def\axxtan{\mathrm{arctan2}}
				The expressions involving the  dihedral angles are the most complicated. Consider the four nuclei labelled 1, a, b, and 4 in \fig{fig:dihedral-atoms}. The dihedral angle between the plane spanned by 1, $a$, and $b$ and that spanned by $a$, $b$, and 4 is shown in \fig{fig:dihedral-atoms} and \fig{fig:dihedral-angle}. We define
				$\normvec{\vect{v}}$ as the unit vector obtained by normalization of $\vect{v}$,
				$\normvec{\vect{v}}$ $=$ $\vect{v}/\vert \vect{v} \vert$. The orientation of the $z$ axis  is defined by $\ez$ $=$ $ \normvec{\vect{r}_{ab}}$. We need two further axes $x$ and $y$ (whose orientations are defined by the unit vectors $\vect{e}_{x}$ and $\vect{e}_{y}$, respectively) which, together with $z$, form a right-handed axis system.
				
\begin{figure}[htbp!]
																	\centering
																	\tdplotsetmaincoords{0}{0}
																	\begin{tikzpicture}[tdplot_main_coords]
																		\draw[thick] (0,0,0)  -- (0,0,1) ;
																		\draw[thick,->] (0,0,1) -- (0.353553,0.612372,1.707107) node[right]{$\vect{r}_{1a}$};
																		\draw[thick,->] (0,0,0) -- (-0.700629,-0.509037,-0.866025) node[left]{$\vect{r}_{4b}$};
																		\draw[thick,->] (0,0,0) -- (-0.700629,-0.509037,-0.866025);
																		\draw[thick,->] (0,0,0) -- (0.573576,-0.819152, 0) node[below right]{$\vect{e}_{y} = \normvec{\ez \times \vect{r}_{4b}}$};
																	\end{tikzpicture}
																	\caption{The $y$ axis.}
																	\label{fig:dihedral-cross}	
																\end{figure}
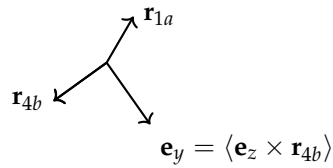

				To simplify the discussion, we take the $xy$ plane to be horizontal and the $z$ axis to be vertical.
				We aim at obtaining the $x$ and $y$ components of $\vect{r}_{1a}$ in order to determine the ``horizontal''  angle $\theta$ it makes with $\vect{r}_{4b}$ as shown in \fig{fig:dihedral-angle}.  We require the $x$ axis to be directed along the horizontal component (i.e., the component perpendicular to $\ez$)
				of $\vect{r}_{4b}$. Thus, the $y$ axis is perpendicular to the plane defined by $\ez$ and $\vect{r}_{4b}$, so that we have  $\vect{e}_{y} = \normvec{\ez \times \vect{r}_{4b}}$ and therefore $\vect{e}_{x} = \ez \times \vect{e}_{y} = \ez \times \normvec{\ez \times \vect{r}_{4b}}$ as shown in \fig{fig:dihedral-cross}. The unit vector $\vect{e}_{x'} =  \ez \times \normvec{\ez \times \vect{r}_{1a}}$ defines an $x'$ axis in the 1--$a$--$b$ plane; this axis is analogous to the $x$ axis in the 4--$a$--$b$ plane. The dihedral angle between the two planes is the angle between the $x$ and $x'$ axes.

				The $x$ and $y$ components of $\vect{e}_{x'}$  are $\vect{e}_{x} \cdot \vect{e}_{x'}$ and  $\vect{e}_{y} \cdot  \vect{e}_{x'}$, respectively. In order to obtain the dihedral angle in the range  $[-\pi,\pi]$ with the correct sign, we use the standard trigonometric
				function\footnote{For $(x,y)$ $=$ $(r\,\cos\varphi, r\,\sin\varphi)$, the function
				$\axxtan (y,x) = \varphi \in [-\pi,\pi]$.}
				$\axxtan$ to obtain
				\begin{eqnarray}
			\theta =	\tau_{41} & = &
				\axxtan(\vect{e}_{y} \cdot \vect{e}_{x'}, \vect{e}_{x} \cdot \vect{e}_{x'})
				\nonumber\\ & = &
				\axxtan(\vect{e}_{y} \cdot  \ez \times \normvec{\ez \times \vect{r}_{1a}},\vect{e}_{x} \cdot  \ez \times \normvec{\ez \times \vect{r}_{1a}})
				\end{eqnarray}
				which, written more explicitly, is
				\begin{equation}
					\begin{split}
				\theta =	\tau_{41}  =
						\axxtan [& \normvec{\ez \times \vect{r}_{4b}} \cdot (\ez \times \normvec{\ez \times \vect{r}_{1a}}),  \\
						&(\ez \times \normvec{\ez \times \vect{r}_{4b}}) \cdot (\ez \times \normvec{\ez \times \vect{r}_{1a}}) ] \\
						=	\axxtan [& \ez \cdot ( \normvec{\ez \times \vect{r}_{1a}} \times \normvec{\ez \times \vect{r}_{4b}}),  \\
							&(\ez \times \normvec{\ez \times \vect{r}_{1a}}) \cdot (\ez \times \normvec{\ez \times \vect{r}_{4b}}) ]
						\end{split}
						\label{eq:died-def}
						\end{equation}
						where the last expression emphasizes the equivalence of nuclei 1 and 4.

						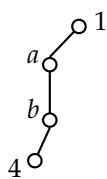
\begin{figure}[htbp!]
																	\centering
																	\tdplotsetmaincoords{70}{110}
																	\begin{tikzpicture}[tdplot_main_coords]
																		\draw[thick,o-o] (0,0,0) node[above left]{$b$}  -- (0,0,1) node[left]{$a$};

																		\draw[thick,-o] (0,0,1) -- (0.353553,0.612372,1.707107) node[right]{1};
																		\draw[thick,-o] (0,0,0) -- (-0.700629,-0.509037,-0.866025) node[left]{4};
																	\end{tikzpicture}
																	\caption{Four nuclei 1, 4, $a$, and $b$. The dihedral angle
																	$\theta$ is the angle between
																	the 1--$a$--$b$ and 4--$a$--$b$ planes. We define the positive direction of the angle by the right hand rule with the thumb pointing in the $\vect{r}_{ab}$ direction.}
																	\label{fig:dihedral-atoms}
																\end{figure}
														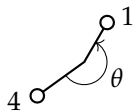
\begin{figure}[htbp!]
																	\centering
																	\tdplotsetmaincoords{0}{0}
																	\begin{tikzpicture}[tdplot_main_coords]
																		\draw[thick] (0,0,0)  -- (0,0,1) ;
																		\draw[thick,-o] (0,0,1) -- (0.353553,0.612372,1.707107) node[right]{1};
																		\draw[thick,-o] (0,0,0) -- (-0.700629,-0.509037,-0.866025) node[left]{4};
																		\tdplotdrawarc[->]{(0,0,0)}{0.3}{-145}{60}{anchor = west}{$\theta$};
																	\end{tikzpicture}
																	\caption{Top down view of \fig{fig:dihedral-atoms} with the dihedral angle (with the defined direction) marked as $\theta$.}
																	\label{fig:dihedral-angle}
																\end{figure}

We define the dihedral angles $\theta_{ij}$ and $\tau_{ij}$ in terms of Eq.~\eqref{eq:died-def} and, by means of this equation,
we can determine their transformation properties under the generating operations of \GTS.
						 We note that generally $E^*\axxtan(y,x) = \axxtan(-y,x) = 2\pi -  \axxtan(y,x).$
						 The dihedral angle $\theta_{23}$, for example, is defined by
						\begin{equation}
							\begin{split}
								\theta_{23} = \axxtan [& \ez \cdot ( \normvec{\ez \times \vect{r}_{3a}} \times \normvec{\ez \times \vect{r}_{2a}}),  \\
								&(\ez \times \normvec{\ez \times \vect{r}_{3a}}) \cdot (\ez \times \normvec{\ez \times \vect{r}_{2a}}) ]
							\end{split}
							\end{equation}
							Under (123) we obtain
							\begin{equation}
								\begin{split}
								\theta_{23}' =
									\axxtan [& \ez \cdot ( \normvec{\ez \times \vect{r}_{2a}} \times \normvec{\ez \times \vect{r}_{1a}}),  \\
									&(\ez \times \normvec{\ez \times \vect{r}_{2a}}) \cdot (\ez \times \normvec{\ez \times \vect{r}_{1a}}) ] = \theta_{12}.
								\end{split}
								\end{equation}
								To obtain the effect of (14)(26)(35)$(ab)^*$, we initially apply (14)(26)(35)$(ab)$ with the result
								\begin{equation}
									\begin{split}
										\axxtan [& -\ez \cdot ( \normvec{-\ez \times \vect{r}_{5b}} \times \normvec{-\ez \times \vect{r}_{6b}}),  \\
										&(-\ez \times \normvec{-\ez \times \vect{r}_{5b}}) \cdot (-\ez \times \normvec{-\ez \times \vect{r}_{6b}}) ].
									\end{split}
									\end{equation}
									Applying $E^*$  reverses $\ez$, and by swapping  $\vect{r}_{5b}$ and $\vect{r}_{6b}$ we obtain the final result:
									\begin{equation}
										\begin{split}
								\theta_{23}' =			\axxtan [& -\ez \cdot ( \normvec{-\ez \times \vect{r}_{6b}} \times \normvec{-\ez \times \vect{r}_{5b}}),  \\
											&(-\ez \times \normvec{-\ez \times \vect{r}_{6b}}) \cdot (-\ez \times \normvec{-\ez \times \vect{r}_{5b}}) ] = \theta_{56}.
										\end{split}
										\end{equation}
										where  the positive direction of the dihedral angle is in the sense of the proton numbering.
										
										Finally, applying (14)(25)(36)$(ab)$ to $\theta_{23}$ gives
										\begin{equation}
											\begin{split}
											\theta_{23}' =	\axxtan [& -\ez \cdot ( \normvec{-\ez \times \vect{r}_{6b}} \times \normvec{-\ez \times \vect{r}_{5b}}),  \\
												&(-\ez \times \normvec{\ez \times \vect{r}_{6b}}) \cdot (-\ez \times \normvec{-\ez \times \vect{r}_{5b}}) ] = \theta_{56}.
											\end{split}
											\end{equation}
											For $\tau_{41}$, which is defined by going counterclockwise from 4 to 1, the equation is
											\begin{equation}
												\begin{split}
													\axxtan [& \ez \cdot ( \normvec{\ez \times \vect{r}_{1a}} \times \normvec{\ez \times \vect{r}_{4b}}),  \\
													&(\ez \times \normvec{\ez \times \vect{r}_{1a}}) \cdot (\ez \times \normvec{\ez \times \vect{r}_{4b}}) ] = \tau_{41}.
												\end{split}
												\end{equation}
												Under operation (123)(456), this becomes
												\begin{equation}
													\begin{split}
												\tau_{41}' =		\axxtan [& \ez \cdot ( \normvec{\ez \times \vect{r}_{3a}} \times \normvec{\ez \times \vect{r}_{6b}}),  \\
														&(\ez \times \normvec{\ez \times \vect{r}_{3a}}) \cdot (\ez \times \normvec{\ez \times \vect{r}_{6b}}) ] = \tau_{63}
													\end{split}
													\end{equation}
													To determine the effect of (14)(26)(35)$(ab)^*$, we first apply (14)(26)(35)$(ab)$ to $\tau_{41}$
and obtain
													\begin{equation}
														\begin{split}
															\axxtan [& -\ez \cdot ( \normvec{-\ez \times \vect{r}_{4b}} \times \normvec{-\ez \times \vect{r}_{1a}}),  \\
															&(-\ez \times \normvec{-\ez \times \vect{r}_{4b}}) \cdot (-\ez \times \normvec{-\ez \times \vect{r}_{1a}}) ].
														\end{split}
														\end{equation}
														Applying $E^*$ reverses $\ez$. After swapping  the order of $\vect{r}_{1a}$ and $\vect{r}_{4b}$, we have
														\begin{equation}
															\begin{split}
															\tau_{41}' =	
																\axxtan [& -\ez \cdot ( \normvec{\ez \times \vect{r}_{1a}} \times \normvec{\ez \times \vect{r}_{4b}}),  \\
																&(\ez \times \normvec{\ez \times \vect{r}_{1a}}) \cdot (\ez \times \normvec{\ez \times \vect{r}_{4b}}) ] = 2\pi - \tau_{41}.
															\end{split}
															\end{equation}
															Finally, applying (14)(25)(36)$(ab)$ to $\tau_{41}$ gives
															\begin{equation}
																\begin{split}
																\tau_{41}' =
																	\axxtan [& -\ez \cdot ( \normvec{-\ez \times \vect{r}_{4b}} \times \normvec{-\ez \times \vect{r}_{1a}}),  \\
																	&(-\ez \times \normvec{-\ez \times \vect{r}_{4b}}) \cdot (-\ez \times \normvec{-\ez \times \vect{r}_{1a}}) ] = \tau_{41}.
																\end{split}
																\end{equation}

\section{Transformation of $\tau$ under the generating operations of \GTS}
\setcounter{equation}{0}
\label{sect:appenix_b}

The torsional coordinate $\tau$ is defined by
Eq.~(\ref{eq:taudef}):
\begin{equation}
\tau =
\frac13 ( \tau_{41} + \tau_{62} + \tau_{53}),
\end{equation}
and we will investigate how this coordinate transforms under the generating operations of \GTS, $R_2^{(+)} = (123)(456)$,  $R_2^{(-)} = (132)(456)$,  $R_4^{(+)} = (14)(26)(35)(ab)^*$, and $R_4^{(-)} = (14)(25)(36)(ab)$ used in the \Trove\ calculations. The transformation properties of the dihedral angles $\tau_{ij}$ are derived as outlined in \ref{sec:appendixB}.

The operation (123)(456) permutes the protons 1, 2, 3, 4, 5, 6 to the positions previously occupied by the protons labelled
3, 1, 2, 6, 4, 5, and the transformed value of $\tau$ is given by
\begin{equation}
\tau^\prime =
\frac13 ( \tau_{63} +\tau_{51} + \tau_{42})
\end{equation}
where
\begin{align}
\tau_{63} & =  \tau_{62} + \theta_{23} \\
	\tau_{51} &  = \tau_{53} + \theta_{31} \\
\tau_{42} & =  \tau_{41} + \theta_{12}. \\
\end{align}
For example, $\tau_{51}$ $=$  $\tau_{53}$ $+$ $\theta_{31}$,  the plus sign coming about because the positive direction of rotation for the $\tau_{ij}$ dihedral angles (proton 1 $\rightarrow$ 2 $\rightarrow$ 3) is the same as that of
$\theta_{12}$,  $\theta_{23}$, and  $\theta_{31}$.
Hence
\begin{align}
 \tau^\prime
 & = \frac13 \left(
 \tau_{41} + \tau_{62}  +
\tau_{53}
+ \left[ \theta_{12} + \theta_{23} + \theta_{31} \right] \right) \\
 & = \frac13 \left(
 \tau_{41} + \tau_{62} + \tau_{53}
 + 2\pi \right) = \tau + 2\pi/3
\end{align}
or, equivalently, $\tau^\prime$ $=$ $\tau - 4\pi/3$ as given in \tabl{tab:torsion_transformations}, to ensure that $(123)^3(456)^3 = E$.

After carrying out the operation (132)(456), the protons 1, 2, 3, 4, 5, 6 are found at the positions previously occupied by the protons labelled
2, 3, 1, 6, 4, 5, respectively. Consequently, the transformed value of $\tau$ is given by
\begin{equation}
\tau^\prime =
\frac13 ( \tau_{62} + \tau_{53} + \tau_{41})
= \tau.
\end{equation}

For $(14)(26)(35)(ab)^*$,
\begin{align}
\tau^\prime & =
\frac13 ( \tau_{14} + \tau_{26} + \tau_{35}) \\
 & =
\frac13 ( -\tau_{41} - \tau_{62} - \tau_{53}) = -\tau
\end{align}
or, equivalently, $2\pi - \tau$ as given in \tabl{tab:torsion_transformations}.

Finally, for (14)(25)(36)$(ab)$
\begin{align}
\tau^\prime & =
\frac13 ( -\tau_{14}  -\tau_{26} - \tau_{35}) \\
 & =
\frac13 ( \tau_{41} + \tau_{62} + \tau_{53}) = \tau.
\end{align}
The transformation properties derived here are summarized in
\tabl{tab:torsion_transformations}.

\bibliography{C2H6symm}

\end{document}